\documentclass[11pt, oneside]{article}   	% use "amsart" instead of "article" for AMSLaTeX format

\usepackage[margin=1in]{geometry}

\usepackage{graphicx}				% Use pdf, png, jpg, or eps§ with pdflatex; use eps in DVI mode
\usepackage{setspace}

\usepackage{float}

\usepackage[toc,page]{appendix}

\usepackage{amssymb}
\usepackage{bbm}

\def\bu{\bm{u}}
\def\bv{\bm{v}}
\def\bw{\bm{w}}

\def\ba{\bm{a}}
\def\bb{\bm{b}}

\def\bd{\bm{d}}

\def\one{\bm{1}}

\usepackage{mathrsfs}
\usepackage[utf8]{inputenc} % allow utf-8 input
\usepackage[T1]{fontenc}    % use 8-bit T1 fonts
\usepackage{hyperref}       % hyperlinks
\usepackage{url}            % simple URL typesetting
\usepackage{booktabs}       % professional-quality tables
\usepackage{amsfonts}       % blackboard math symbols
\usepackage{nicefrac}       % compact symbols for 1/2, etc.

\usepackage[utf8]{inputenc}
\usepackage[english]{babel}
 
\usepackage{amsmath,amsfonts,amssymb,amsthm,epsfig,epstopdf,titling,url,array,bm}

\theoremstyle{plain}

\theoremstyle{definition}

\theoremstyle{remark}

\usepackage{algorithmic,algorithm}

\usepackage{comment}

\usepackage{amsmath}
\usepackage{amsfonts}
\usepackage{amsmath,graphicx,centernot}

\usepackage{amssymb}
\usepackage{amsfonts,amssymb}
\usepackage[usenames, dvipsnames]{color}

\usepackage{mathrsfs}

\usepackage{color}
\usepackage[usenames,dvipsnames,svgnames,table]{xcolor}

\usepackage{natbib}
\usepackage{gensymb}

\hypersetup{
     colorlinks   = true,
     citecolor    = black,
     allcolors = black
}

\usepackage{csquotes}

\usepackage{authblk}

\usepackage{indentfirst}

\providecommand{\keywords}[1]
{
  \textbf{Keywords:} #1
}
\title{Sparse Logistic Tensor Decomposition for Binary Data}

\author{Jianhao Zhang\quad }
\author{\quad Yoonkyung Lee}
\affil{Department of Statistics,  The Ohio State University}

\date{}							% Activate to display a given date or no date

\begin{document}

\maketitle
%\tableofcontents

\begin{abstract}

 Tensor data are increasingly available in many application domains.
 We develop several tensor decomposition methods for binary tensor
 data. Different from classical tensor decompositions for continuous-valued 
data with squared error loss, we formulate logistic tensor
decompositions for binary data with a Bernoulli likelihood.
To enhance the interpretability of estimated factors and improve
their stability further, we propose sparse formulations of logistic tensor
decomposition by considering $\ell_{1}$-norm and $\ell_{0}$-norm
regularized likelihood. To handle the resulting optimization problems, we develop
computational algorithms which combine the strengths of tensor power
method and majorization-minimization (MM) algorithm.
Through simulation studies, we demonstrate the utility of our methods in
analysis of binary tensor data. To illustrate the effectiveness of the
proposed methods, we analyze a  dataset concerning nations and their
political relations and perform co-clustering of estimated factors
to find associations between the nations and political relations.

\end{abstract}


\section{Introduction}

As a natural generalization of vectors and matrices, tensors have
appeared frequently as a data form in many fields including social
networks \citep{nickel2011three}, recommender systems
\citep{bi2018multilayer} and genomics \citep{wang2019three}. As a
result, tensor decomposition has attracted interests from machine
learning and statistics with applications in chemometrics
\citep{bro1997parafac}, computer vision and signal processing. See
\cite{kolda2009tensor} for a comprehensive review. 
In general, there are two approaches to decomposition of tensor data: CP decomposition and Tucker decomposition. 
CP decomposition is the abbreviation of canonical decomposition
(CANDECOMP) and parallel factor (PARAFAC) analysis, which were proposed
independently in psychometrics by \cite{harshman1970foundations}
and \cite{carroll1970analysis}.  
Tucker decomposition
\citep{tucker1966some} is a generalization of singular value decomposition for higher-order data, which includes CP decomposition as a special case.

There has been a growing body of literature that
extends tensor decomposition methods for real valued data to
other types such as tensors with binary outcomes or counts
 for dimensionality reduction and latent factor modeling.
As a closely related problem, various types of principal component analysis (PCA) and
matrix factorization methods have been developed for discrete non-Gaussian matrix
data
\citep{collins2002generalization, deLeeuw2006, udell2016generalized,
  landgraf2020generalized, landgraf2020dimensionality}.
Extending the matrix factorization approach in
\cite{collins2002generalization} to binary tensor data, \cite{mavzgut2014dimensionality}
considered a Tucker decomposition of the logit parameter tensor,
and \cite{wang2020learning} considered a CP decomposition of the logit parameter tensor
with max-norm constraint and investigated its statistical optimality.
More generally, \cite{hong2020generalized} proposed a CP decomposition
of the natural parameter tensor for exponential family data.

In this paper, we focus on binary tensor data and consider settings
where sparse latent factors are desired for modeling the underlying
logit parameter tensor.
Taking collaborative filtering
as an example,  how users interact with items in
different contexts can be organized in the form of a tensor
with user, item and context as three modes. Presence or absence
of a user's interaction with an item in each context then makes up a
binary tensor. Relations between the users and items could be
context-specific, and they may involve a small subset of users, items,
or contexts, rendering such relational factors sparse.
%%%
Benefits of sparse factors and principal components for high dimensional
matrix data have been well understood.
Similar to sparse PCA \citep{jolliffe2003modified, zou2006sparse,
  shen2008sparse}, sparsity or regularization of factors is often desired in
tensor decompositions. Sparsity in estimated factor matrices can
provide a concise description of the latent structure and improved understanding of
the latent factors in relation to the observable features.  For real valued tensors, 
\cite{allen2012sparse} and \cite{sun2017provable} proposed sparse
tensor decomposition methods based on the  CP decomposition with an
$\ell_{1}$-norm penalty and $\ell_{0}$-norm constraint on factor
matrices, respectively. Besides,  \cite{madrid2017tensor} considered
sparse tensor decomposition with generalized lasso penalties on factor
matrices to obtain smoothly varying factors.
\cite{zhang2019optimal} proposed a 
sparse tensor singular value decomposition based on the Tucker
decomposition and studied its statistical optimality.

To handle binary tensor data efficiently, we combine
dimensionality reduction with regularization and selection of features
and consider sparse decomposition of a logit parameter tensor.
We propose a formulation of sparse logistic tensor decomposition by
imposing an $\ell_{1}$-norm penalty or $\ell_{0}$-norm constraint on
the factor matrices in the CP decomposition of a centered logit
parameter tensor.
Our approach naturally extends the sparse tensor decomposition
\citep{allen2012sparse, sun2017provable} to binary data and also
extends the sparse logistic PCA \citep{lee2010sparse, lee2014biclustering} to higher-order data.

Rank-one components in the decomposition of the underlying logit tensor
generally correspond to multiplicative interactions
among different modes. For this reason, sparse factors are
well suited for modeling more local patterns of interactions involving only a subset
of features along each mode. Such patterns can reveal interesting
co-clustering structures between different modes. For binary matrices,
\cite{lee2014biclustering} demonstrated the idea of co-clustering
with a biclustering algorithm. Recently, \cite{li2020generalized}
proposed a more general co-clustering analysis framework for
exponential family tensor data.

Computationally, binary tensor decomposition entails maximization
of the likelihood of a logit parameter tensor under a Bernoulli distribution assumption 
on the binary entries.
Incorporating an $\ell_{1}$-norm penalty or $\ell_{0}$-norm constraint on
the factors in the decomposition for encouraging sparsity leads to regularization of the likelihood.
To solve the resulting optimization problems, we develop several
novel computational algorithms.
In a nutshell, we combine the strengths of tensor power method for
tensor decompositions and majorization-minimization (MM) algorithms, 
which have been successfully applied in logistic PCA and exponential
family PCA for matrix data. By majorizing the negative log likelihood with a
quadratic function, we turn the logistic CP decomposition problems
with binary data into iterative applications of a plain CP decomposition with 
real-valued data. Thereby, we could make use of the 
tensor power method and its adaptations to sparse tensor
decompositions for analysis of binary data.
In particular, we adopt the tensor power method with alternating
rank-one updates \citep{anandkumar2014guaranteed} in the MM approach
to logistic tensor decomposition. Further, we incorporate
the truncated power method \citep{yuan2013truncated, sun2017provable}
for the $\ell_{0}$-norm constrained logistic tensor decomposition
and the soft-thresholding power method \citep{witten2009penalized, allen2012sparse} 
for the $\ell_{1}$-norm penalized logistic tensor decomposition.
We illustrate the utility of the proposed algorithms for analysis of
binary tensor data.

%%%%%
The rest of the paper is organized as follows.
Section \ref{sec:prelim} reviews tensor decomposition for real-valued
tensor data.  In Section \ref{sec:lcpd}, we introduce logistic CP decomposition for binary data
using a Bernoulli likelihood and present sparse  logistic CP
decomposition using a regularized likelihood in Section
\ref{sec:slcpd}. In addition, Sections \ref{sec:lcpd} and
\ref{sec:slcpd}  include MM-based computational
algorithms for logistic tensor decomposition and sparse counterpart,  respectively.
Section \ref{sec:missing} regards an extension of logistic CP decomposition for
handling missing data and tensor completion.
In Section \ref{sec:tuning},  we discuss several criteria for choosing the rank of
tensor decomposition and tuning parameters. 
We present simulation studies in Section \ref{sec:simulate}  and demonstrate the
effectiveness of sparse logistic CP decomposition with  an application to nations data in
Section \ref{sec:nations}.  We summarize our contributions and list several directions for further
investigation in Section \ref{sec:conclusion}.

\section{Preliminaries} \label{sec:prelim}

This section provides a technical background of tensor
decomposition. Throughout the paper we focus on third-order tensor data,
which are common in many applications. Methods for higher-order tensors can be developed similarly.

\subsection{Notation}

For $p\in \mathbb{N}$, we use $[p]$ to denote the index set $\{1,\dots, p\}$.
For two tensors $\mathcal{X}$ and $\mathcal{Y}\in \mathbb{R}^{p_{1}\times p_{2} \times p_{3}}$, the inner product of $\mathcal{X}$ and $\mathcal{Y}$ is defined as 
$\langle \mathcal{X},\mathcal{Y}\rangle= \sum_{\omega\in [p_{1}]\times [p_{2}] \times [p_{3}]}\mathcal{X}(\omega)\mathcal{Y}(\omega)$. This induces the Frobenius norm of $\mathcal{X}$ as $\|\mathcal{X}\|_{F}=\sqrt{\langle \mathcal{X},\mathcal{X}\rangle}$, similar to the Frobenius norm of a matrix.

A tensor can be transformed into a matrix or \textit{matricized} by unfolding it in a given mode.  
The mode-$n$ matricization of a tensor $\mathcal{X}\in \mathbb{R}^{p_{1}\times p_{2} \times p_{3}}$ is denoted by $X_{(n)}$ for $n\in [3]$. For example, $X_{(1)}\in \mathbb{R}^{p_{1}\times p_{-1}}$ with $p_{-1}=p_{2}p_{3}$ is a matrix whose columns are the mode-$1$ fibers of $\mathcal{X}$. Multiplication of a tensor by a matrix in mode $n$ is called the mode-$n$ matrix product and denoted by $\times_n$. For example, 
the mode-$1$ matrix product of a tensor $\mathcal{X}\in \mathbb{R}^{p_{1}\times p_{2} \times p_{3}}$ and a matrix $U\in \mathbb{R}^{q\times  p_{1}}$ is denoted by $\mathcal{X}\times_{1}U\in \mathbb{R}^{q\times p_{2} \times p_{3}}$.

The outer product of vectors $\ba\in\mathbb{R}^{I}$ and $\bb\in\mathbb{R}^{J}$ is denoted by $\ba
\circ \bb\in\mathbb{R}^{I\times J}$.
For a vector $\bv\in\mathbb{R}^{p}$, $\|\bv\|_{2}$ refers to the Euclidean norm, $\|\bv\|_{1}$ the $\ell_{1}$-norm,  and $\|\bv\|_{0}$ the number of non-zero entries in $\bv$.
The Kronecker product of matrices $A\in\mathbb{R}^{I\times J}$ and $B\in\mathbb{R}^{K\times L}$ is denoted by $A\otimes B\in\mathbb{R}^{IK\times JL}$ with $a_{ij}B$ in the $ij$th block.
The Khatri-Rao product of matrices $A\in\mathbb{R}^{I\times K}$ and $B\in\mathbb{R}^{J\times K}$ is the columnwise Kronecker product of $A$ and $B$ denoted by $A\odot B=[\ba_1\otimes \bb_1 ~\ba_2\otimes \bb_2 ~\ldots~ \ba_K\otimes \bb_K] \in\mathbb{R}^{IJ\times K}$.
The Hadamard product of matrices $A\in\mathbb{R}^{I\times J}$ and $B\in\mathbb{R}^{I\times J}$ is the elementwise product of $A$ and $B$ denoted by $A* B=[a_{ij}b_{ij}]\in\mathbb{R}^{I\times J}$. The Hadamard product of two tensors can be defined analogously.

The following property of the Kronecker product will be useful. Let $\mathcal{X}\in \mathbb{R}^{p_{1}\times p_{2}\cdots \times p_{N}}$ and $A^{(n)}\in \mathbb{R}^{q_{n}\times p_{n}}$ for each $n\in[N]$. Then 
$\mathcal{Y}=\mathcal{X}\times_{1}A^{(1)}\times_{2}A^{(2)}\cdots\times_{N}A^{(N)}$ is equivalent to 
$Y_{(n)}=A^{(n)}X_{(n)}(A^{(N)}\otimes\cdots A^{(n+1)}\otimes A^{(n-1)}\cdots A^{(1)})^{T}$ for every $n\in[N]$.

% \section{Previous Work}
\subsection{Tensor Decomposition}

We briefly review tensor decomposition for real-valued tensor data. The idea of a CP decomposition \citep{carroll1970analysis,  harshman1970foundations}  is to factorize a tensor $\mathcal{X}\in\mathbb{R}^{p_{1}\times p_{2}\times p_{3}}$ into a sum of rank-one component tensors of the form:
$$\mathcal{X}\approx\sum_{r\in[R]}d_{r}\cdot \bu_{r}\circ \bv_{r}\circ \bw_{r},$$
where $\bu_{r}\in\mathbb{R}^{p_{1}}$, $\bv_{r}\in\mathbb{R}^{p_{2}}$, $\bw_{r}\in\mathbb{R}^{p_{3}}$, $d_{\max}=d_{1}\geq \cdots\geq d_{R}=d_{\min}>0$, and $\bu_{r}^{T}\bu_{r}=1$, $\bv_{r}^{T}\bv_{r}=1$, $\bw_{r}^{T}\bw_{r}=1$ for $r\in[R]$. Here $R$ is the rank of the tensor $\mathcal{X}$. This can be formulated as a minimization problem with squared error loss:
$$\underset{\bd, U, V, W}{\min}\quad\|\mathcal{X}-\sum_{r\in[R]}d_{r}\cdot \bu_{r}\circ \bv_{r}\circ \bw_{r}\|_{F}^{2},$$
where $\bd=(d_{1},\dots, d_{R})^{T}\in\mathbb{R}^{R}$ is a vector of
weight parameters, and $U=[\bu_{1},\dots,
\bu_{R}]=[u_{ir}]\in\mathbb{R}^{p_{1}\times R}$, $V=[\bv_{1},\dots,
\bv_{R}]=[v_{jr}]\in\mathbb{R}^{p_{2}\times R}$, and
$W=[\bw_{1},\dots, \bw_{R}]=[w_{kr}]\in\mathbb{R}^{p_{3}\times R}$ are
the factor matrices. It's worth noting that this CP decomposition has the property of essential uniqueness. That is, the columns of $U, V$ and $W$ are determined up to joint permutation.

\cite{kruskal1977three, kruskal1989rank} provided a sufficient
condition for the uniqueness of a three-way CP decomposition up to permutation and rescaling of rank-one tensors. 
Kruskal's condition is $$k_{U}+k_{V}+k_{W}\geq 2R+2,$$
where $k_{U}, k_{V}$ and $k_{W}$ are the Kruskal ranks of the matrices $U, V$  and $W$.

The Tucker decomposition \citep{tucker1966some} aims at approximating a tensor $\mathcal{X}\in\mathbb{R}^{p_{1}\times p_{2}\times p_{3}}$ with a reduced core tensor $\mathcal{S}\in\mathbb{R}^{R_{1}\times R_{2} \times R_{3}}$ and factor matrices $U\in\mathbb{R}^{p_{1}\times R_{1}}$, $V\in\mathbb{R}^{p_{2}\times R_{2}}$, and  $W\in\mathbb{R}^{p_{3}\times R_{3}}$ as follows:
$$\mathcal{X}\approx\mathcal{S} \times_{1} U\times_{2} V\times_{3} W.$$
Again this can be formulated as an optimization problem with squared error loss:
$$\underset{\mathcal{S}, U, V, W}{\min}\quad \|\mathcal{X}-\mathcal{S} \times_{1} U\times_{2} V\times_{3} W\|_{F}^{2},$$
where $U^{T}U=I_{R_{1}}$, $V^{T}V=I_{R_{2}}$, and $W^{T}W=I_{R_{3}}$. Here $R_{1},R_{2}$ and $R_{3}$ are the numbers of components in the factor matrices $U, V$ and  $W$, respectively. The CP decomposition can be viewed as a special case of the Tucker decomposition when the core tensor $\mathcal{S}$ is super-diagonal and $R_{1}=R_{2}=R_{3}=R$. However, the Tucker decomposition doesn't have uniqueness since we can multiply factor matrices by nonsingular matrices and define a new core tensor and new factor matrices.

A comprehensive review of tensor decomposition is available in \cite{kolda2009tensor}.
In this paper, we focus on the CP decomposition for tensors as it is a more natural choice for defining latent factors, and it is also more amenable to computation. 
The use of squared error loss can be regarded as an implicit normal distribution assumption for real-valued tensor data. We will employ an alternative loss for binary tensor data.

\section{CP Decomposition for Binary Data} \label{sec:lcpd}

\subsection{Logistic CP Decomposition}

To handle tensors with dichotomous outcomes in many applications, we consider a binary tensor $\mathcal{X}=(x_{ijk})\in\{0,1\}^{p_{1}\times p_{2} \times p_{3}}$, where each entry $x_{ijk}$ encodes one of the two types of outcomes (e.g., absence or presence)  with 0 or 1. We posit a generative model for the tensor and assume that $x_{ijk}$ are realizations of mutually independent  Bernoulli random variables with probability $p_{ijk}$, or $x_{ijk}\sim \text{Bernoulli}(p_{ijk})$.

For a Bernoulli random variable with probability parameter $p$, the 
probability mass function is $P(X=x)=p^x(1-p)^{1-x}$, and $\log P(X=x) = x\log\frac{p}{1-p} +\log(1-p)$. Reparametrizing with the logit parameter $\theta := \log\frac{p}{1-p}$,  the log likelihood of $\theta$ based on $x$
is given by $\ell(x;\theta) = x\theta -\log(1+\exp(\theta))$.

Letting $\theta_{ijk}:=\log\frac{p_{ijk}}{1-p_{ijk}}$ for individual data entries $x_{ijk}$ in the binary tensor $\mathcal{X}$, we derive the log likelihood of
the logit parameter tensor $\Theta=(\theta_{ijk})\in\mathbb{R}^{p_{1}\times p_{2} \times p_{3}}$ as follows:
\begin{eqnarray*}
\ell(\mathcal{X};\Theta)
&=&\sum_{i,j,k} \big\{ x_{ijk}\theta_{ijk} -\log(1+\exp(\theta_{ijk})) \big\}\\
&=& \langle\mathcal{X},\Theta \rangle -\langle \one_{p_{1}p_{2}p_{3}}, \log(\one_{p_{1}p_{2}p_{3}}+\exp(\Theta)) \rangle,
\end{eqnarray*}
where $\one_{p_{1}p_{2}p_{3}}:=\one_{p_{1}}\circ \one_{p_{2}}\circ\one_{p_{3}}\in\mathbb{R}^{p_{1}\times p_{2} \times p_{3}}$ is the tensor with all entries equal to one, and $\log(\cdot)$ and $\exp(\cdot)$ are taken as element-wise operators with tensors.
Naturally extending the exponential family PCA for a data matrix in \cite{collins2002generalization} to a higher-order tensor, 
we consider a CP decomposition of the  logit  parameter tensor
$\Theta$ rather than the binary data tensor $\mathcal{X}$ itself and call it \textit{logistic tensor decomposition}.

We include an offset term $\mu\in\mathbb{R}$ in logistic CP decomposition taken as an overall logit parameter value and consider the following decomposition:
\begin{equation}\label{theta}
\Theta=\mu\one_{p_{1}p_{2}p_{3}}+\sum_{r\in[R]}d_{r}\cdot \bu_{r}\circ \bv_{r}\circ \bw_{r}
\end{equation}
or $\theta_{ijk}=\mu+\sum_{r\in[R]}d_{r}u_{ir}v_{jr}w_{kr}$,  where the multiplicative part has rank $R$. Standard logistic CP decomposition in the literature assumes $\mu=0$. 
For notational convenience, we use $\Theta_{c}$ to refer to $\sum_{r\in[R]}d_{r}\cdot \bu_{r}\circ \bv_{r}\circ \bw_{r}$, the portion of $\Theta$ adjusted by the offset.

To find a CP decomposition of $\Theta$ of rank $R$ given the binary tensor $\mathcal{X}$, we  maximize the log likelihood or equivalently 
minimize the negative log likelihood and formulate 
 a \textit{logistic CP decomposition} problem as follows:
\begin{equation}\label{lcpd}
  \begin{split}
\underset{\mu,\bd, U, V, W} {\min}\quad&-\langle\mathcal{X},\Theta \rangle +\langle \one_{p_{1}p_{2}p_{3}}, \log(\one_{p_{1}p_{2}p_{3}}+\exp(\Theta)) \rangle\\
\text{s.t.} 
\quad& \Theta=\mu\one_{p_{1}p_{2}p_{3}}+\sum_{r\in[R]}d_{r}\cdot \bu_{r}\circ \bv_{r}\circ \bw_{r},\\
\quad&  \bu_{r}^{T}\bu_{r}=1, \bv_{r}^{T}\bv_{r}=1,\bw_{r}^{T}\bw_{r}=1,
\mbox{ and } d_{r}>0 \mbox{ for } r\in[R],
\end{split}
\end{equation}
where $\bd \in \mathbb{R}^{r}$, $U \in\mathbb{R}^{p_{1}\times R}, V \in\mathbb{R}^{p_{2}\times R}$ and $W \in\mathbb{R}^{p_{3}\times R}$ are the weight vector and factor matrices as defined before.
While the objective function in  \eqref{lcpd},
$-\ell(\mathcal{X};\Theta)$, is convex in $\Theta$, it is not convex in the factor matrices jointly, and this leads to  
 a non-convex optimization problem with possibly multiple local optima.
Further, the objective function in \eqref{lcpd} is convex in each factor when the other two factors are fixed, but the unit norm constraint on each factor makes the problem non-convex.

 \subsection{Majorization-Minimization Approach}
 
For logistic PCA and exponential family PCA involving similar optimization problems, Majorization-Minimization (MM) algorithms \citep{hunter2004tutorial} have been used successfully. See  \cite{deLeeuw2006, lee2010sparse, lee2014biclustering, landgraf2020generalized} for example.
To solve the logistic CP decomposition problem in \eqref{lcpd}, 
we propose to majorize the objective function with a quadratic loss function and apply state-of-the-art algorithms for CP decomposition iteratively.

To majorize the negative likelihood, we first rewrite
$P(X=x)=p^x(1-p)^{1-x}=\sigma(q\theta)$ with the sigmoid function
$\sigma(x)=\text{logit}^{-1}(x)=\{1+\exp(-x)\}^{-1}$ and $q=2x-1$, and use the following tight and uniform quadratic majorization of $-\log\sigma(x) $
from \cite{jaakkola2000bayesian, deLeeuw2006}:
\begin{equation*}
\begin{split}
-\log\sigma(x) &\leq -\log\sigma(y)+(\sigma(y)-1)(x-y)+\frac{2\sigma(y)-1}{4y}(x-y)^{2}\\
&\leq -\log\sigma(y)+(\sigma(y)-1)(x-y)+\frac{1}{8}(x-y)^{2},\end{split}
\end{equation*}
where the equalities hold when $x=y$. We will focus on the uniform
bound (the second inequality) for computational convenience and leave
the tight bound (the first inequality) for future study.

Let $\Theta^{[m]}$ be the estimate of $\Theta$ obtained in the $m$th step of the MM  algorithm. Then, applying the above majorization to $-\log\sigma(q_{ijk}\theta_{ijk})$ at $\theta_{ijk}^{[m]}$ with $q_{ijk}=2x_{ijk}-1$ and completing the square,  we have
\[
  -\ell(\mathcal{X};\Theta) =-\sum_{i,j,k}\log\sigma(q_{ijk}\theta_{ijk})
  \leq -\sum_{i,j,k}\log\sigma(q_{ijk}\theta_{ijk}^{[m]})
  +\frac{1}{8} \sum_{i,j,k} (\theta_{ijk}-z_{ijk}^{[m]})^{2},
  \]
where 
$z_{ijk}^{[m]}=\theta_{ijk}^{[m]}+4(x_{ijk}-\sigma(\theta_{ijk}^{[m]}))$
or in the form of tensor
\begin{equation}
\label{newx}
 \mathcal{Z}^{[m]}
=\Theta^{[m]}+4(\mathcal{X}-\sigma(\Theta^{[m]})).
\end{equation}

In other words,  up to a constant depending on  $\Theta^{[m]}$, the function $\frac{1}{8}\|\mathcal{Z}^{[m]}-\Theta\|_{F}^{2}$ majorizes $-\ell(\mathcal{X};\Theta)$ at $\Theta^{[m]}$, and this turns our problem into
a simple CP decomposition problem with $\mathcal{Z}^{[m]}$ in the next iteration:
\begin{equation}\label{mm1}
\begin{split}
  \underset{\mu,\bd, U, V, W}{\min}
  &\quad \frac{1}{8}\|\mathcal{Z}^{[m]}-\Theta\|_{F}^{2}\\
\text{s.t.} 
&\quad \Theta=\mu\one_{p_{1}p_{2}p_{3}}+\sum_{r\in[R]}d_{r}\cdot \bu_{r}\circ \bv_{r}\circ \bw_{r},\\
& \quad \bu_{r}^{T}\bu_{r}=1, \bv_{r}^{T}\bv_{r}=1,\bw_{r}^{T}\bw_{r}=1,\mbox{ and }d_{r}>0 \mbox{ for } r\in[R].
\end{split}
\end{equation}
As the number of iterations grows, $-\ell(\mathcal{X};\Theta^{[m]})$ decreases and it converges to a local minimum of $-\ell(\mathcal{X};\Theta)$, the loss function for logistic tensor decomposition as $m\rightarrow \infty$.
 
There exist various approaches for a CP decomposition of real-valued tensors. Among those, we will consider the Alternating Least Squares (ALS) method \citep{harshman1970foundations, carroll1970analysis} and the Tensor Power (TP) method \citep{allen2012sparse} to solve the CP decomposition problem \eqref{mm1} in each iteration.

\subsubsection{Alternating Least Squares Method}

The ALS approach optimizes one factor matrix while treating all the other factor matrices as constants and alternates this optimization procedure over each of the factor matrices repeatedly until some convergence criterion is satisfied.

To solve the CP decomposition problem \eqref{mm1} at the $m$th step, we update parameters $(\mu,\bd, U, V, W)$ in a block coordinate-wise manner. Given $(\bd, U, V, W)$ at step $m$, we compute 
$\Theta_{c}^{[m]}=\sum_{r\in[R]}\widehat{d}^{[m]}_{r}\cdot \widehat{\bu}^{[m]}_{r}\circ \widehat{\bv}^{[m]}_{r}\circ \widehat{\bw}^{[m]}_{r}$ first and update $\mu$ by taking the average of $\mathcal{Z}^{[m]}-\Theta_{c}^{[m]}$. With this updated $\widehat{\mu}^{[m+1]}$, we define 
\begin{equation}\label{newZc}
\mathcal{Z}^{[m]}_{c}=\mathcal{Z}^{[m]}-\widehat{\mu}^{[m+1]}\one_{p_{1}p_{2}p_{3}}
\end{equation}
as offset adjusted working variables. To update $\bd, U, V$, and  $W$, we minimize 
\begin{equation}
\label{newmm}
\|\mathcal{Z}_{c}^{[m]}-\sum_{r\in[R]}d_{r}\cdot \bu_{r}\circ \bv_{r}\circ \bw_{r}\|_{F}^{2}.
\end{equation}

Given factor matrices $V$ and $W$, we define $A:=U\text{diag}(\bd)$, and rewrite the above problem in matrix form as a linear least squares problem for $A$ as follows:
\begin{equation*}
\underset{A}{\min} \quad \|Z_{c(1)}^{[m]}-A(W\odot V)^{T}\|_{F}^{2}.
\end{equation*}
Then it can be shown that  
$$\widehat{A}=Z_{c(1)}^{[m]}[(W\odot V)^{T}]^{\dagger},$$ where
$[(W\odot V)^{T}]^{\dagger}$ indicates the Moore-Penrose
pseudo-inverse of $(W\odot V)^{T}\in\mathbb{R}^{R\times p_{2}p_{3}}$
\citep{golub1996matrix}. To avoid the pseudo-inverse of a large
matrix, we can rewrite the above solution
as $$\widehat{A}=Z_{c(1)}^{[m]}[(W\odot
V)^{T}]^{\dagger}=Z_{c(1)}^{[m]}(W\odot V)(W^{T}W*
V^{T}V)^{\dagger},$$  where $W^{T}W*V^{T}V\in\mathbb{R}^{R\times R}$
is typically smaller. %but sometimes this version is not suggested due to numerical ill-conditioning.
To obtain $\widehat{U}$ and $\widehat{\bd}$, we let $\widehat{d}_{r}=\|\widehat{\ba}_{r}\|_{2}$ and $\widehat{\bu}_{r}=\widehat{\ba}_{r}/\|\widehat{\ba}_{r}\|_{2}$ for $r\in[R]$.
Similarly, we could define $B:=V\text{diag}(\bd)$ and $C:=W\text{diag}(\bd)$ and obtain
$$\widehat{B}=Z_{c(2)}^{[m]}[(U\odot W)^{T}]^{\dagger}=Z_{c(2)}^{[m]}(U\odot W)(U^{T}U*W^{T}W)^{\dagger}$$ and
$$\widehat{C}=Z_{c(3)}^{[m]}[(U\odot V)^{T}]^{\dagger}=Z_{c(3)}^{[m]}(U\odot V)(U^{T}U*V^{T}V)^{\dagger}$$ given other factor matrices.
We normalize each column of $\widehat{A}, \widehat{B},$ and $\widehat{C}$ to unit length and update $\widehat{U}, \widehat{V},$ and $\widehat{W}$ alternatingly. Here we use $\text{Normalize}(U)$ to denote the matrix with normalized columns of matrix $U$.

This MM approach with the ALS method (MM-ALS) is summarized in
Algorithm \ref{mmals} of Appendix \ref{app:algorithm}. While this approach is easy to implement, it may take many iterations to converge, and there is no guarantee for convergence to a global minimum or even a stationary point of problem \eqref{mm1} according to \cite{kolda2009tensor}.
%\red{[Is there a better/more direct reference for this fact?]}
Moreover, inversions of $R\times R$ matrices appear many times in Algorithm \ref{mmals}, so the ALS method will be computational expensive for large rank $R$. The results of MM-ALS method may contain local optima. We make comparisons with other methods in a simulation study later.

\subsubsection{Tensor Power Method with Clustering}

As a related approach to ALS, we consider iterative rank-one approximations of $\mathcal{Z}_{c}^{[m]}$ known as the tensor power method \citep{allen2012sparse} to solve the rank-$R$ tensor decomposition problem \eqref{newmm}. 
It is related to the power method for eigendecomposition \citep{golub1996matrix}. 
For a rank-one problem, the logit parameter tensor has representation of 
\begin{equation}\label{newtheta}
\Theta=\mu\one_{p_{1}p_{2}p_{3}}+d\cdot \bu\circ \bv\circ \bw.
\end{equation}
Focusing on rank-one vectors, we aim to solve the following approximation problem:
\begin{equation}\label{md}
 \begin{split}
 \underset{d, \bu, \bv, \bw}{\min}\quad& \|\mathcal{Z}_{c}^{[m]}-d\cdot \bu\circ \bv\circ \bw\|_{F}^{2}\\
 \text{s.t.} 
% &\; \Theta=\mu\one_{p_{1}p_{2}p_{3}}+d\cdot \bu\circ \bv\circ \bw,\\
\quad  & \bu^{T}\bu=1, \bv^{T}\bv=1,\bw^{T}\bw=1, \mbox{ and }d>0.\\
\end{split}
\end{equation}

Given $\bu\in\mathbb{R}^{p_{1}}, \bv\in\mathbb{R}^{p_{2}}$ and $\bw\in\mathbb{R}^{p_{3}}$, the minimizer $d\in\mathbb{R}$ is analytically identified as 
$d=\mathcal{Z}_{c}^{[m]}\times_{1}\bu^{T}\times_{2}\bv^{T}\times_{3}\bw^{T}$, and this  allows us to rewrite the squared error objective function as
\begin{equation*}
\|\mathcal{Z}_{c}^{[m]}-d\cdot \bu\circ \bv\circ \bw\|_{F}^{2}
=\|\mathcal{Z}_{c}^{[m]}\|_{F}^{2}-\|\mathcal{Z}_{c}^{[m]}\times_{1}\bu^{T}\times_{2}\bv^{T}\times_{3}\bw^{T}\|_{F}^{2}.
\end{equation*}
Then we can recast the above problem \eqref{md} as
\begin{equation*}
\begin{split}
\underset{\bu, \bv, \bw}{\max}\quad&\mathcal{Z}_{c}^{[m]}\times_{1}\bu^{T}\times_{2}\bv^{T}\times_{3}\bw^{T}\\
\text{s.t.} \quad&\bu^{T}\bu=1, \bv^{T}\bv=1, \bw^{T}\bw=1.\\
\end{split}
\end{equation*}
See \citet{kolda2009tensor} for reference. Given $\bv$ and $\bw$, \eqref{md} can be rewritten as the following subproblem for $\bu$:
\begin{equation*}
\begin{split}
\underset{\bu}{\max}\quad& \bu^{T}(\mathcal{Z}_{c}^{[m]}\times_{2}\bv^{T}\times_{3}\bw^{T})\\
\text{s.t.} \quad& \bu^{T}\bu=1.\\
\end{split}
\end{equation*}
\cite{allen2012sparse} showed that the above problem has explicit solution of 
\begin{equation}
\begin{split}\label{u1}
\widehat{\bu}=\text{Normalize}(\mathcal{Z}_{c}^{[m]}\times_{2}\bv^{T}\times_{3}\bw^{T}),
\end{split}
\end{equation} which satisfies the Karush–Kuhn–Tucker (KKT) conditions \citep{boyd2004convex}.

Again alternating among the three factors, we can update one factor at a time given the other factors to maximize the objective function. For instance, we first update $\bu$ given $\bv$ and $\bw$, and then update $\bv$ and $\bw$ respectively in a similar way.
Each subproblem in matricized form can be solved explicitly, and the solutions are given by
\begin{equation*}
\begin{split}
\widehat{\bu}&=\text{Normalize}(\mathcal{Z}_{c}^{[m]}\times_{2}\bv^{T}\times_{3}\bw^{T}),\\
\widehat{\bv}&=\text{Normalize}(\mathcal{Z}_{c}^{[m]}\times_{1}\bu^{T}\times_{3}\bw^{T}),\\
\widehat{\bw}&=\text{Normalize}(\mathcal{Z}_{c}^{[m]}\times_{1}\bu^{T}\times_{2}\bv^{T}).
\end{split}
\end{equation*}
Then $d$ is updated with $\widehat{d}= \mathcal{Z}_{c}^{[m]}\times_{1}\widehat{\bu}^{T}\times_{2}\widehat{\bv}^{T}\times_{3}\widehat{\bw}^{T}$.  Finally,  the logit parameter tensor  is updated with
  $\Theta^{[m+1]}=\widehat{\mu}^{[m+1]}\one_{p_{1}p_{2}p_{3}}+\widehat{d} \cdot \widehat{\bu}\circ \widehat{\bv}\circ \widehat{\bw}$, which then gives $\mathcal{Z}^{[m+1]}$ in  \eqref{newx}.

For rank $R$ component tensors, we repeat this rank-one decomposition multiple times with different initializations and produce $R$ rank-one component tensors to combine.
By repeating the tensor power method with differential initializations
for $L$ times, we can obtain $L$ tuples stored in
$S=\{(\widehat{d}_{\tau},\widehat{\bu}_{\tau},\widehat{\bv}_{\tau},\widehat{\bw}_{\tau}),\tau\in[L]\}$.
Focusing on large estimates of $\widehat{d}_{\tau}$ and removing too
similar tuples, we further cluster those $L$ tuples into $R$ clusters
to produce $R$ distinct rank-one component tensors. Then we reorder the $R$ components with decreasing magnitude of $\widehat{d}_{j}$, which are the final output of our algorithm. 
We summarize this MM approximation with Tensor Power method (MM-TP) in
Algorithm \ref{mm}.

The Tensor Power method (TP) can be viewed as a rank-one version of ALS. It only updates one column of each factor matrix in each iteration and does not require matrix inversion, which greatly reduces computational complexity compared to ALS \citep{anandkumar2014guaranteed}. Also,  as a greedy method, the first few estimated factors by the tensor power method typically explain more deviance than ALS method \citep{allen2012sparse}.

Initialization is important for non-convex problems. In order to avoid
local optima, Algorithm \ref{mm} contains a loop running for $L$
different initializations. Because good initial values are not known
in advance, we need to identify them. As suggested in
\cite{anandkumar2014guaranteed}, we develop an algorithm which
clusters $L$ tuples in
$S=\{(\widehat{d}_{\tau},\widehat{\bu}_{\tau},\widehat{\bv}_{\tau},\widehat{\bw}_{\tau}),\tau\in[L]\}$
into $R$ clusters
$\{(\widehat{d}_{j},\widehat{\bu}_{j},\widehat{\bv}_{j},\widehat{\bw}_{j}),
j\in[R]\}$ to obtain the final estimates. This clustering algorithm is
described in Algorithm \ref{clust}. Defining 
$\widehat{\Theta}_{c}=\sum_{j\in[R]}\widehat{d}_{j}\cdot
\widehat{\bu}_{j}\circ \widehat{\bv}_{j}\circ \widehat{\bw}_{j}$ and
fixing this portion of $\Theta$ in \eqref{lcpd}, we can obtain the
final estimate $\widehat{\mu}$ as the solution to \eqref{lcpd}.

\begin{algorithm}[]  
\caption{MM-Tensor Power algorithm for logistic CP decomposition}
\small
\label{mm}
\begin{algorithmic}[1]
\STATE \textbf{input:} tensor $\mathcal{X}$, number of initializations $L$, and rank $R$.

\STATE Initialize with $\widehat{\mu}_{\tau}^{[0]}$ and $(\widehat{d}_{\tau}^{[0]}, \widehat{\bu}_{\tau}^{[0]},\widehat{\bv}_{\tau}^{[0]},\widehat{\bw}_{\tau}^{[0]})$ where $\tau\in[L]$. Set $m=0$.

\FOR {$\tau=1$ {\bfseries to} $L$}

\REPEAT

\STATE Compute $\mathcal{Z}^{[m]}$ in $\eqref{newx}$.

\STATE Update $\widehat{\mu}^{[m+1]}$.

\STATE Compute $\mathcal{Z}_{c}^{[m]}=\mathcal{Z}^{[m]}-\widehat{\mu}^{[m+1]}\one_{p_{1}p_{2}p_{3}}$.

\REPEAT

\STATE $\bu_{\tau}=\text{Normalize}(\mathcal{Z}_{c}^{[m]}\times_{2}(\bv_{\tau})^{T}\times_{3}(\bw_{\tau})^{T})$

%$\|\mathcal{X}^{[m]}\times_{2}\bv^{T}\times_{3}\bw^{T}\|_{2}$

\STATE $\bv_{\tau}=\text{Normalize}(\mathcal{Z}_{c}^{[m]}\times_{1}(\bu_{\tau})^{T}\times_{3}(\bw_{\tau})^{T})$

%$\|\mathcal{X}^{[m]}\times_{1}\bu^{T}\times_{3}\bw^{T}\|_{2}$

\STATE $\bw_{\tau}=\text{Normalize}(\mathcal{Z}_{c}^{[m]}\times_{1}(\bu_{\tau})^{T}\times_{2}(\bv_{\tau})^{T})$

%$\|\mathcal{X}^{[m]}\times_{1}\bu^{T}\times_{2}\bv^{T}\|_{2}$

\UNTIL{converge}

\STATE Update $(\widehat{d}_{\tau}^{[m+1]}, \widehat{\bu}_{\tau}^{[m+1]}, \widehat{\bv}_{\tau}^{[m+1]}, \widehat{\bw}_{\tau}^{[m+1]})$.

\STATE $m\leftarrow m+1$

\UNTIL{converge}

\STATE Return $\widehat{\mu}_{\tau}$ and $(\widehat{d}_{\tau},\widehat{\bu}_{\tau},\widehat{\bv}_{\tau},\widehat{\bw}_{\tau})$.

\ENDFOR

 \STATE Cluster $\{(\widehat{d}_{\tau},\widehat{\bu}_{\tau},\widehat{\bv}_{\tau},\widehat{\bw}_{\tau}),\tau\in[L]\}$ into $R$ clusters $\{(\widehat{d}_{j},\widehat{\bu}_{j},\widehat{\bv}_{j},\widehat{\bw}_{j}), j\in[R]\}$ 
by Algorithm \ref{clust}.
 
 \STATE \textbf{output:} $\widehat{\mu}$ and $R$ clusters $\{(\widehat{d}_{j},\widehat{\bu}_{j},\widehat{\bv}_{j},\widehat{\bw}_{j}), j\in[R]\}$.

  \end{algorithmic} 
 \end{algorithm}

\section{Sparse CP Decomposition for Binary Data} \label{sec:slcpd}

\subsection{Sparse Logistic CP Decomposition}

Based on the formulation of logistic tensor decomposition,
we consider two approaches which can produce sparse factor matrices
with many zero entries. Similar to sparse logistic PCA and sparse
tensor decomposition, when appropriate, sparse factor matrices can describe the latent
structure more concisely, and nonzero entries can indicate important variables in each mode.
To obtain sparse factor matrices for logistic tensor decomposition, we could add a penalty or constraint on the factor matrices. For example, the $\ell_{1}$-norm penalty and $\ell_{0}$-norm penalty have been successfully applied in the problems of penalized matrix decomposition \citep{witten2009penalized, yuan2013truncated} and penalized tensor decomposition \citep{allen2012sparse, sun2017provable}.

Based on the logistic CP decomposition \eqref{lcpd}, we propose the following \textit{sparse logistic CP decomposition} (SLCPD):
\begin{equation}\label{plcpd}
\begin{split}
\underset{\mu,\bd, U,V,W}{\min}\quad&-\langle\mathcal{X}, \Theta\rangle +\langle \one_{p_{1}p_{2}p_{3}}, \log(\one_{p_{1}p_{2}p_{3}}+\exp(\Theta)) \rangle\\
\text{s.t.}
\quad& \Theta=\mu\one_{p_{1}p_{2}p_{3}}+\sum_{r\in[R]}d_{r}\cdot \bu_{r}\circ \bv_{r}\circ \bw_{r},\\
 \quad& \bu_{r}^{T}\bu_{r}=1, \bv_{r}^{T}\bv_{r}=1,\bw_{r}^{T}\bw_{r}=1,d_{r}>0,\\
% &\;\|\bu_{r}\|_{1}\leq c_{1r},\|\bv_{r}\|_{1}\leq c_{2r},\|\bw_{r}\|_{1}\leq c_{3r}, r\in[R]\\ 
\quad &p_{1}(\bu_{r})\leq t_{1r},p_{2}(\bv_{r})\leq t_{2r},p_{3}(\bw_{r})\leq t_{3r}, r\in[R],\\  
\end{split}
\end{equation}
where $p_{i}(\cdot)$  for $i=1,2,3$ are penalty functions for factors, 
and $t_{ir}$ are tuning parameters. 
We consider two types of sparsity inducing penalty for each factor: the $\ell_{1}$-norm and $\ell_{0}$-norm penalty functions for $p_{i}(\cdot)$.
For the $\ell_{1}$-norm constrained formulation, 
$p_{1}(\bu_{r})=\|\bu_{r}\|_{1}$,
$p_{2}(\bv_{r})=\|\bv_{r}\|_{1}$ and
$p_{3}(\bw_{r})=\|\bw_{r}\|_{1}$.
For the $\ell_{0}$-norm constrained formulation, 
$p_{1}(\bu_{r})=\|\bu_{r}\|_{0}$,
$p_{2}(\bv_{r})=\|\bv_{r}\|_{0}$ and
$p_{3}(\bw_{r})=\|\bw_{r}\|_{0}$. 
This formulation naturally extends sparse logistic PCA \citep{lee2010sparse} and binary matrix biclustering \citep{lee2014biclustering} to higher-order binary tensors.

To solve problem \eqref{plcpd}, we could update $U, V$ and $W$ in an
alternative manner similar to ALS. When $V$ and $W$ are fixed, we
could solve a regularized non-convex problem for $U$. And we could solve for $V$ and $W$ in an analogous manner. However, this regularized alternating least squares approach cannot guarantee that the solution is the global minimizer of the problem \citep{allen2012sparse}. Instead, we consider a tensor power method and update each factor in an iterative block-wise manner.

 \subsection{Majorization-Minimization Approach with $\ell_{1}$-norm Constraint}

In order to simplify problem \eqref{plcpd} with the $\ell_{1}$-norm constraints and obtain a simple analytic solution, we relax the original non-convex equality constraints (e.g., $\bu^{T}\bu=1$), and consider the tensor decomposition problem with convex inequality constraints (e.g., $\bu^{T}\bu\leq1$) \citep{allen2012sparse}. 
Although the objective function is not convex in factor matrices jointly, it is convex in each factor matrix individually with all other factor matrices fixed.

Similar to logistic CP decomposition, we consider a rank-one problem to avoid local minima and the MM algorithm. 
For a rank-one problem,
in the $m$th step of MM algorithm with $\mathcal{Z}^{[m]}_{c}$ defined in \eqref{newZc}, we have the following relaxation: 
\begin{equation}\label{pmd}
\begin{split}
\underset{d, \bu, \bv, \bw}{\min}\quad&\|\mathcal{Z}_{c}^{[m]}-d\cdot \bu\circ \bv\circ \bw\|_{F}^{2}\\
\text{s.t.} 
%&\; \Theta=\mu\one_{p_{1}p_{2}p_{3}}+d\cdot \bu\circ \bv\circ \bw ,\\
\quad & \bu^{T}\bu\leq1, \bv^{T}\bv\leq1,\bw^{T}\bw\leq1, d>0,\\
\quad& \|\bu\|_{1}\leq c_{1}, \|\bv\|_{1}\leq c_{2}, \|\bw\|_{1}\leq c_{3}, \\
\end{split}
\end{equation}
where $c_{i}\geq 0$ for $i\in[3]$ are tuning parameters.
%$c_{1}, c_{2}$ and $c_{3}\geq 0$ are tuning parameters.

The constrained formulation \eqref{pmd} produces a feasible solution
if $1\leq c_{i}\leq \sqrt{p_{i}}$  
%$1\leq c_{1}\leq \sqrt{p_{1}}, 1\leq c_{2}\leq \sqrt{p_{2}}$ and $1\leq c_{3}\leq \sqrt{p_{3}}$,
and reduces to the un-regularized version when $c_{i}=\sqrt{p_{i}}$.
%$c_{1}=\sqrt{p_{1}}, c_{2}=\sqrt{p_{2}}, c_{3}=\sqrt{p_{3}}$. 
%If $c_{1}$, $c_{2}$ and $c_{3}$ are chosen appropriately,
If $c_{i}$ are chosen appropriately,
the solution to the relaxed problem still solves the original problem with the $\ell_{2}$-norm constraints. See \cite{witten2009penalized} for detailed arguments.
Given $\bv$ and $\bw$, the relaxed formulation in \eqref{pmd} can be rewritten as a subproblem for $\bu$:
\begin{equation*}
\begin{split}
\underset{\bu}{\max}\quad& \bu^{T}(\mathcal{Z}_{c}^{[m]}\times_{2}\bv^{T}\times_{3}\bw^{T})\\
\text{s.t.} \quad & \bu^{T}\bu\leq1, \|\bu\|_{1}\leq c_{1}.\\
\end{split}
\end{equation*}
This subproblem has explicit solution of 
\begin{equation}
\begin{split}\label{u2}
\widehat{\bu}=\text{Normalize}(S(\mathcal{Z}_{c}^{[m]}\times_{2}\bv^{T}\times_{3}\bw^{T},\lambda_{1})).
\end{split}
\end{equation}
 Here $S(\cdot,\lambda)=\text{sign}(\cdot)(|\cdot|-\lambda)_{+}$ is
 the soft-thresholding operator, and $\lambda_{1}$ is the smallest
 positive value such that $\|\bu\|_{1}\leq c_{1}$. The value of
 $\lambda_{1}$ can be chosen by a binary search
 \citep{witten2009penalized}. We can update $\bv$ and $\bw$ in a
 similar manner.  This MM approximation with  Tensor Soft-thresholding
 Power method (MM-TSP) is summarized in Algorithm \ref{mmtp} of
 Appendix \ref{app:algorithm}.
 %, which can be found in the appendix. 

 \subsection{Majorization-Minimization Approach with $\ell_{0}$-norm Constraint}

We propose to solve problem \eqref{plcpd} with the $\ell_{0}$-norm constraints in a manner similar to the tensor power method, and consider iterative rank-one sparse approximations of $\mathcal{Z}_{c}^{[m]}$ in \eqref{newZc}. 
For a rank-one problem, in the $m$th step of MM algorithm, we have the following problem:
\begin{equation}\label{pmd1}
\begin{split}
\underset{d, \bu,\bv,\bw}{\min}\quad &\|\mathcal{Z}_{c}^{[m]}-d\cdot \bu\circ \bv\circ \bw\|_{F}^{2}\\
\text{s.t.} 
\quad& \bu^{T}\bu=1, \bv^{T}\bv=1,\bw^{T}\bw=1, d>0,\\
\quad&\|\bu\|_{0}\leq s_{1},\|\bv\|_{0}\leq s_{2},\|\bw\|_{0}\leq s_{3},\\
\end{split}
\end{equation}
where $s_{i}\le p_i$ for $i\in [3]$ are tuning parameters.
%$s_{1}\le p_1$, $s_{2}\le p_2$ and $s_{3}\le p_3$ are tuning parameters.

The constrained formulation \eqref{pmd1} produces  a feasible solution
if $1\leq s_{i}\leq p_{i}$.
%$1\leq s_{1}\leq p_{1}, 1\leq s_{2}\leq p_{2}$ and $1\leq s_{3}\leq p_{3}$.
It reduces to the un-regularized problem without any constraint in
each factor when $s_{i}=p_{i}$.
%$s_{1}=p_{1}, s_{2}=p_{2}$ and $s_{3}=p_{3}$.
Inspired by \cite{yuan2013truncated, sun2017provable}, we could apply the tensor truncated power method in solving the above problem. 
Given $\bv$ and $\bw$, the constrained problem can be rewritten as a subproblem for $\bu$:\begin{equation*}
\begin{split}
\underset{\bu}{\max}\quad&\bu^{T}(\mathcal{Z}_{c}^{[m]}\times_{2}\bv^{T}\times_{3}\bw^{T})\\
\text{s.t.} \quad&\bu^{T}\bu=1, \|\bu\|_{0}\leq s_{1},\\
\end{split}
\end{equation*}
where $s_{1}$ denotes the number of non-zero entries.
This subproblem has explicit solution of 
\begin{equation}
\begin{split}\label{u3}
\widehat{\bu}=\text{Normalize}(T(\mathcal{Z}_{c}^{[m]}\times_{2}\bv^{T}\times_{3}\bw^{T},s_{1})).
\end{split}
\end{equation}
 Here $T(\cdot,s)$ is the truncation operator which keeps the largest
 $\lfloor s\rfloor$ entries of a vector in the absolute value and
 truncates the remaining entries to zero.  We can update $\bv$ and
 $\bw$ in a similar manner. This MM approximation with Tensor Truncated
 Power method (MM-TTP) is summarized in Algorithm \ref{mmttp} of Appendix.

 \section{Missing Data and Tensor Completion} \label{sec:missing}
 
In practice, missing data is common. To handle missing data,  we extend our algorithms.
Given data tensor $\mathcal{X}$  of size $p_{1}\times p_{2}\times p_{3}$, we let $\Omega=\{(i,j,k)\in[p_{1}]\times [p_{2}] \times [p_{3}]|x_{ijk}\;\text{is observed}\}$ denote the index set of observed entries. 
%and $\ell_{0}$-norm problem \eqref{clcpd}.
Given $\Omega\subseteq [p_{1}]\times [p_{2}] \times [p_{3}]$, we can define the projection operation $\mathcal{P}_\Omega: \mathbb{R}^{p_{1}\times p_{2}\times p_{3}} \mapsto \mathbb{R}^{p_{1}\times p_{2}\times p_{3}}$ as follows:
\begin{equation*}
\mathcal{P}_{\Omega}(\mathcal{X})=
\left\{
\begin{aligned}
& x_{ijk} & \text{if }(i,j,k)\in\Omega \\
&0  & \text{if } (i,j,k)\notin\Omega.
\end{aligned}
\right.
\end{equation*}
$\mathcal{P}_\Omega$ replaces the missing entries in the data tensor $\mathcal{X}$ with zeros, and leaves the observed entries unchanged. Let $\mathcal{H}=(h_{ijk})\in\mathbb{R}^{p_{1}\times p_{2}\times p_{3}}$ be a masking tensor such that $\mathcal{P}_{\Omega}(\mathcal{X})=\mathcal{H}*\mathcal{X}$, where $*$ denotes the Hadamard product of two tensors. Then for partially observed data, we can redefine the following rank-$R$ logistic CP decomposition problem: 
\begin{equation}\label{mlcpd}
  \begin{split}
\underset{\mu,\bd, U, V, W} {\min}\quad&-\langle\mathcal{H}*\mathcal{X},\Theta\rangle +\langle \mathcal{H}, \log(\one_{p_{1}p_{2}p_{3}}+\exp(\Theta)) \rangle\\
\text{s.t.} 
\quad&\Theta=\mu\one_{p_{1}p_{2}p_{3}}+\sum_{r\in[R]}d_{r}\cdot \bu_{r}\circ \bv_{r}\circ \bw_{r},\\
\quad& \bu_{r}^{T}\bu_{r}=1, \bv_{r}^{T}\bv_{r}=1,\bw_{r}^{T}\bw_{r}=1,
\mbox{ and } d_{r}>0 \mbox{ for } r\in[R].
\end{split}
\end{equation}
To solve the above problem, we modify the previous algorithms by  introducing new working variables. We define new working variables $\mathcal{Y}^{[m]}=(y_{ijk}^{[m]})$ by filling in the missing values with fitted values based on the current estimate of logit parameter tensor $\Theta$ as follows:
\begin{equation}\label{varz}
\begin{split}
y_{ijk}^{[m]}&=
\left\{
\begin{aligned}
& z_{ijk}^{[m]} & (i,j,k)\in\Omega\\
& \theta_{ijk}^{[m]} & (i,j,k)\not\in\Omega,
\end{aligned}
\right.
\end{split}
\end{equation}
where $\mathcal{Z}^{[m]}=\Theta^{[m]}+4(\mathcal{X}-\sigma(\Theta^{[m]}))$.
In the $m$th step of MM approximation, the objective function of problem \eqref{lcpd} turns into 
$\frac{1}{8}\|\mathcal{Y}^{[m]}-\Theta\|_{F}^{2}$.
For sparse logistic CP decomposition, we could also replace the working variables $\mathcal{Z}^{[m]}$ with $\mathcal{Y}^{[m]}$ in the MM approximation of regularized problem \eqref{plcpd}.

Once we have $\widehat{\Theta}$ from observed data, we can use it for missing value prediction.
 After estimating $\widehat{\Theta}$ from
observed data as in \eqref{mlcpd}, we could predict the missing entries of $\mathcal{X}$ by using $\widehat{\mathcal{P}}=\text{logit}^{-1}(\widehat{\Theta})$, where $\widehat{\mathcal{P}}=(\widehat{p}_{ijk})$ is the tensor 
with estimated probabilities of Bernoulli random variables. For probability $\widehat{p}_{ijk}\geq 0.5$, we  could impute missing $x_{ijk}$ with one and zero otherwise.
Similar ideas of such tensor completion for continuous data and binary
data have been investigated in \cite{acar2011scalable} and \cite{wang2020learning}.

\section{Selecting Rank and Tuning Parameters} \label{sec:tuning}

Selecting an appropriate rank for tensor decomposition is an issue of
practical importance. However, there has been few discussions in the
literature. \cite{allen2012sparse, sun2017provable, wang2020learning}
derived a BIC heuristic to select the rank and degree of sparsity for
CP decomposition. The consistency of BIC in binary tensor
decomposition is unknown, but similar problems have been investigated
by \cite{shi2019determining} in the context of relational learning. Alternatively, we could use cross-validation to choose the rank and tuning parameters, but cross-validation can be slow to carry out for high-dimensional
tensors. \cite{shen2008sparse, witten2009penalized, udell2016generalized} have used cross-validation to select the rank and sparsity tuning parameters for matrix decomposition problems. 
In this section, we investigate  AIC, BIC, cross-validation and
explained deviance as possible approaches to select the rank and
tuning parameters. These approaches are illustrated with simulated
data in Appendix \ref{appsim}. 

\subsection{BIC and AIC}

As the tuning parameters $c_i$ or $s_i$ decrease, the estimated factor
matrices become sparser, and the model for the underlyng logit parameter tensor
becomes simpler and easier to interpret. To reach a balance between model complexity and goodness of fit, we adopt the Bayesian information criterion (BIC) to select the optimal penalty parameter in sparse logistic tensor decomposition.

Given a prespecified set of rank values and penalty parameter values $c_i$ or cardinality values $s_i$, we choose the combination of parameters
$(\widehat{R},\widehat{c}_{1},\widehat{c}_{2}, \widehat{c}_{3})$ or
$(\widehat{R},\widehat{s}_{1},\widehat{s}_{2}, \widehat{s}_{3})$ which
minimizes the BIC criterion for sparse logistic CP decomposition $\widehat{\Theta}$ in \eqref{plcpd}:
\begin{equation}\label{bicdef}
\text{BIC}:=-2\ell(\mathcal{X}; \widehat{\Theta})+\log (p_{1}p_{2}p_{3})\times(1+\|\widehat{U}\|_{0}+\|\widehat {V}\|_{0}+\|\widehat {W}\|_{0}-2R).
\end{equation}
Here $\|\widehat{U}\|_{0}$ is the number of nonzero entries in matrix $\widehat{U}$ when
the penalty parameter is $\widehat{c}_{1}$ or cardinality parameter is
$\widehat{s}_{1}$, and  $\|\widehat{V}\|_{0}$ and $\|\widehat{W}\|_{0}$ are defined analogously.
Note that $\hat{\mu}$ has $df=1$, $\widehat{\bd}$ has $df=R$ and there are $3R$
constraints on $(\widehat{U}, \widehat{V}, \widehat{W})$, %$(\bu,\bv,\bw)$,
so the overall $df$ of the logit tensor model can be taken as $1+\|\widehat{U}\|_{0}+\|\widehat{V}\|_{0}+\|\widehat{W}\|_{0}-2R$.
This is analogous to the way the model degrees of freedom is defined
for sparse logistic PCA.

For the case with missing data, letting $\Omega$ denote the index set
of observed entries, we use $|\Omega|$ rather than $p_{1}p_{2}p_{3}$,
and the log likelihood for the observed entries in $\Omega$ is defined as $\ell(\mathcal{X}_{\Omega};\Theta):=\sum_{ijk\in\Omega}\ell(x_{ijk};\theta_{ijk})$ in the BIC. This leads to the following extended BIC:
\begin{equation*}
\begin{split}
\text{BIC}_{\Omega}:=-2\ell(\mathcal{X}_{\Omega}; \widehat{\Theta})+\log (|\Omega|)\times(1+\|\widehat{U}\|_{0}+\|\widehat{V}\|_{0}+\|\widehat{W}\|_{0}-2R).
\end{split}
\end{equation*}
Note that for fully observed data, $|\Omega|=p_{1}p_{2}p_{3}$ and $\ell(\mathcal{X}_{\Omega};\Theta)=\ell(\mathcal{X};\Theta)$, and the extended BIC reduces to \eqref{bicdef}.

To select the optimal tuning parameters, we could also consider minimizing the Akaike information criterion (AIC) for tensor decomposition:
\begin{equation}\label{aicdef}
\begin{split}
\text{AIC}:=-2\ell(\mathcal{X}; \widehat{\Theta})+2\times(1+\|\widehat{U}\|_{0}+\|\widehat{V}\|_{0}+\|\widehat{W}\|_{0}-2R).
\end{split}
\end{equation}

Given a fixed rank $R$, we first search for the optimal tuning parameters $(\widehat{c}_{1},\widehat{c}_{2}, \widehat{c}_{3})$ or $(\widehat{s}_{1},\widehat{s}_{2}, \widehat{s}_{3})$ by BIC/AIC, and then given the tuning parameter values, we seek the best rank $R$ which minimizes BIC/AIC.

\subsection{Cross-validation}

Cross-validation has been proven to be useful in selecting tuning
parameters in many settings. We could also select the rank and
sparsity penalty parameters  by an approach similar to cross-validation since our algorithms can handle missing data.
However, compared with BIC or AIC, cross-validation is  computationally more expensive. Many metrics could be used for cross-validation in binary tensor decomposition. 
We cross-validate each tuning parameter value by minimizing the negative log likelihood, $-\ell(\mathcal{X};\Theta)$ in this paper.

For a $5$-fold cross-validation of rank $R$, we randomly split binary tensor entries into 5 folds: 4 folds are used for training and 1 fold is used for testing, where nonzero entries and zero entries are split separately with the same ratio.
For a fixed rank $R$, we treat the test data as missing data and
estimate $\Theta$ by minimizing the negative log likelihood
$-\ell(\mathcal{X}_{\text{train}};\Theta)$ with the training data $\mathcal{X}_{\text{train}}$ only. Then for evaluation of $\widehat{\Theta}$, we calculate the negative log likelihood $-\ell(\mathcal{X}_{\text{test}};\widehat{\Theta})$ using the test data $\mathcal{X}_{\text{test}}$. 
We repeat the above process for five times and obtain the average negative log likelihood for each rank. A similar process can be used to select the levels of sparsity in factor matrices.

\subsection{Explained Deviance}

Alternatively, we could also use the explained deviance for determining the rank and tuning parameter values analogous to the use of explained total variance in standard PCA. 
The deviance of estimated logit parameter tensor $\widehat{\Theta}$ based on  data $\mathcal{X}$ is defined as $D(\mathcal{X}; \widehat{\Theta})=-2(\ell(\mathcal{X};\widehat{\Theta})-\ell(\mathcal{X};\Theta_{S})),$
where $\Theta_{S}$ is the logit parameter tensor of the saturated model.
For binary tensor $\mathcal{X}$,
$\Theta_{S}=\text{logit}(\mathcal{X})$, where
$\text{logit}(x)=\log(\frac{x}{1-x})$ is taken elementwise, and thus $\ell(\mathcal{X};\Theta_{S})=0$. This leads to $D(\mathcal{X}; \widehat{\Theta})=-2\ell(\mathcal{X}; \widehat{\Theta})$,  and we have
\begin{equation*}
\begin{split}
D(\mathcal{X}; \widehat{\Theta})
:=-2\langle \mathcal{X},\widehat{\Theta}\rangle+2\langle\one_{p_{1}p_{2}p_{3}},\log(\one_{p_{1}p_{2}p_{3}}+\exp(\widehat{\Theta}))\rangle.\\
\end{split}
\end{equation*}
For partially observed data $\mathcal{X}_{\Omega}$, the deviance can be expressed as \begin{equation*}
\begin{split}
D(\mathcal{X}_{\Omega}; \widehat{\Theta}):=-2\langle \mathcal{H}*\mathcal{X},\widehat{\Theta}\rangle+2\langle \mathcal{H},\log(\one_{p_{1}p_{2}p_{3}}+\exp(\widehat{\Theta}))\rangle
\end{split}
\end{equation*}
using the masking tensor $\mathcal{H}$ defined previously.
%In simulation study, we generate the tensor without main effect $\mu$. 

Let $\widehat{\Theta}_{0}:=\widehat{\mu}\one_{p_{1}p_{2}p_{3}}$ as an estimated tensor with offset term $\mu$ only,  and for $r\in[R]$,  let 
$\widehat{\Theta}_{r}:=\widehat{\mu}\one_{p_{1}p_{2}p_{3}}+\sum_{i=1}^{r}\widehat{d}_{i}\cdot\widehat{\bu}_{i}\circ \widehat{\bv}_{i}\circ \widehat{\bw}_{i}$ with the first $r$ components. We call $D(\mathcal{X}_{\Omega}; \widehat{\Theta}_{0})$ the null deviance
and define the cumulative percentage of explained deviance of the first $r$  components as
 $$1-\frac{D(\mathcal{X}_{\Omega}; \widehat{\Theta}_{r})}{D(\mathcal{X}_{\Omega}; \widehat{\Theta}_{0})}.$$  Similarly, we define the marginal percentage of explained deviance by the $r$th component as $$\frac{D(\mathcal{X}_{\Omega}; \widehat{\Theta}_{r-1})-D(\mathcal{X}_{\Omega}; \widehat{\Theta}_{r})}{D(\mathcal{X}_{\Omega}; \widehat{\Theta}_{0})}.$$
These criteria extend the proportion of total variance explained in real-valued tensors \citep{allen2012sparse} to binary tensors. The same criteria have been considered in the context of binary matrix factorization \citep{landgraf2020dimensionality}.

We could also define the marginal deviance of the $r$th component as
$$D_{r}:=D(\mathcal{X}; \widehat{\Theta}_{(r)})
=-2\langle \mathcal{X},\widehat{\Theta}_{(r)}\rangle+2\langle\one_{p_{1}p_{2}p_{3}},\log(\one_{p_{1}p_{2}p_{3}}+\exp(\widehat{\Theta}_{(r)}))\rangle,$$
where $\widehat{\Theta}_{(r)}:=\widehat{\mu}\one_{p_{1}p_{2}p_{3}}+\widehat{d}_{r}\cdot\widehat{\bu}_{r}\circ \widehat{\bv}_{r}\circ \widehat{\bw}_{r}$. As index $r$ corresponds to weight $d_{r}$ ordered from largest to smallest, typically the $r$th marginal deviance will increase as $r$ increases. Therefore the first component with largest weight $d_{1}$ will have the smallest marginal deviance $D_{1}$, and the last component with smallest weight $d_{R}$ will have the largest marginal deviance $D_{R}$.

\section{Simulation Study} \label{sec:simulate}

We compare the proposed $\ell_{0}$-norm constrained logistic tensor decomposition with Tensor Truncated Power (TTP) method, $\ell_{1}$-norm constrained logistic tensor decomposition with Tensor Soft-thresholding Power (TSP) method, and un-regularized logistic tensor decomposition with Tensor Power (TP) method and Alternating Least Squares (ALS) method. We have implemented all methods in R \citep{r-core} using the \texttt{rTensor} package \citep{li2018rtensor} for efficient tensor computations.
Appendix \ref{apppara} describes implementational details including initialization and termination of the proposed algorithms as well as the clustering procedure.

\subsection{Simulation Setup}

To generate binary tensor data $\mathcal{X}$ with sparse logit parameters, we first specify the underlying logit parameter tensor $\Theta^{*}$ of size $p_{1}\times p_{2}\times p_{3}$.
We consider the following four scenarios for the size and rank of
$\Theta^{*}\label{setup}$: \\
I. $p_{1}=1000, p_{2}=10, p_{3}=10, \text{and}\; R=1$; II. $p_{1}=1000, p_{2}=10, p_{3}=10, \text{and}\; R=2$;\\
III. $p_{1}=1000, p_{2}=100, p_{3}=10, \text{and}\; R=1$; IV. $p_{1}=1000, p_{2}=100, p_{3}=10, \text{and}\; R=2$.

In all simulation settings, we keep the level of sparsity equal in each dimension by setting the cardinality of nonzero entries as $p_{0j}=0.2 p_{j}$ for $j=1,2,3$. With fixed dimensionality $(p_{1},p_{2},p_{3})$ and true rank $R$, we first generate independent and identically distributed standard Gaussian entries for three factor matrices $U\in\mathbb{R}^{p_{1}\times R}, V\in\mathbb{R}^{p_{2}\times R}$ and $W\in\mathbb{R}^{p_{3}\times R}$. Then to induce sparsity in the factor matrices with fixed cardinality parameters $(p_{01}, p_{02}, p_{03})$, we truncate some entries in each column of $U, V$ and  $W$ to zero. Finally, we normalize each column of $U, V$ and $W$ to get $U^{*}, V^{*}$ and $W^{*}$. 

To specify the weights $d_{1}^{*},\dots, d_{R}^{*}$ properly, we first consider  their null values when 
$\Theta=(0)$ or $\mathcal{P}=(1/2)$, taken as the baseline noise
level, and then determine their actual values proportionally. To find
such null values, we first generate a $p_{1}\times p_{2}\times p_{3}$
binary tensor whose entries are mutually independent realizations from a
Bernoulli distribution with $p=1/2$. We carry out a rank-$R$ logistic CP decomposition \eqref{lcpd} of the binary tensor and calculate the average of $R$ weights denoted by $d_{b}$. We repeat this process for 100 times and take the mean of $d_{b}$ as the baseline noise level. Then using $d_{b}$, we could define the signal-to-noise ratio (SNR) as $\text{SNR}_{r}=d_{r}^{*}/d_{b}$ to determine the weights $d_{r}^{*}$ for $r\in[R]$. 

We consider different combinations of signal-to-noise ratio values: $\text{SNR}=(5,3)$ when $R=2$, and $\text{SNR}=3$ when $R=1$.
 With specified weights, we define the logit parameter tensor as
$$\Theta^{*}=\mu^{*}\one_{p_{1}p_{2}p_{3}}+\sum_{r\in[R]}d_{r}^{*}\cdot \bu_{r}^{*}\circ \bv_{r}^{*}\circ \bw_{r}^{*},$$
which extends the spiked tensor model \citep{montanari2014statistical} to binary data.
The overall logit parameter $\mu^{*}$ is set to zero.
Because of the sparsity in $(\bu_{r}^{*}, \bv_{r}^{*},\bw_{r}^{*})$ for $r\in[R]$, $\Theta^{*}$ is also sparse.
Finally, we generate $x_{ijk}$ from $\text{Bernoulli}(p_{ijk}^{*})$, where $p_{ijk}^{*}=\text{logit}^{-1}(\theta_{ijk}^{*})$
for $i\in[p_{1}]$, $j\in[p_{2}]$, $k\in[p_{3}]$, and obtain a binary tensor $\mathcal{X}=(x_{ijk})$ with the corresponding probability tensor $\mathcal{P}^{*}=(p_{ijk}^{*})$.

As for tuning parameters in this simulation study, we parameterize $c_{j}=\sqrt{p_{j}}\times c$ ($j=1,2,3$) for the $\ell_{1}$-norm constraint and $s_{j}=p_{j}\times s$ for the $\ell_{0}$-norm constraint.
To make the $\ell_{1}$-norm and $\ell_{0}$-norm problems well-defined,
we vary the ratio $c\in [\underset{i}{\max}\frac{1}{\sqrt{p_{i}}},1]$
for the $\ell_{1}$-norm constraint and ratio $s\in
[\underset{i}{\max}\frac{1}{p_{i}},1]$ for the $\ell_{0}$-norm
constraint. For the numerical results in Table \ref{simutable}, we considered a prespecified set of rank values $\{1,\dots,4\}$  and a
range of values for the ratio parameters $c$ and $s$ and tuned the
parameters using AIC.
%$\lambda_{j}=10^{\{-1,-0.9,\dots,0\}}$
%\textcolor{red}{For case 2, we need larger $L$.}
In simulation settings where  true factors $\bu^{*},\bv^{*}$ and
$\bw^{*}$ are known, we could set $(s_{1},
s_{2},s_{3})=(\|\bu^{*}\|_{0},\|\bv^{*}\|_{0},\|\bw^{*}\|_{0})$ and
$(c_{1}, c_{2},
c_{3})=(\|\bu^{*}\|_{1},\|\bv^{*}\|_{1},\|\bw^{*}\|_{1})$ as optimal
tuning parameters in the $\ell_{0}$-norm and $\ell_{1}$-norm problems,
respectively.

\subsection{True Positive Rate and False Positive Rate}

% Due to the sparse pattern in true logit parameters,
When the true logit parameters are sparse,
we are interested in recovering the sparse pattern and selecting important nonzero features in the latent factors. The selection performance can be measured by the true positive rate (TPR):
 the proportion of correctly estimated non-zeros in the true parameter
 and the false positive rate (FPR): the proportion of true zeros that are incorrectly estimated to be nonzero.
For an estimated factor matrix $\widehat{U}$, the TPR and FPR are defined as
% $$\text{TPR}_{U}=\frac{\#\{(i,j): \widehat{\bu}_{ij}\neq 0 \; \text{and}\; U_{ij}\neq0\}}{\#\{(i,j): U_{ij}\neq 0\}},$$
% $$\text{FPR}_{U}=\frac{\#\{(i,j): \widehat{\bu}_{ij}\neq 0 \; \text{and}\; U_{ij}=0\}}{\#\{(i,j): U_{ij}=0\}},$$ 
  $$\text{TPR}_{\widehat{U}}=\frac{1}{r}\sum_{r\in[R]}\frac{|\{i: (\widehat{\bu}_{r})_{i}\neq 0 \; \text{and}\; (\bu_{r}^{*})_{i}\neq0\}|}{|\{i: (\bu_{r}^{*})_{i}\neq 0\}|}$$ and 
   $$\text{FPR}_{\widehat{U}}=\frac{1}{r}\sum_{r\in[R]}\frac{|\{i: (\widehat{\bu}_{r})_{i}\neq 0 \; \text{and}\; (\bu_{r}^{*})_{i}=0\}|}{|\{i: (\bu_{r}^{*})_{i}= 0\}|} ,$$ respectively.
 For $\widehat{V}$ and $\widehat{W}$, 
    $\text{TPR}_{\widehat{V}}, \text{TPR}_{\widehat{W}}, \text{FPR}_{\widehat{V}}$ and $\text{FPR}_{\widehat{W}}$ can be defined analogously.
 Then the overall TPR and FPR for $\widehat{\Theta}$ can be defined as 
$\text{TPR}(\widehat{\Theta})=(\text{TPR}_{\widehat{U}}+\text{TPR}_{\widehat{V}}+\text{TPR}_{\widehat{W}})/3$ and 
$\text{FPR}(\widehat{\Theta})=(\text{FPR}_{\widehat{U}}+\text{FPR}_{\widehat{V}}+\text{FPR}_{\widehat{W}})/3$.

\subsection{Estimation Errors}

%\subsubsection{Simulation Data}

To evaluate the accuracy of $\widehat{\Theta}=\widehat{\mu}\one_{p_{1}p_{2}p_{3}}+\sum_{r=1}^{R}\widehat{d}_{r}\cdot\widehat{\bu}_{r}\circ \widehat{\bv}_{r}\circ \widehat{\bw}_{r}$ in recovering the true 
%Similar to estimate the signal recovery in \cite{allen2012sparse}, we can also measure the estimation of 
logit parameter tensor $\Theta^{*}$, we look at its root mean squared error (RMSE) defined as
%by using the deviation in the Frobenius norm:
$$\text{RMSE}(\widehat{\Theta})=\frac{1}{\sqrt{p_{1}p_{2}p_{3}}}\|\widehat{\Theta}-\Theta^{*}\|_{F}.$$
%For factor matrix, we can measure the accuracy by using the distance between subspace
%$$\text{MSE}(U)=\|\widehat{\bu}\widehat{\bu}^{T}-U\bu^{T}\|_{F}^{2}$$
To measure the quality of the estimated components and weights in tensor decomposition separately, we also calculate the mean vector estimation error and weight estimation error \citep{anandkumar2014guaranteed, sun2017provable}:
%$$\text{Mean Error}=\frac{1}{3}\sum_{r\in[R]}\{\|\widehat{\bu}_{r}-\bu_{r}^{*}\|_{2}+\|\widehat{\bv}_{k}-\bv_{r}^{*}\|_{2}+\|\widehat{\bw}_{k}-\bw_{r}^{*}\|_{2}\},$$ 
$$\text{Mean Error}=\frac{1}{3}\{\text{ME}_{\widehat{U}}+\text{ME}_{\widehat{V}}+\text{ME}_{\widehat{W}}\},$$ and
$$\text{Weight Error}=\frac{\|\widehat{\bd}-\bd^{*}\|_{2}}{\|\bd^{*}\|_{2}},$$ where $\text{ME}_{\widehat{U}}=\frac{1}{R}\sum_{r\in[R]}\min\{\|\widehat{\bu}_{r}-\bu_{r}^{*}\|_{2}, \|\widehat{\bu}_{r}+\bu_{r}^{*}\|_{2}\},$
$\text{ME}_{\widehat{V}}$ and $\text{ME}_{\widehat{W}}$ are defined analogously. 
Operating characteristics of these evaluation metrics are illustrated with simulated data in Appendix \ref{appsim}.

\subsection{Comparisons}

We compare the proposed $\ell_{0}$-norm logistic tensor decomposition with TTP method, $\ell_{1}$-norm logistic tensor decomposition with TSP method, and un-regularized logistic tensor decomposition method with ALS, TP and block relaxation (BR) \citep{wang2020learning} methods by calculating the average mean squared error, mean estimation error, weight estimation error and TPR/FPR over 20 random replicates simulated from the four scenarios. Table \ref{simutable} presents the results with standard error in parentheses.

The columns for TPR and FPR indicate that the use of $\ell_{1}$-norm and $\ell_{0}$-norm constraints can lead to correct identification of nonzero entries in the logit tensor with FPR close to 0 and TPR mostly $80\%$ to $90\%$. Regularized estimates tend to have smaller errors on average in terms of RMSE, mean vector estimation error and weight estimation error. In particular, $\ell_{1}$-norm regularized estimates with TSP method have minimum errors on the whole. Table  \ref{simutable} suggests that sparse logistic tensor decompositions indeed have better performance than their non-sparse counterpart when the true factor matrices are sparse.

\begin{table*}[htbp] \small
\caption{Comparisons of five logistic tensor decomposition methods under four simulation settings. The minimum value for each error measure is highlighted in bold, and the numbers in parentheses are standard errors. }
\label{simutable}

\resizebox{0.9\textwidth}{!}{
\begin{minipage}{\textwidth}
\vskip 0.15in
\begin{center}
\begin{small}
%\begin{sc}
\begin{tabular}{ccccccc}
\toprule
Scenario & Method& $\text{RMSE}(\Theta)$ & Mean Error & Weight Error & TPR & FPR \\
\midrule

   & BR & $0.3683\; (0.0675)$ & $0.8651\;(0.3601)$ & $0.5649\;(0.1469)$ & $1\;(0)$ & $1\;(0)$ \\ 
 & ALS & $0.3255\; (0.0159)$ & $0.5418\;(0.0465)$ & $0.5600\;(0.0272)$ & $1\;(0)$ & $1\;(0)$ \\
 1& TP & $0.3450\;(0.0146)$ & $0.6279\;(0.0413)$ & $0.5590\;(0.0338)$ & $1\;(0)$ & $1\;(0)$ \\    
 & TSP & $\bm{0.1744}\;(0.0312)$ & $\bm{0.2832}\;(0.0265)$ &   $\bm{0.0347}\;(0.0255)$ &$0.8916\;(0.0083)$ & $0.0208\;(0.0041)$ \\
    & TTP & $0.2877\;(0.0028)$ & $0.3922\;(0.0021)$ & $0.5405\;(0.0029)$ & $0.9000\;(0.0000)$ & $0.0250\;(0.0000)$ \\
\midrule        

   & BR & $0.6815\; (0.0592)$ & $0.6911\;(0.1768)$ & $0.6266\;(0.1362)$ & $1\;(0)$ & $1\;(0)$ \\ 
 & ALS & $0.6504\;(0.0309)$ & $0.6932\;(0.1740)$ & $0.6283\;(0.0170)$ & $1\;(0)$  & $1\;(0)$\\
 2     & TP & $0.7353\;(0.0276)$ & $0.5534\;(0.0347)$ & $0.6049\;(0.0238)$ & $1\;(0)$  & $1\;(0)$\\ 
    & TSP & $\bm{0.6491}\;(0.0627)$ & $\bm{0.4150}\;(0.1095)$ & $\bm{0.1994}\;(0.0068)$ & $0.8791\;(0.0291)$ & $0.0208\;(0.0020)$ \\
    & TTP & $0.7195\;(0.0507)$ & $0.6478\;(0.2754)$ & $0.5629\;(0.0413)$ & $0.7083\;(0.2250)$ & $0.0729\;(0.2250)$\\
\midrule    

   & BR & $0.1545\; (0.0112)$ & $0.5192\;(0.0420)$ & $0.2957\;(0.0444)$ & $1\;(0)$ & $1\;(0)$ \\ 
 & ALS & $0.1842\;(0.0102)$ & $0.6017\;(0.0286)$ & $0.4970\;(0.0209)$ & $1\;(0)$ & $1\;(0)$\\
 3 & TP & $0.1914\;(0.0123)$ & $0.6456\;(0.0351)$ & $0.4920\;(0.0296)$ & $1\;(0)$  & $1\;(0)$\\ 
    & TSP& $\bm{0.1451}\;(0.0180)$& $\bm{0.4197}\;(0.0495)$  & $\bm{0.1225}\;(0.0981)$ & $0.7900\;(0.0200)$ & $0.0137\;(0.0070)$\\
    & TTP & $0.1523\;(0.0101)$ & $0.4296\;(0.0149)$ & $0.4289\;(0.0294)$ & $0.8708\;(0.0225)$ & $0.0322\;(0.0225)$\\ 
\midrule

   & BR & $0.3951\; (0.0142)$ & $0.4248\;(0.0164)$ & $0.3081\;(0.0168)$ & $1\;(0)$ & $1\;(0)$ \\
& ALS &  $0.3930\;(0.0179)$ & $0.5733\;(0.0162)$ & $0.5841\;(0.0113)$ & $1\;(0)$ & $1\;(0)$\\
4 & TP & $ 0.4344\;(0.0162)$ & $0.5560\;(0.0158)$ & $0.4561\;(0.0182)$ & $1\;(0)$  & $1\;(0)$\\
    & TSP& $\bm{0.3911}\;(0.0152)$ & $\bm{0.3590}\;(0.0120)$ & $\bm{0.1985}\;(0.0218)$   & $0.8591\;(0.0066)$ & $0.0053\;(0.0038)$\\
    & TTP& $0.4155\;(0.0145)$ & $0.4088\;(0.0172)$ & $0.4207\;(0.0371)$ & $0.9358\;(0.0025)$ & $0.0160 \;(0.0025)$\\ 
\bottomrule
\end{tabular}
%\end{sc}
\end{small}
\end{center}
\vskip -0.1in
\end{minipage}
 }
\end{table*}

We also compare the five methods computationally in terms of the number of iterations and run time. For comparison, we ran all methods on the same data simulated from scenario 3 with rank-one decomposition using the same initialization and repeated the process 10 times. Table \ref{time} shows the average run time in seconds and number of iterations. Their standard errors are in parentheses. The time for clustering is ignored. Computing was done on a laptop with a 2.7 GHz processor and 8 GB of memory.
The times for TSP and TTP methods correspond to the optimal tuning parameters. 
Note that the run time varies with different initializations, 
which result in relatively large standard errors. According to Table \ref{time}, TP is faster than ALS,  and TSP and TTP take more time to converge than the un-regularized TP method. For time per iteration, we find all methods based on tensor power method are faster than ALS. And among all tensor power methods, TTP is the slowest due to truncation.
Notably, the block relaxation approach to logistic tensor decomposition using iteratively reweighted least squares method takes significantly longer than the proposed MM approach.

\begin{table*}[]\small
\caption{Comparison of the average run time (in seconds) and number of iterations for five logistic tensor decomposition methods under scenario 3.}
\label{time}
\vskip 0.15in
\begin{center}
\begin{small}
\begin{sc}
\begin{tabular}{ccrrr}
\toprule
& Method & Time  & Iteration Number & Time per Iteration\\
\midrule
%1   
&  BR & $59.1640 \;(0.8690)$ &     &  \\
 &  ALS & $8.9147 \;(1.4783)$ &  $12.3 \;(4.2322)$  & $2.7853 \;(0.7180)$ \\
 & TP & $6.3479 \;(2.1182)$  &  $25.1  \;(6.7797)$  & $0.2336 \;(0.0221)$\\
    & TSP  & $10.8628 \;(4.4506)$ &  $36.0 \;(11.6961)$  & $0.2545 \;(0.0490)$\\
    & TTP & $10.9305 \;(4.3180)$  &  $12.6  \;(0.7023)$  & $0.9083 \;(0.3570)$\\
\bottomrule
\end{tabular}
\end{sc}
\end{small}
\end{center}
\vskip -0.1in
\end{table*}

\section{Analysis of Nations Data} \label{sec:nations}

This section investigates the efficacy of our methods on transposable binary tensor data. The nations dataset \citep{rummel1968dimensionality} we consider includes 14 countries and 54 binary predicates (e.g. \textit{treaties}, \textit{exports}) representing interactions between countries. \cite{kemp2006learning} thresholded each continuous variable at its mean and created a binary tensor of size $14\times14\times56$. 
This tensor consists of 56 political relations of 14 countries between
1950 and 1965. Each entry in the tensor (\textit{nation, nation,
  relation}) indicates the presence or absence of a political
relation. If nation $i$ and nation $j$ have relation $k$, $x_{ijk}=1$
and otherwise $x_{ijk}=0$.

The relationship between a nation and itself is not well defined, so
we exclude the diagonal elements $x_{iik}$ and treat them as missing
entries. Overall the missing rate is $11.1\%$. This dataset has been
investigated by \cite{kemp2006learning, nickel2011three,
  wang2020learning}. Different from the previous analysis, we
incorporate an offset term $\mu$ for the logit parameter tensor and
impose a sparsity penalty on factor matrices.

The goals for this data analysis are grouping nations and relations,  and identifying potential blocks of nations and relations. For example, we are interested in finding relations that exist significantly for certain groups of nations, or that can help to distinguish different groups of nations. Appendix \ref{appda} gives provides more results of data analysis.

\subsection{Visualization of Factors}

Due to the special structure of the nations data, we consider a
special logistic CP decomposition with the same first two modes $\bu$
and $\bv$. More specifically, we impose the additional constraints
$\bu_{r}=\bv_{r}$ for $r\in[R]$ in standard logistic CP decomposition in \eqref{lcpd}. To maintain this special structure, we keep the original update of $\bu$ and $\bw$ but set $\bv=\bu$ for the update of $\bv$ in the tensor power method.

\begin{figure}[tb]  
    \begin{minipage}[b]{0.45\textwidth}
  \centering    
  \includegraphics[height=65mm,width=70mm]{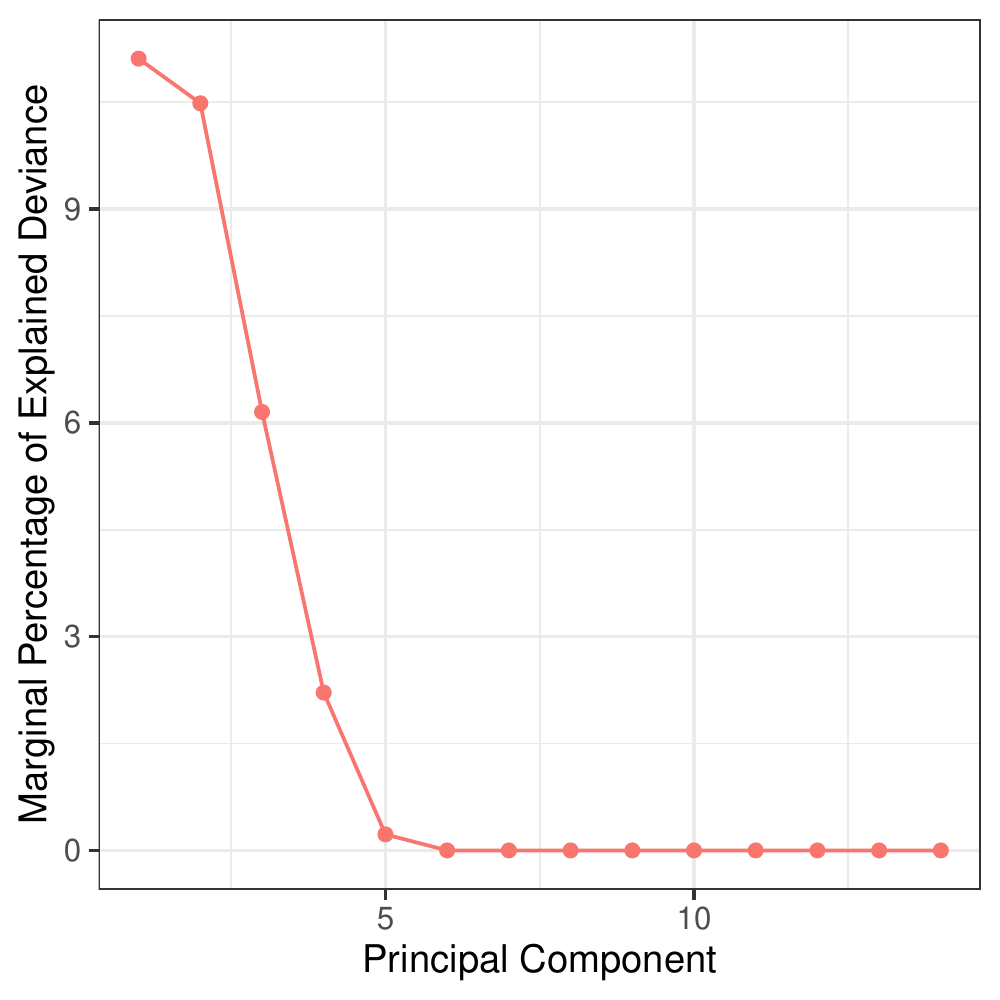}
    \end{minipage}
    \hfill
        \begin{minipage}[b]{0.45\textwidth}
  \centering    
  \includegraphics[height=65mm,width=70mm]{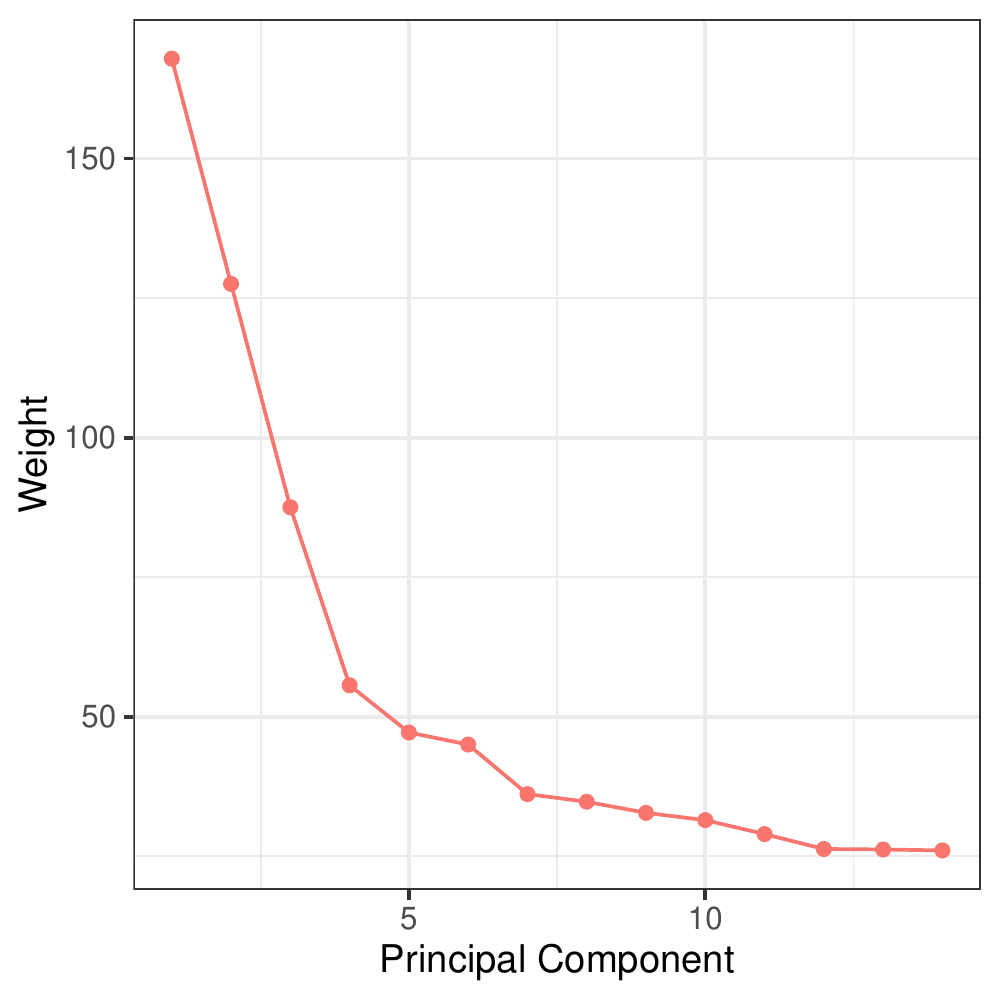}
  \end{minipage}  \hfill    
    \caption{Explained deviance and weight versus the number of principal components for the nations data.} 
    \label{da2}
\end{figure}

 To decide a proper rank of logistic CP decomposition, we fit a rank-$14$ logistic CP decomposition first. We find that there are 3 weights much larger than other weights as shown in Figure \ref{da2} and the offset term $\mu$ is estimated to be  $-1.69$.  In Figure \ref{da2}, the scree plot of marginal explained deviance suggests $R=4$. Based on the information, we conclude that a rank-$4$ logistic CP decomposition is reasonable for the nations data.

\begin{figure}[H]  
  \begin{minipage}[b]{0.45\textwidth}
  \centering
    \includegraphics[width=\textwidth] %\includegraphics[height=50mm,width=50mm]
    {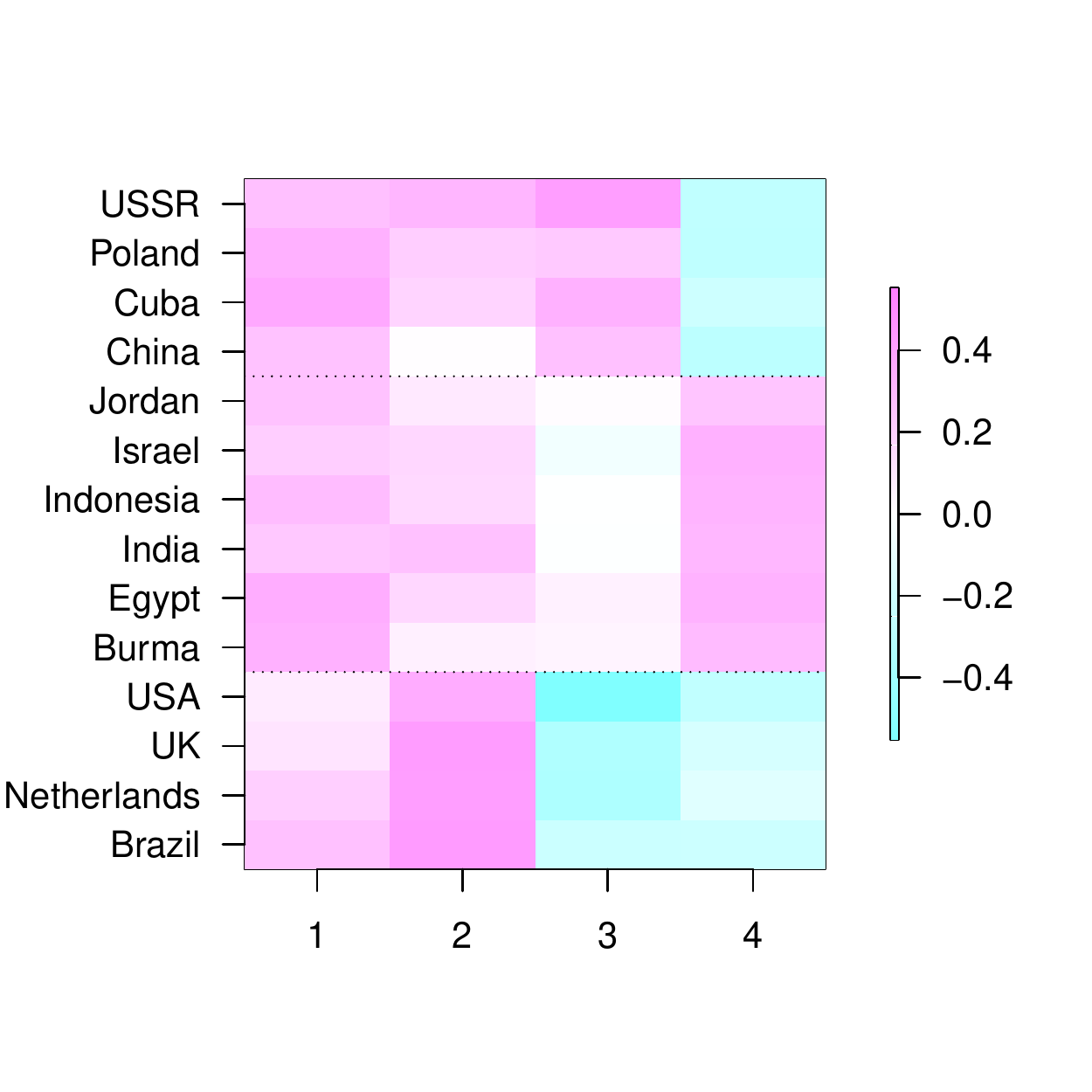}
  \end{minipage}   
    \begin{minipage}[b]{0.45\textwidth}
  \centering
    \includegraphics[width=\textwidth] %\includegraphics[height=50mm,width=50mm]
    {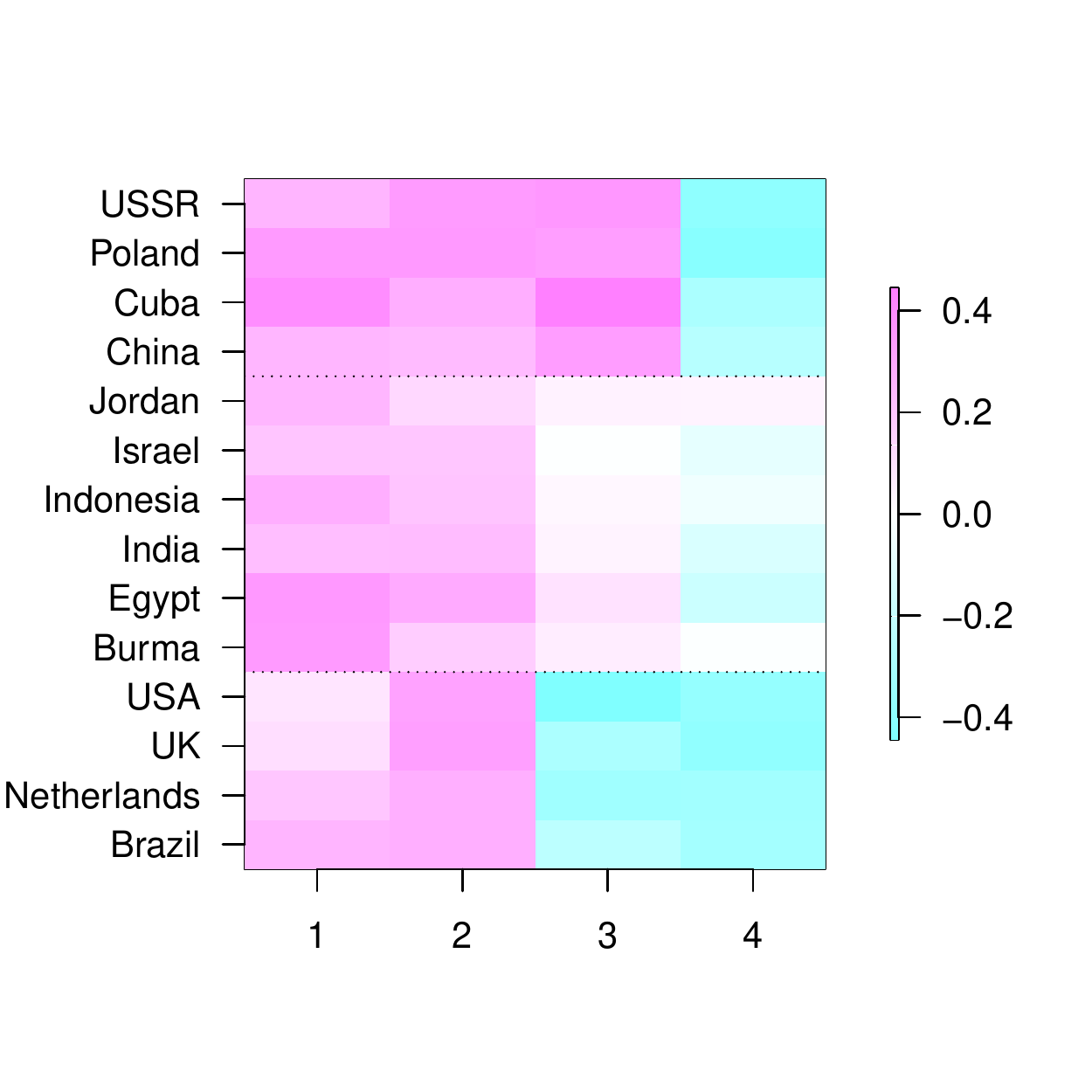}
  \end{minipage}     
  \caption{Heatmap of the estimated nation factors from the  un-regularized model (left) and regularized model (right).} 
 \label{daA1}
    \end{figure}

\begin{figure}[H]
  \centering
   \vspace{-8mm}
    \includegraphics[width=170mm]{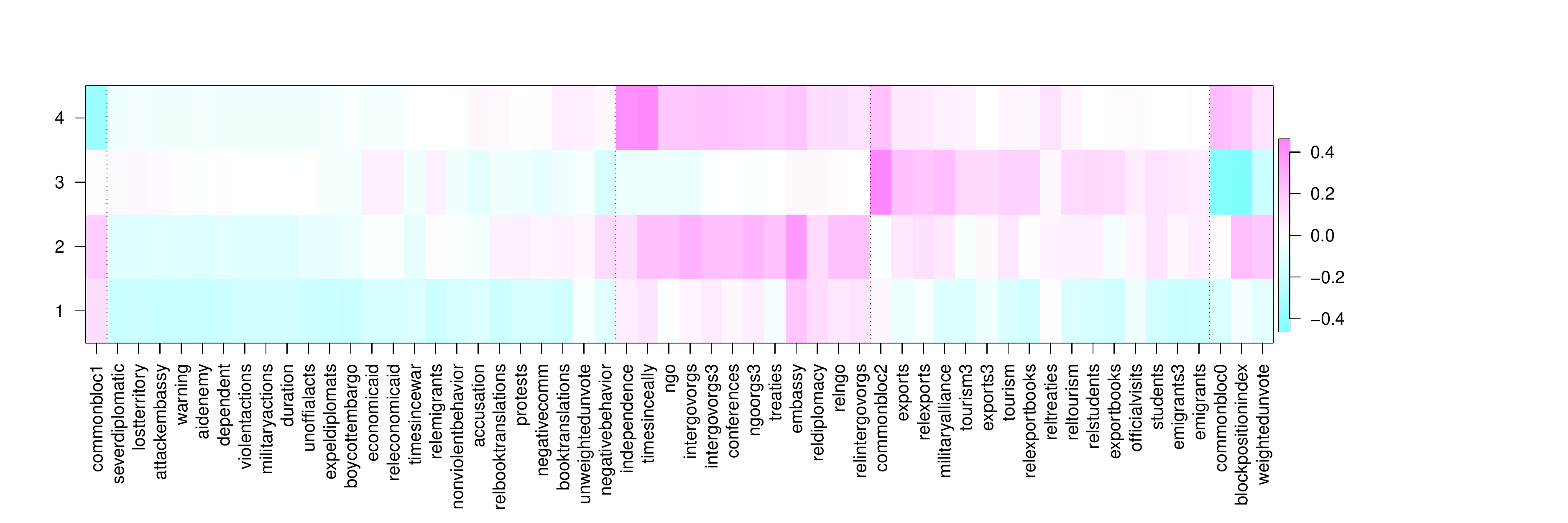}
  \caption{Heatmap of the estimated relation factors without regularization.} 
   \label{daC1}
       \end{figure}

To get a better understanding of factors, we apply $K$-means clustering \citep{macqueen1967some, lloyd1982least} on the estimated factors for the nations and relations, and visualize them in Figures \ref{daA1} and \ref{daC1}. The estimated number of clusters can be determined by BIC criterion, where $K=3$ is chosen for the nation factors, and $K=5$ is chosen for the relation factors.
The three clusters of nations contain communist countries
(\textit{USSR,  Poland, Cuba, China}), western countries (\textit{USA,
  UK, Netherlands, Brazil}), and neutral countries. The relations are grouped into five clusters. Three major clusters regard i) negative/hostile
actions (e.g.,  \textit{warning, protests, accusation, military actions}), ii)
international partnerships through intergovernmental organizations and 
NGOs (e.g., \textit{intergovorgs, relngo, ngo}), and iii) exports and population
exchanges (e.g., \textit{exportbook, exports, students,
  emigrants}). The remaining two minor clusters are defined by the
nation's common bloc membership: opposing common bloc membership (\textit{commmonbloc0,
  blockpositionindex, weightedunvote}) and different common bloc membership (\textit{commmonbloc1}).

\begin{figure}[H]
  \centering
    \vspace{-8mm}
    \includegraphics[width=170mm]{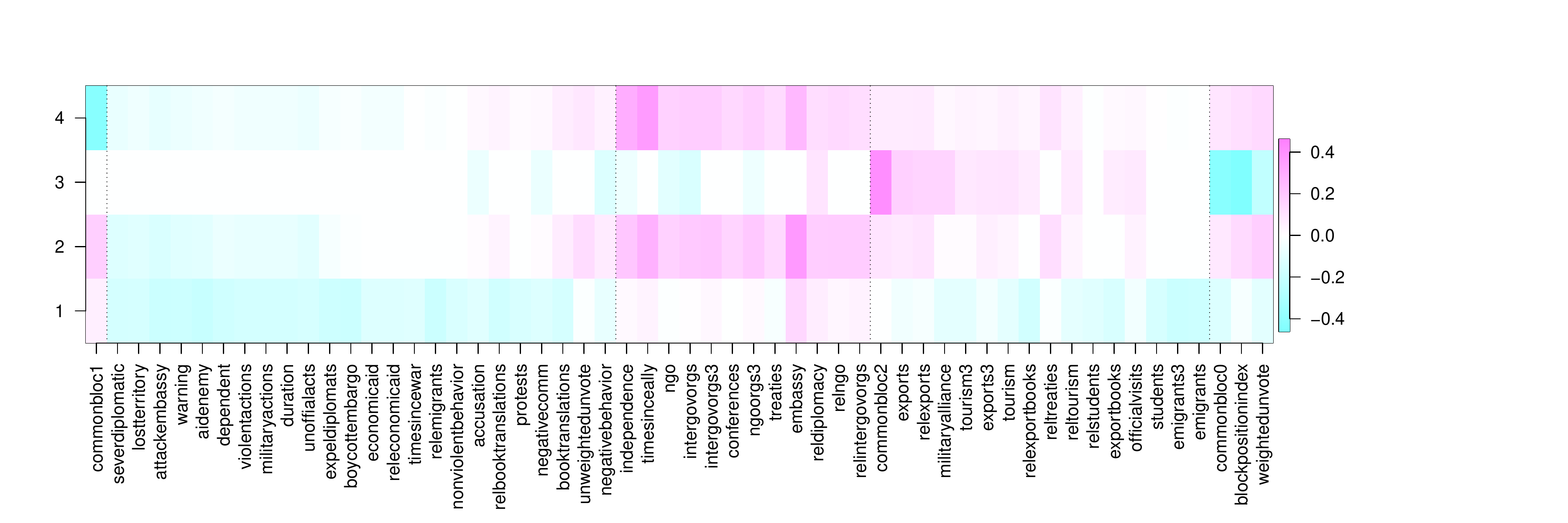}
  \caption{Heatmap of the estimated relation factors with regularization.} 
   \label{daC3}
       \end{figure}

\subsection{Co-clustering of Nations and Relations}

The two-way clustering methods in \cite{lee2010biclustering} and \cite{lee2014biclustering} have been proven to be successful for analysis of continuous and binary matrix data. The core idea of two-way clustering is imposing sparsity inducing penalties on the row score vector $\bu_{r}$ and column loading vector $\bv_{r}$ in the SVD of centered data matrix or centered logit parameter matrix. This could yield a checkerboard-like structure for each rank-one matrix $d_{r}\cdot\bu_{r}\circ\bv_{r}$ for $r\in[R]$.
By penalizing $\bu_{r}$ and $\bv_{r}$ in the $r$th component, the rows with nonzero $u_{ir}$ are naturally clustered together, and the columns with nonzero $v_{jr}$ are naturally clustered together.
So penalization on both the score and loading vectors could simultaneously link sets of rows and sets of columns together, and reveal some desirable row-column association.

More generally, co-clustering methods could cluster related variables in each factor for tensor data. For simplicity of explanation, we focus on the three-way clustering and assume $\bu_{r}, \bv_{r}$ and $\bw_{r}$ are the factors in the CP decomposition. If $\bu_{r}, \bv_{r}$ and $\bw_{r}$ are sparse in the $r$th component, then the non-zero entries in $d_{r}\cdot\bu_{r}\circ\bv_{r}\circ\bw_{r}$ form a sub-tensor for co-clustering, which could help to identify checkerboard-like local patterns of different modes.  It's worth noting that different components may identify different co-clusters, and the identified co-clusters may overlap.  If at least one of $\bu_{r}, \bv_{r}$ and $\bw_{r}$ has the same sign in the $r$th component, then this component is viewed as a global pattern or a main effect.  For continuous tensor data, 
\cite{allen2012sparse} applied the sparse CP decomposition in clustering multi-way microarray data. For the analysis of the nations data, we could apply our methods for clustering associated nations and relations.

The heatmap for the estimated nation factors is displayed in the left panel of Figure \ref{daA1}. It doesn't have a sparse pattern. By contrast, the heatmap for the relation factors in Figure \ref{daC1} does suggest a potential benefit of sparsity because many entries are close to zero. Therefore we fit a rank-$4$ sparse logistic CP decomposition with an $\ell_{0}$-norm constraint on the relation factors $W$. More specifically, this model imposes the constraints $\bu_{r}=\bv_{r}$ and $\|\bw_{r}\|_{0}\leq s_{3r}$, where tuning parameters $s_{3r}$ control the number of nonzero entries in $\bw_{r}$ for $r\in[R]$. The sparse estimated relation factors are presented in Figure \ref{daC3}, where the nonzero entries reveal important relations in each component.

Based on the estimated nation factors $\bu_{r}$ and relation factors $\bw_{r}$, we could build rank-one tensors $d_{r}\cdot\bu_{r}\circ\bv_{r}\circ\bw_{r}$ for $r\in[R]$. For visualization,  we display rank-one matrices $d_{r}\cdot\bu_{r}\circ\bw_{r}$ for $r\in[R]$.  
%For nation versus nation, we ordered $\bu$ and $\bv$ in an increasing order. 
With given $r\in[R]$, $\bu_{r}$ and $\bw_{r}$ split the nations and
relations into two or three clusters according to the sign of the
entries, and therefore produce the clusters of nations and relations.

\begin{figure}[H]
   \centering
     \vspace{-5mm}
    \includegraphics[width=150mm]%\includegraphics[height=40mm,width=120mm]
    {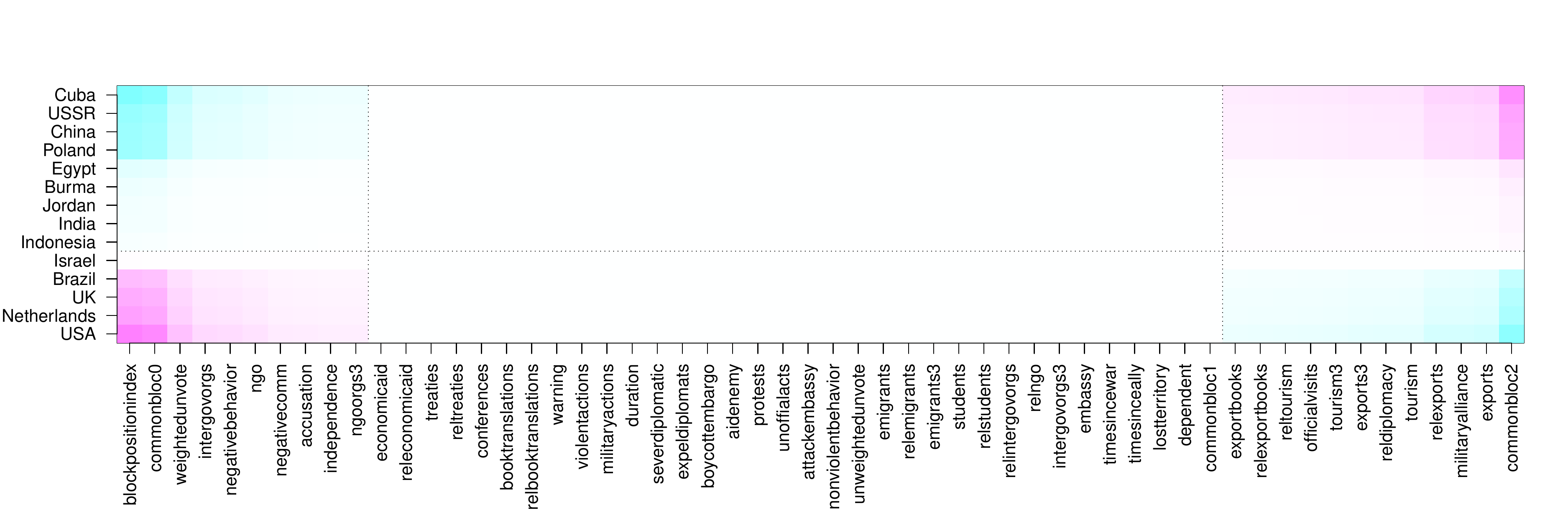}
  \caption{Heatmap of relation versus nation  for component 3 with regularization.} 
 \label{hmap_3}
     \end{figure}

Figure \ref{hmap_3}  shows the heatmap of $d_{r}\cdot\bu_{r}\circ\bw_{r}$ for component 3. 
In the heatmap, the entries of $\bu_{r}$ and $\bw_{r}$ are arranged in increasing order.
For the $x$-axis of the heatmap, factor $\bw_{r}$ is displayed with entries in increasing order from left to right. For the $y$-axis of the heatmap, factor $\bu_{r}$ is displayed with entries in increasing order from bottom to top. 

For component 3 shown in Figure \ref{hmap_3}, negative values of the
relation factor are associated with opposing common bloc membership,
and positive values are associated with common bloc membership and
resulting economic and cultural relations through exports and tourism. The nations are
clearly separated into three groups.
The nations with positive values are countries in the Communist bloc, and the nations
 with negative values are countries in the  Western bloc. Neutral
 countries have almost zero values. We may as well consider imposing
 sparsity on this nations factor.  As a form of interaction between nations and
 political relations, this component captures opposite political interactions between communist and western countries. 
 It reveals a natural partition of the countries and clustering of
 relations as shown in Figure \ref{hmap_3}. 
 This co-clustering suggested by the component is sensible, and the countries in the same
 cluster tend to share similar relation patterns.

While component 3 reflects a strong interaction between nations
and political relations, the first two components mostly indicate 
main effects of the relations. The heatmaps of other components can be
found in Appendix \ref{appda}.

\section{Conclusions and Discussion} \label{sec:conclusion}

In this paper, we have proposed several novel tensor decomposition
methods for binary tensor data using the CP decomposition of a logit
parameter tensor.
We have mainly focused on three-way tensors in the paper,
but similar methods can be developed for higher-order data.
Starting with logistic CP decomposition, we have 
incorporated an $\ell_{1}$-norm or $\ell_{0}$-norm constraint on
factors into the tensor decomposition formulation. To estimate factor matrices in logistic CP
decomposition, we have developed computational algorithms that combine MM algorithm and variants of
tensor power method. By imposing sparsity constraints on the factor matrices,
we could identify and select important features in each factor.
Sparse logistic CP decompositions can capture local multi-way
interactions and therefore facilitate co-clustering of entities in different modes.
Such co-clusters can reveal interesting associations between different modes.

There are several directions worth further investigation.
As a structural element in logistic tensor decomposition, we have considered
 a constant offset term only. 
However, main effects along each mode are likely to be significant systematic
elements in many applications as evidenced in the 
 nations data analysis as well. From a modeling point of view,
 including additive main effects in the decomposition and using a small number
 of  sparse rank-one tensors for multiplicative interactions will be a fruitful direction
 for extension. A similar logistic ANOVA model has been proposed for
 binary matrix data \citep{jung2014biomarker}.

 As another extension, we could generalize the current formulation with CP decomposition for binary data to a Tucker decomposition and develop a corresponding regularized version. 
Besides, we could replace the $\ell_{0}$-norm and $\ell_{1}$-norm
penalties with general penalties such as fused lasso
\citep{tibshirani2005sparsity} in certain applications. For example,
when one mode of a given tensor represents time points, smoothness in
temporal factors might be desired.

Throughout the paper we only discuss the logit link function for
binary data, but we could also develop logistic CP decompositions
with the probit link function (i.e., $\theta=\Phi^{-1}(p)$ using the cdf of
standard normal distribution $\Phi$). Similar to the logit link, we can use
the quadratic majorization of $-\log\Phi(x) $ from \cite{deLeeuw2006} to devise MM algorithms.

On the theoretical front, it is of interest to extend the work of
\cite{montanari2014statistical} to binary tensors and investigate
conditions on the signal-to-noise ratio to recover true factor
matrices from an observed binary tensor with high probability.
Moreover, the optimality of model selection approaches in binary tensor
decomposition is still unknown, and it is worth investigating the consistency of AIC,
BIC or other information criteria.

Last but not least, we could develop similar methods for tensor
decompositions in the natural parameter space for other types of
exponential family data.
For example, tensor data with counts or ratings as entries are common in
recommender systems. Sparse Poisson or multinomial CP decompositions will be
useful extensions of the current work.

\section*{Acknowledgments} 
This research was supported in part by the National Science Foundation Grants DMS-15-13566 and DMS-20-15490.

%\begin{supplement}
%\stitle{Title of Supplement A}
%\sdescription{Short description of Supplement A.}
%\end{supplement}
%\begin{supplement}
%\stitle{Title of Supplement B}
%\sdescription{Short description of Supplement B.}
%\end{supplement}

%\begin{acks}[Acknowledgments]
%The authors would like to thank the anonymous referees, an Associate
%Editor and the Editor for their constructive comments that improved the
%quality of this paper.
%\end{acks}

\renewcommand\refname{References}
\bibliographystyle{apalike}
  % \bibliographystyle{imsart-nameyear} 
%\bibliography{slds}
   \bibliography{mybib}

\newpage

   \section*{Appendix}

\begin{appendix}

\section{Algorithms} \label{app:algorithm}
%\label{appC}
We summarize the proposed algorithms for logistic tensor decomposition
with alternating least squares
method, tensor soft-thresholding power method and tensor truncated
power method in Algorithms \ref{mmals}, \ref{mmtp} and \ref{mmttp}.

\begin{algorithm}[H]  
\caption{MM-Alternating Least Squares algorithm for logistic CP decomposition}
\small
\label{mmals}

\begin{algorithmic}[1]

\STATE \textbf{input:} tensor $\mathcal{X}$ and rank $R$.

\STATE Initialize  with $\widehat{\mu}^{[0]}$ and $(\widehat{\bd}^{[0]},\widehat{U}^{[0]}, \widehat{V}^{[0]}, \widehat{W}^{[0]})$. Set $m=0$.

\REPEAT
%\FOR {$m=1$ {\bfseries to} $M$}

\STATE Compute $\mathcal{Z}^{[m]}$ in $\eqref{newx}$.

\STATE Update $\widehat{\mu}^{[m+1]}$.

\STATE Compute $\mathcal{Z}_{c}^{[m]}=\mathcal{Z}^{[m]}-\widehat{\mu}^{[m+1]}\one_{p_{1}p_{2}p_{3}}$.

\STATE Initialize  with $(\widehat{\bd}^{[m]}, \widehat{U}^{[m]}, \widehat{V}^{[m]}, \widehat{W}^{[m]})$.

 \REPEAT
\STATE $U\leftarrow \text{Normalize}(Z_{c(1)}^{[m]}[(W\odot V)^{T}]^{\dagger})$

\STATE
$V\leftarrow \text{Normalize}(Z_{c(2)}^{[m]}[(U\odot W)^{T}]^{\dagger})$

\STATE
$W\leftarrow \text{Normalize}(Z_{c(3)}^{[m]}[(U\odot V)^{T}]^{\dagger})$.

\STATE $\bd \leftarrow \text{column norms of}\; Z_{c(3)}^{[m]}[(U\odot V)^{T}]^{\dagger}$

\UNTIL{converge}

\STATE Update $(\widehat{\bd}^{[m+1]},\widehat{U}^{[m+1]}, \widehat{V}^{[m+1]}, \widehat{W}^{[m+1]})$.

%\STATE Update $\Theta^{[m+1]}$.

\STATE $m\leftarrow m+1$

\UNTIL{converge}
%\ENDFOR

%\STATE Normalize columns of $\widehat{W}$, and store their norms as $\widehat{\bd}$

\STATE \textbf{output:} $\widehat{\mu}$ and $(\widehat{\bd},\widehat{U}, \widehat{V}, \widehat{W})$.

 \end{algorithmic} 
 \end{algorithm}

\begin{algorithm}[H] 
\small 
\caption{MM-Tensor Soft-thresholding Power algorithm for sparse logistic CP decomposition}
\label{mmtp}
\begin{algorithmic}[1]
\STATE  \textbf{input:} tensor $\mathcal{X}$, number of initializations $L$, rank $R$, and penalization vector $(c_{1},c_{2},c_{3})$.

\STATE Initialize with $\widehat{\mu}_{\tau}^{[0]}$ and $(\widehat{\bu}_{\tau}^{[0]},\widehat{\bv}_{\tau}^{[0]},\widehat{\bw}_{\tau}^{[0]})$ where $\tau\in[L]$. Set $m=0$.

\FOR {$\tau=1$ {\bfseries to} $L$}

\REPEAT

\STATE Compute $\mathcal{Z}^{[m]}$ in $\eqref{newx}$.

\STATE Update $\widehat{\mu}^{[m+1]}$.

\STATE Compute $\mathcal{Z}_{c}^{[m]}=\mathcal{Z}^{[m]}-\widehat{\mu}^{[m+1]}\one_{p_{1}p_{2}p_{3}}$.

\REPEAT

\STATE $\bu_{\tau}=\text{Normalize}(S(\mathcal{Z}_{c}^{[m]}\times_{2}(\bv_{\tau})^{T}\times_{3}(\bw_{\tau})^{T},\lambda_{1}))$

\STATE $\bv_{\tau}=\text{Normalize}(S(\mathcal{Z}_{c}^{[m]}\times_{1}(\bu_{\tau})^{T}\times_{3}(\bw_{\tau})^{T},\lambda_{2}))$

\STATE $\bw_{\tau}=\text{Normalize}(S(\mathcal{Z}_{c}^{[m]}\times_{1}(\bu_{\tau})^{T}\times_{2}(\bv_{\tau})^{T},\lambda_{3}))$

%$\|\mathcal{X}^{[m]}\times_{1}\bu^{T}\times_{2}\bv^{T}\|_{2}$

%\ENDFOR
\UNTIL{converge}

\STATE Update $(\widehat{d}_{\tau}^{[m+1]}, \widehat{\bu}_{\tau}^{[m+1]}, \widehat{\bv}_{\tau}^{[m+1]}, \widehat{\bw}_{\tau}^{[m+1]})$.

\STATE $m\leftarrow m+1$

\UNTIL{converge}
%\ENDFOR

\STATE Return $\widehat{\mu}_{\tau}$ and $(\widehat{d}_{\tau},\widehat{\bu}_{\tau},\widehat{\bv}_{\tau},\widehat{\bw}_{\tau})$.

\ENDFOR

 \STATE Cluster $\{(\widehat{d}_{\tau},\widehat{\bu}_{\tau},\widehat{\bv}_{\tau},\widehat{\bw}_{\tau}),\tau\in[L]\}$ into $R$ clusters $\{(\widehat{d}_{j},\widehat{\bu}_{j},\widehat{\bv}_{j},\widehat{\bw}_{j}), j\in[R]\}$ 
by Algorithm \ref{clust}.

\STATE  \textbf{output:} $\widehat{\mu}$ and $R$ clusters $\{(\widehat{d}_{j},\widehat{\bu}_{j},\widehat{\bv}_{j},\widehat{\bw}_{j}), j\in[R]\}$.
%\UNTIL{Converge}

  \end{algorithmic} 
 \end{algorithm}

\begin{algorithm}[H]  
\small
\caption{MM-Tensor Truncated Power algorithm for sparse logistic CP decomposition}
\label{mmttp}
\begin{algorithmic}[1]
\STATE \textbf{input:} tensor $\mathcal{X}$, number of initializations $L$, rank $R$, and cardinality vector $(s_{1},s_{2},s_{3})$.

\STATE Initialize with $\widehat{\mu}_{\tau}^{[0]}$ and $(\widehat{\bu}_{\tau}^{[0]},\widehat{\bv}_{\tau}^{[0]},\widehat{\bw}_{\tau}^{[0]})$ where $\tau\in[L]$. Set $m=0$.

\FOR {$\tau=1$ {\bfseries to} $L$}

\REPEAT

\STATE Compute $\mathcal{Z}^{[m]}$ in $\eqref{newx}$.

\STATE Update $\widehat{\mu}^{[m+1]}$.

\STATE Compute $\mathcal{Z}_{c}^{[m]}=\mathcal{Z}^{[m]}-\widehat{\mu}^{[m+1]}\one_{p_{1}p_{2}p_{3}}$.

%\FOR {$n=1$ {\bfseries to} $N_{m}$}
\REPEAT

\STATE 
$\bu_{\tau}=\text{Normalize}(T(\mathcal{Z}_{c}^{[m]}\times_{2}(\bv_{\tau})^{T}\times_{3}(\bw_{\tau})^{T},s_{1}))$

%\text{Normalize}(

\STATE 
$\bv_{\tau}=\text{Normalize}(T(\mathcal{Z}_{c}^{[m]}\times_{1}(\bu_{\tau})^{T}\times_{3}(\bw_{\tau})^{T},s_{2}))$

\STATE
$\bw_{\tau}=\text{Normalize}(T(\mathcal{Z}_{c}^{[m]}\times_{1}(\bu_{\tau})^{T}\times_{2}(\bv_{\tau})^{T},s_{3}))$

\UNTIL{converge}

\STATE Update $(\widehat{d}_{\tau}^{[m+1]}, \widehat{\bu}_{\tau}^{[m+1]}, \widehat{\bv}_{\tau}^{[m+1]}, \widehat{\bw}_{\tau}^{[m+1]})$.

\STATE $m\leftarrow m+1$

\UNTIL{converge}
%\ENDFOR

\STATE Return $\widehat{\mu}_{\tau}$ and $(\widehat{d}_{\tau},\widehat{\bu}_{\tau},\widehat{\bv}_{\tau},\widehat{\bw}_{\tau})$.

\ENDFOR

 \STATE Cluster $\{(\widehat{d}_{\tau},\widehat{\bu}_{\tau},\widehat{\bv}_{\tau},\widehat{\bw}_{\tau}),\tau\in[L]\}$ into $R$ clusters $\{(\widehat{d}_{j},\widehat{\bu}_{j},\widehat{\bv}_{j},\widehat{\bw}_{j}), j\in[R]\}$ 
by Algorithm \ref{clust}.
 
 \STATE \textbf{output:} $\widehat{\mu}$ and $R$ clusters $\{(\widehat{d}_{j},\widehat{\bu}_{j},\widehat{\bv}_{j},\widehat{\bw}_{j}), j\in[R]\}$.
  
  \end{algorithmic} 
 \end{algorithm}

\section{Simulation Study}\label{appsim}

\subsection{BIC and AIC}

\begin{figure}[H]
  \centering
      \begin{minipage}[b]{0.45\textwidth}
    \includegraphics[height=65mm,width=70mm]{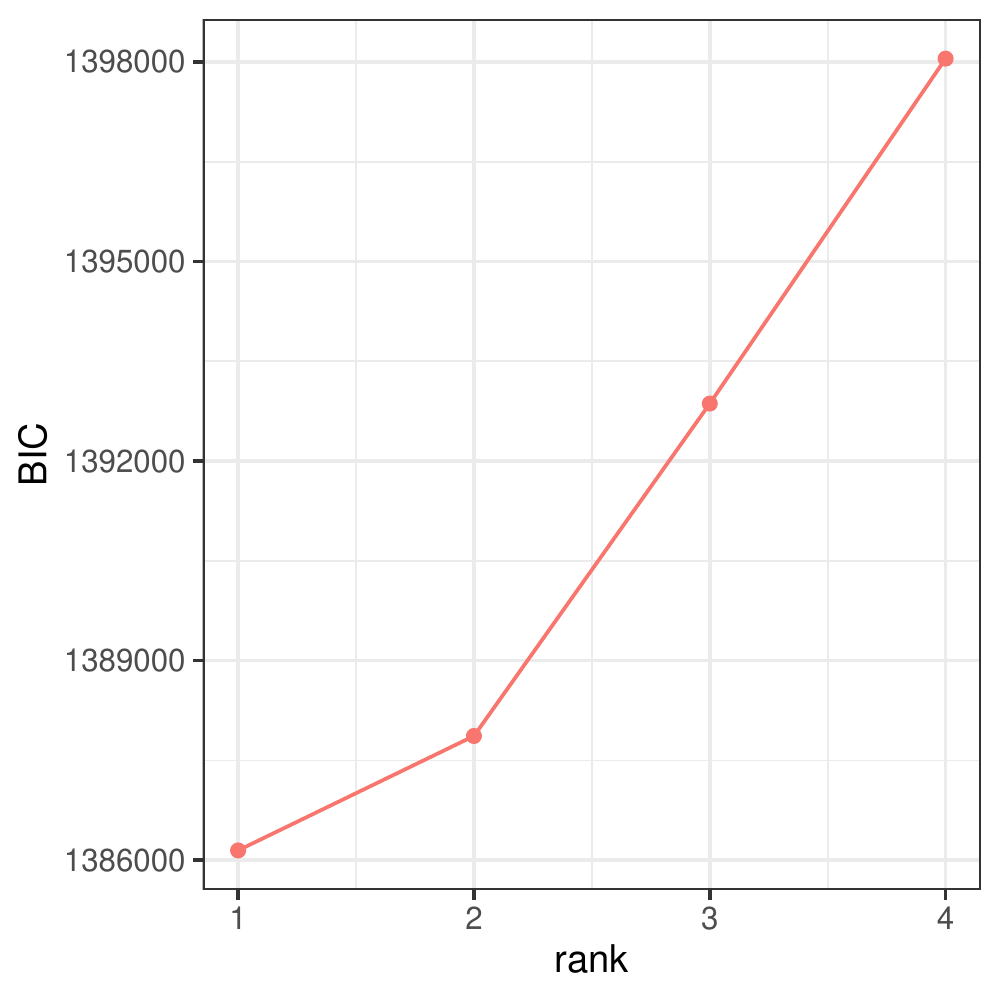}
  \end{minipage}
  \hfill    
  \begin{minipage}[b]{0.45\textwidth}
    \includegraphics[height=65mm,width=70mm]{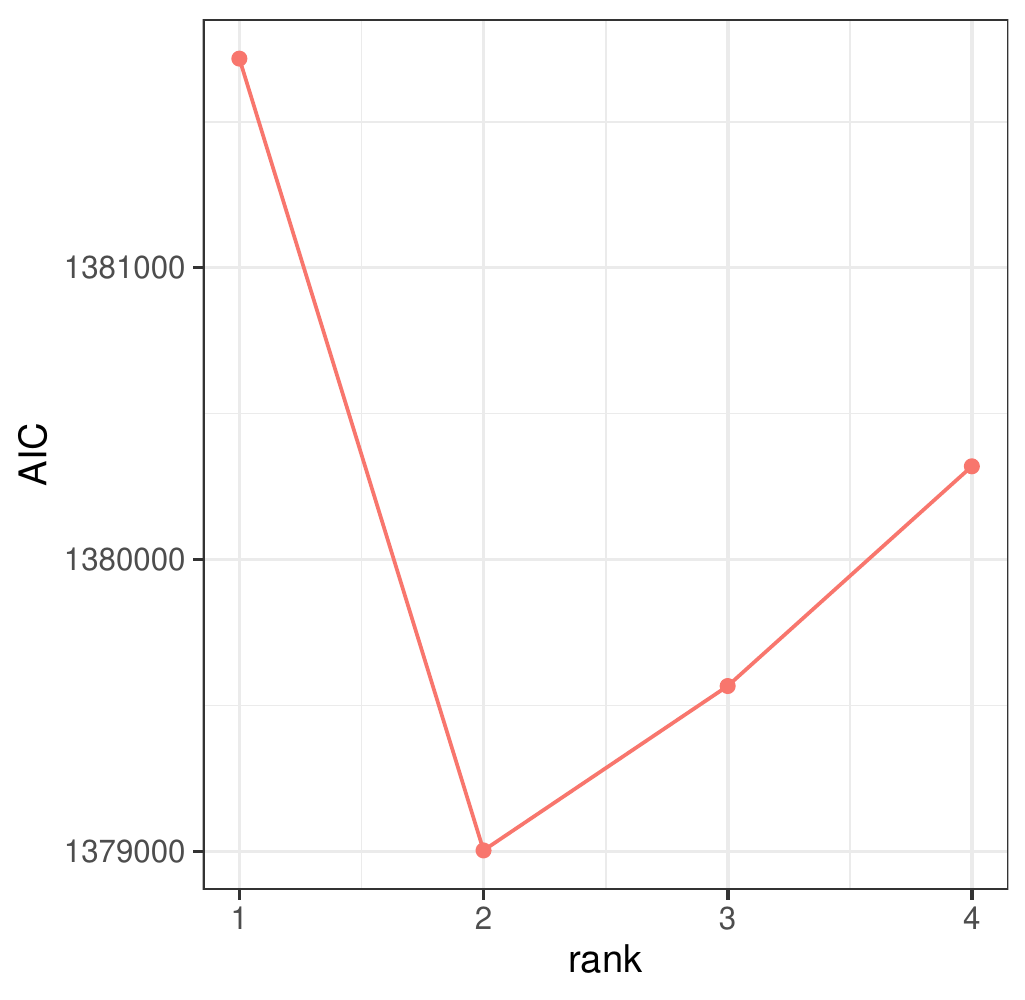}
    \end{minipage}
     \caption{ BIC and AIC versus rank $R$ for simulated data with $p_{1}=1000, p_{2}=100, p_{3}=10$ and $R=2$.}        
  \label{rank}  
    \end{figure}

\begin{figure}[H]
  \centering
  \begin{minipage}[b]{0.45\textwidth}
    \includegraphics[height=65mm,width=70mm]{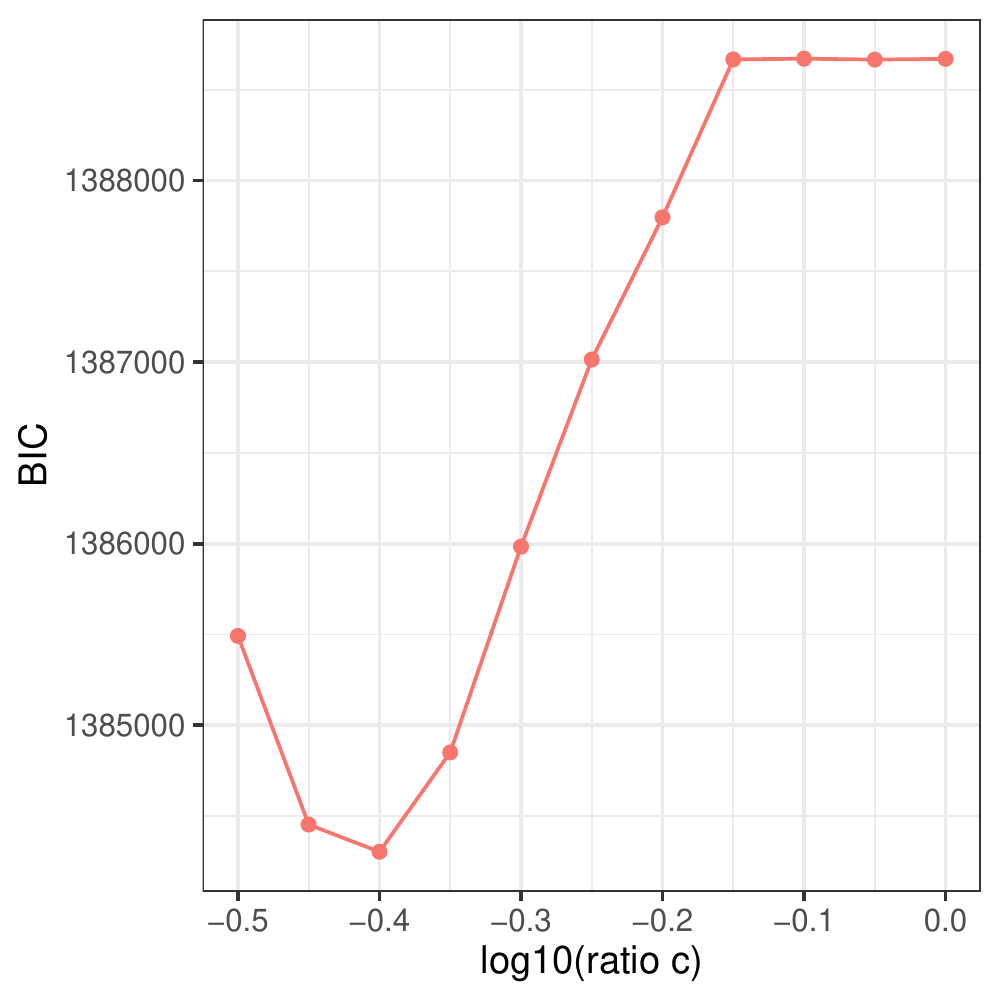}
    \end{minipage}
  \hfill
  \begin{minipage}[b]{0.45\textwidth}
    \includegraphics[height=65mm,width=70mm]{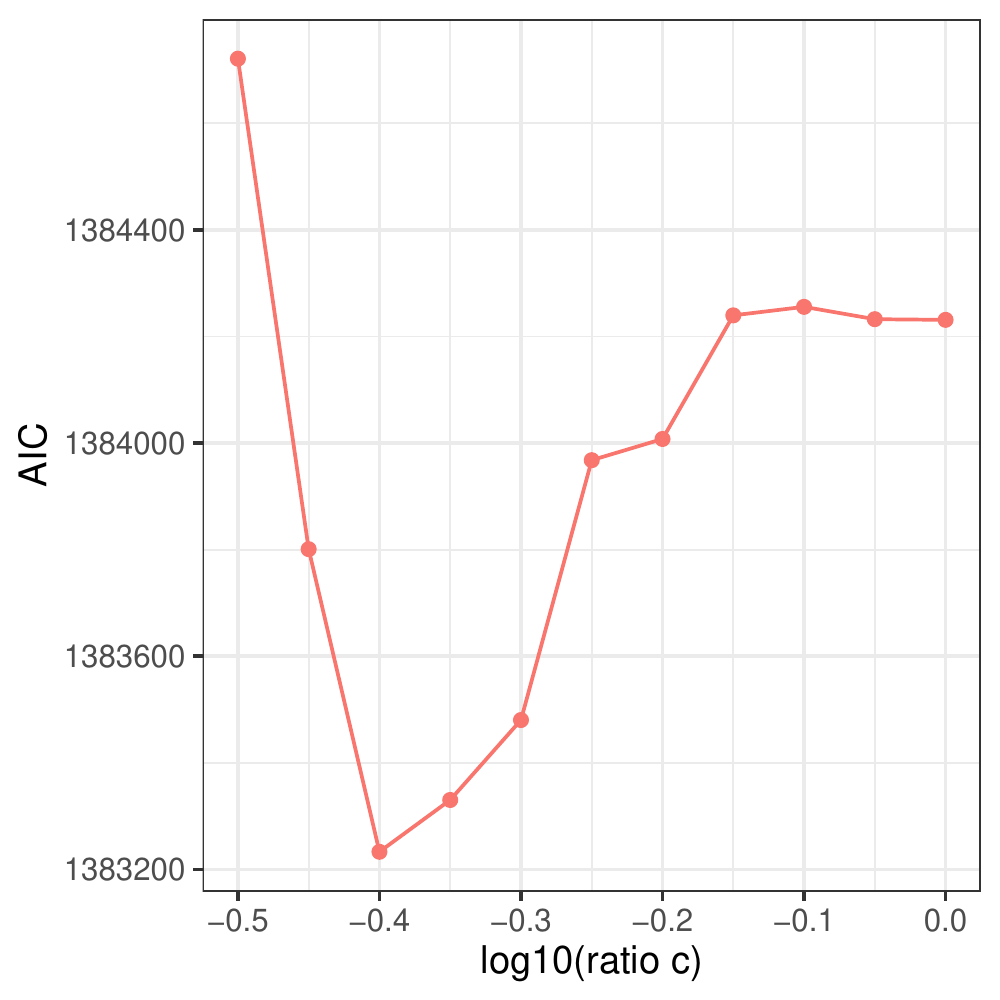}
  \end{minipage}
     \caption{ BIC and AIC versus ratio $c$ when rank $R$ is fixed at $2$ for simulated data with $p_{1}=1000, p_{2}=100, p_{3}=10$ and $R=2$.}        
  \label{ratioc}  
    \end{figure}

\begin{figure}[H]
  \centering
      \begin{minipage}[b]{0.45\textwidth}
    \includegraphics[height=65mm,width=70mm]{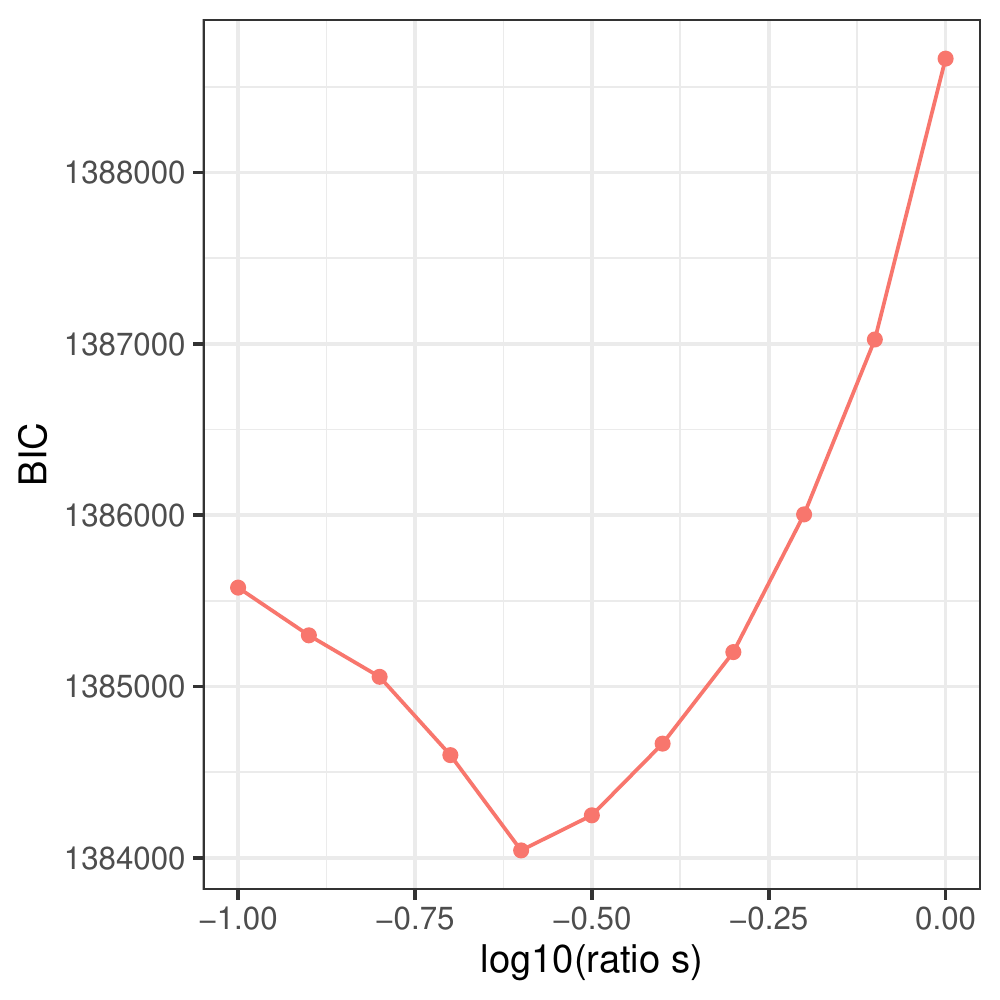}
    \end{minipage}
  \hfill
  \begin{minipage}[b]{0.45\textwidth}
    \includegraphics[height=65mm,width=70mm]{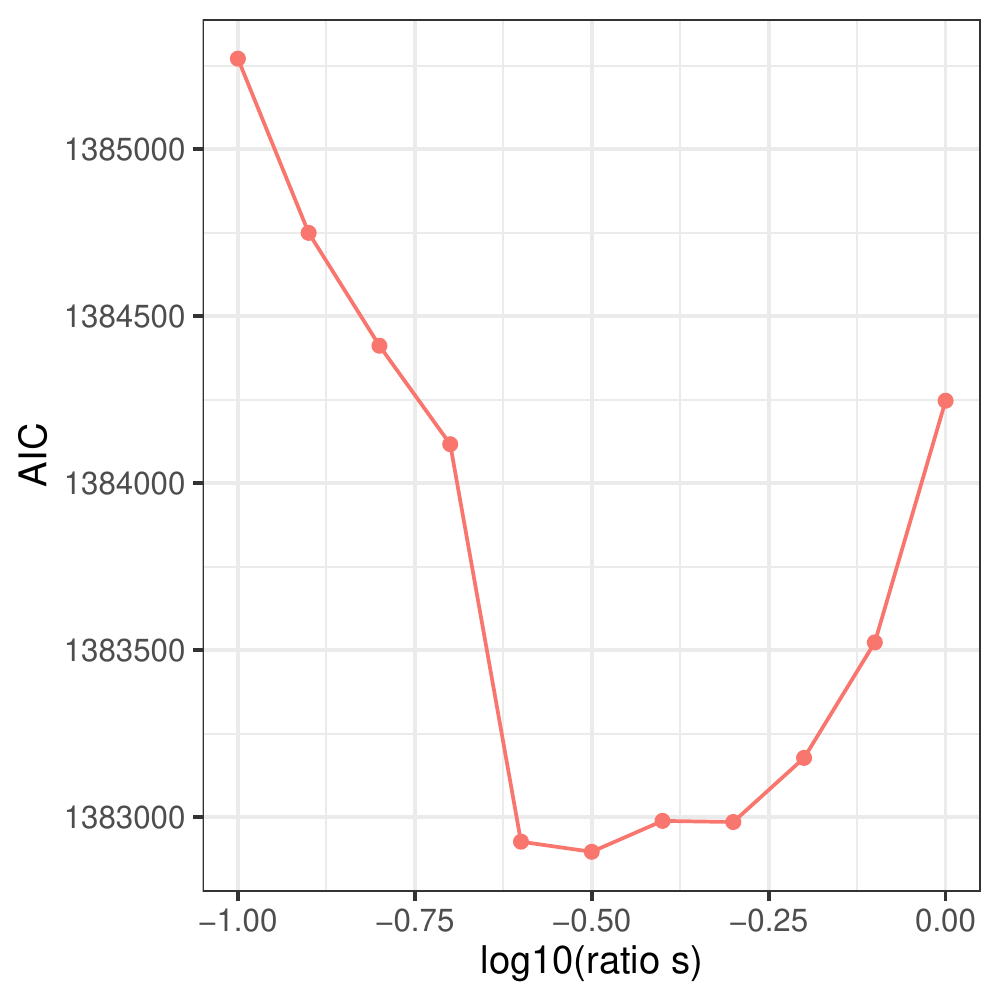}
  \end{minipage}
     \caption{BIC and AIC versus ratio $s$ when rank $R$ is fixed at $2$ for simulated data with $p_{1}=1000, p_{2}=100, p_{3}=10$ and $R=2$.}        
  \label{ratios}  
    \end{figure}

For illustration, we simulated binary data of size $p_{1}=1000, p_{2}=100$ and $p_{3}=10$ from a rank-2 logit tensor. Due to the sparsity in factors, the underlying logit tensor is also very sparse. 
The level of sparsity in each dimension is set to be equal.
The details of the simulation setting can be found in Section \ref{setup}.
In this setting, we could express the tuning parameters 
 $c_{i}=c\sqrt{p_{i}}$ or $s_{i}=sp_{i}$ using a common ratio parameter $c$ or $s$.
For feasibility, the ratio parameters $c$ and $s$ should satisfy $\underset{i}{\max}\frac{1}{\sqrt{p_{i}}}\leq c\leq1$ and $\underset{i}{\max}\frac{1}{p_{i}}\leq s\leq1$, respectively.
Figure \ref{rank} shows how BIC and AIC change with rank $R$ for the simulated data.  It shows that AIC selects the true rank ($R=2$) correctly while BIC chooses a smaller rank ($R=1$).
For the same data, Figures \ref{ratioc} and \ref{ratios} show how BIC and AIC change with ratio $c\in [\underset{i}{\max}\frac{1}{\sqrt{p_{i}}},1]$ and ratio $s\in [\underset{i}{\max}\frac{1}{p_{i}},1]$  given rank $R=2$. Clearly, regularized models have a smaller BIC/AIC than the un-regularized model, and the optimal tuning parameter can be selected by minimizing BIC and AIC.  Our limited experiments suggest that AIC tends to be more accurate than BIC in selecting the rank $R$.

\subsection{Cross-validation}

\begin{figure}[H]
   \begin{minipage}[b]{1\textwidth}
  \centering
    \includegraphics[height=64mm,width=80mm]{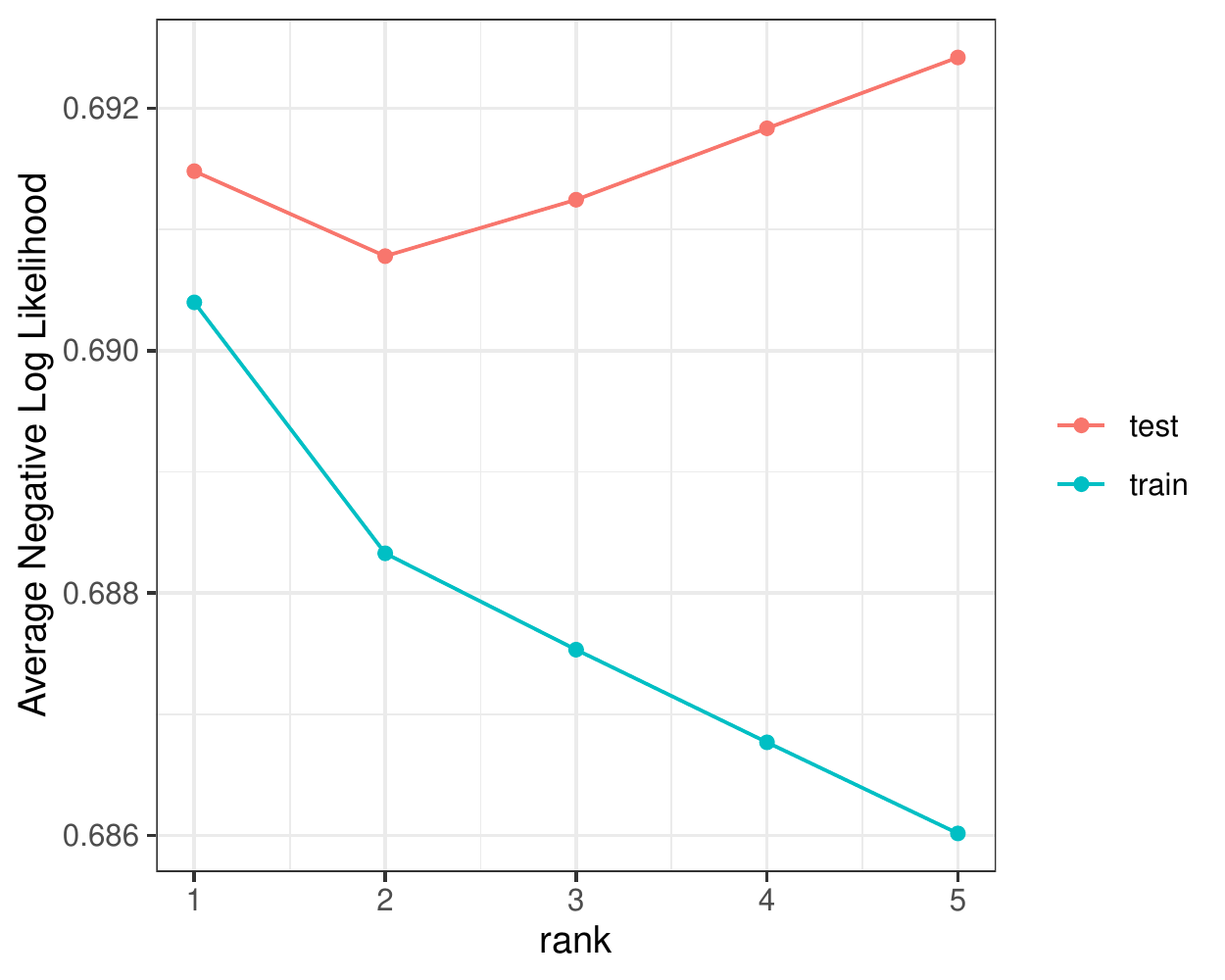}
  \end{minipage}
 \caption{Training error and test error measured in terms of the average negative log likelihood versus rank $R$ for simulated data with $p_{1}=1000, p_{2}=100, p_{3}=10$ and $R=2$. }
       \label{cvr}
            \end{figure}

\begin{figure}[H]
  \begin{minipage}[b]{0.45\textwidth}
  \centering
   \includegraphics[height=64mm,width=80mm]{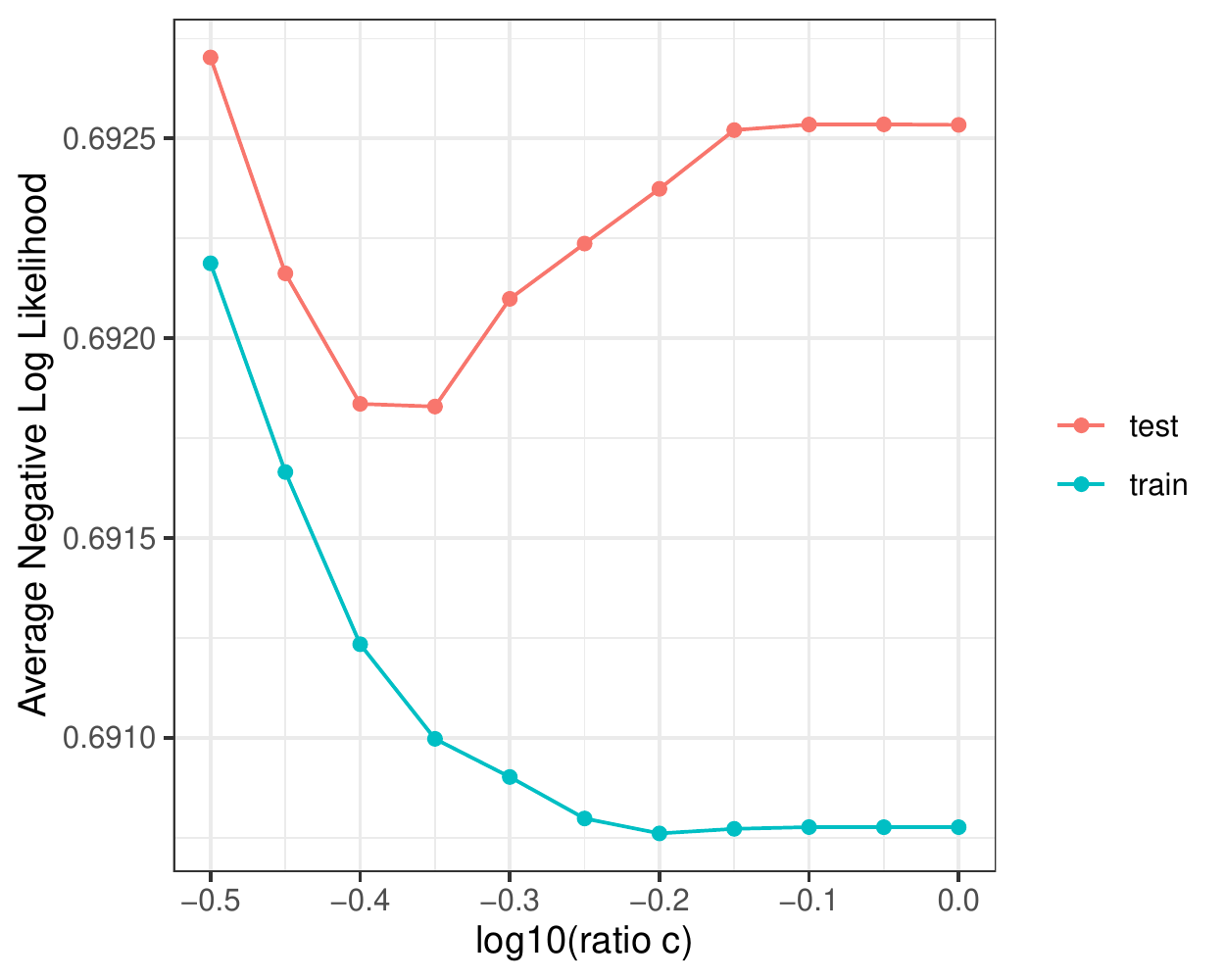}
  \end{minipage}\hfill
  \begin{minipage}[b]{0.45\textwidth}
  \centering
  \includegraphics[height=64mm,width=80mm]{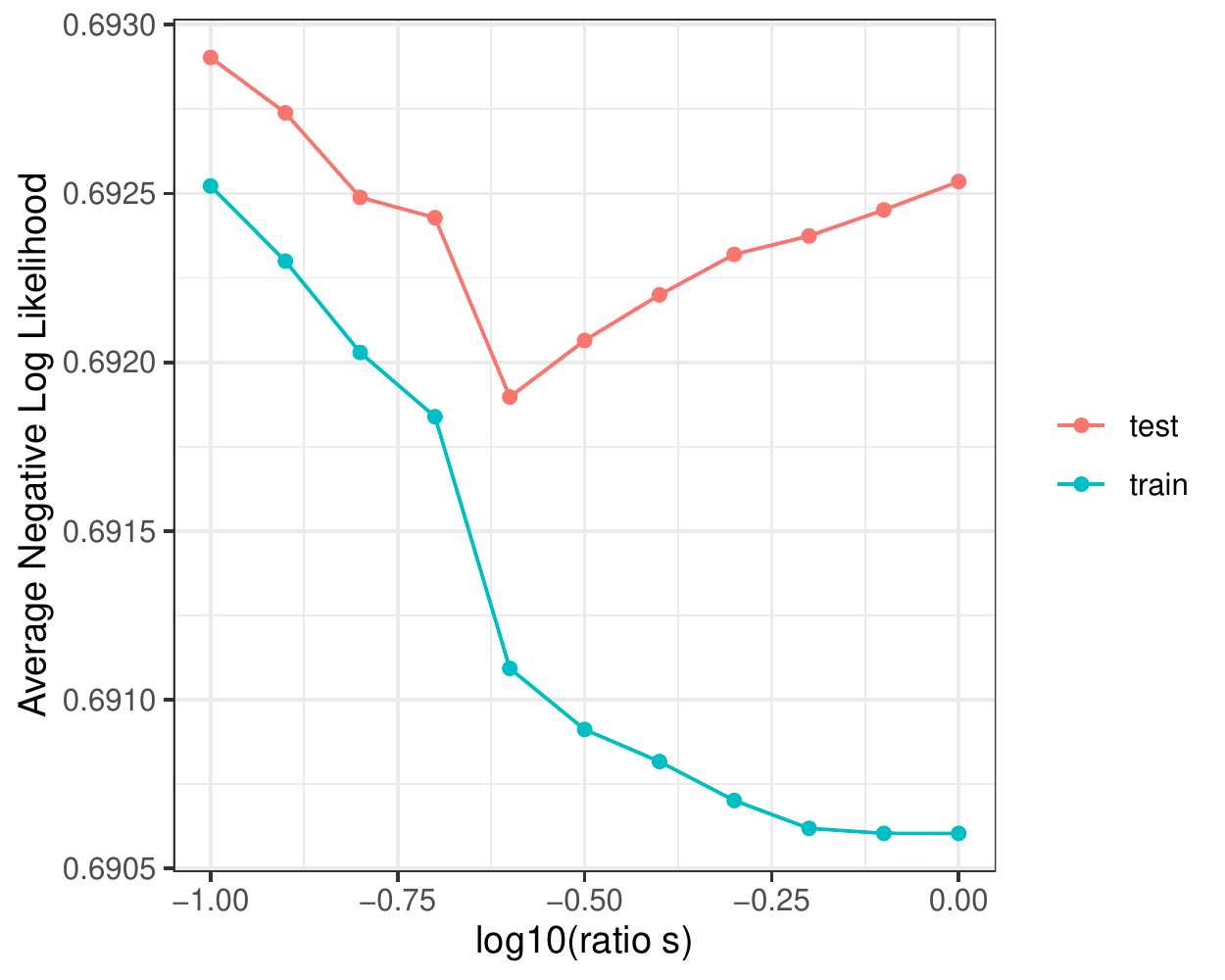}
  \end{minipage}
    \caption{Training error and test error measured in terms of the average negative log likelihood versus ratio $c$ (left) and ratio $s$ (right) for simulated data with $p_{1}=1000, p_{2}=100, p_{3}=10$ and $R=2$.} 
    \label{cv}
\end{figure}

We apply cross-validation to the same simulated data used for BIC and AIC.
Figures \ref{cvr} and \ref{cv} show the average negative log likelihood over training data (training error) and that over test data (test error). The test error is minimized when the rank is $2$.
Also we find the selected ratio $c$ and ratio $s$ values are close to those from BIC and AIC.

\subsection{Explained Deviance}

\begin{figure}[H]
  \begin{minipage}[b]{0.45\textwidth}
  \centering
    \includegraphics[height=60mm,width=70mm]{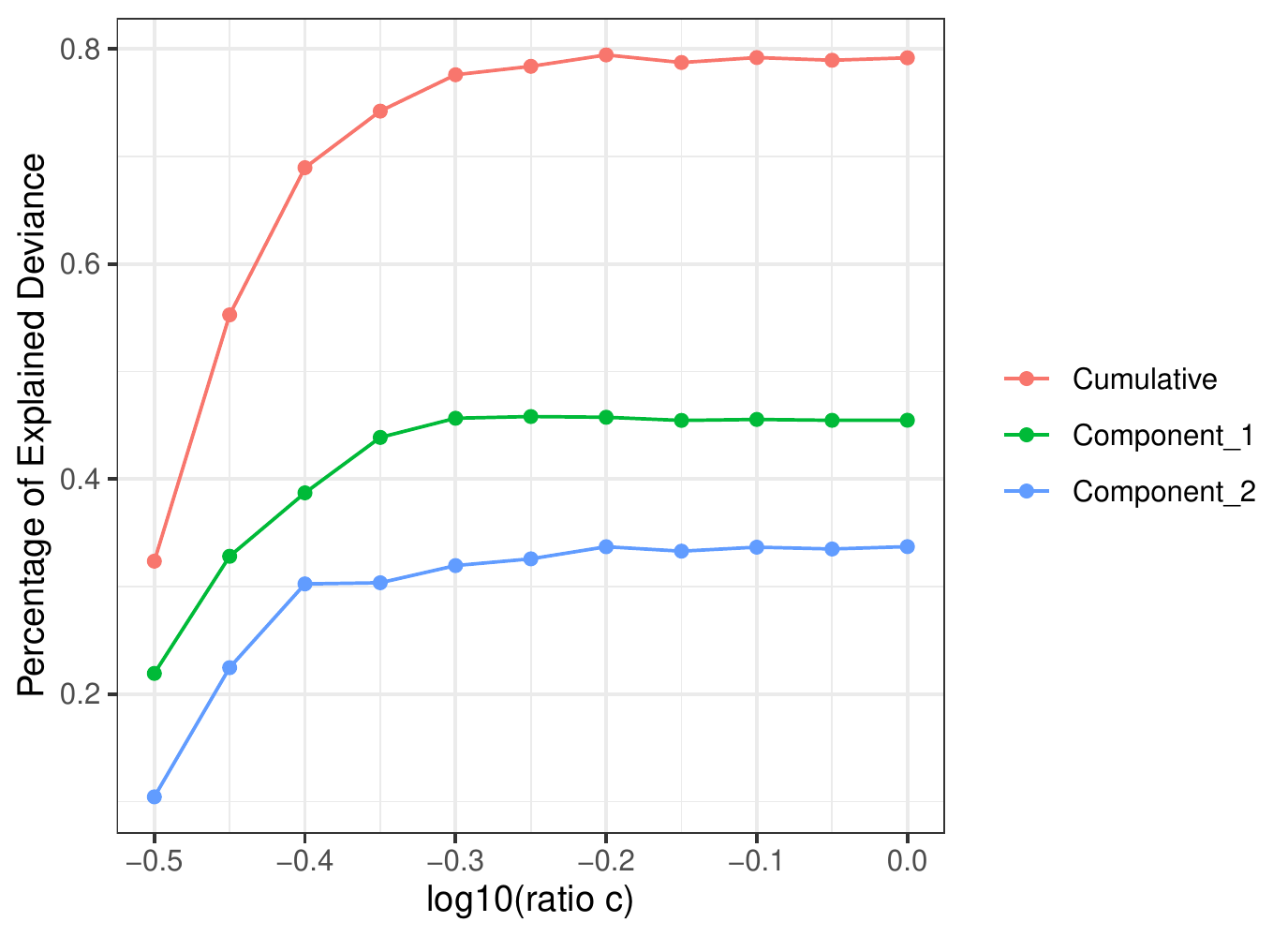}
  \end{minipage}\hfill
  \begin{minipage}[b]{0.45\textwidth}
  \centering
   \includegraphics[height=60mm,width=70mm]{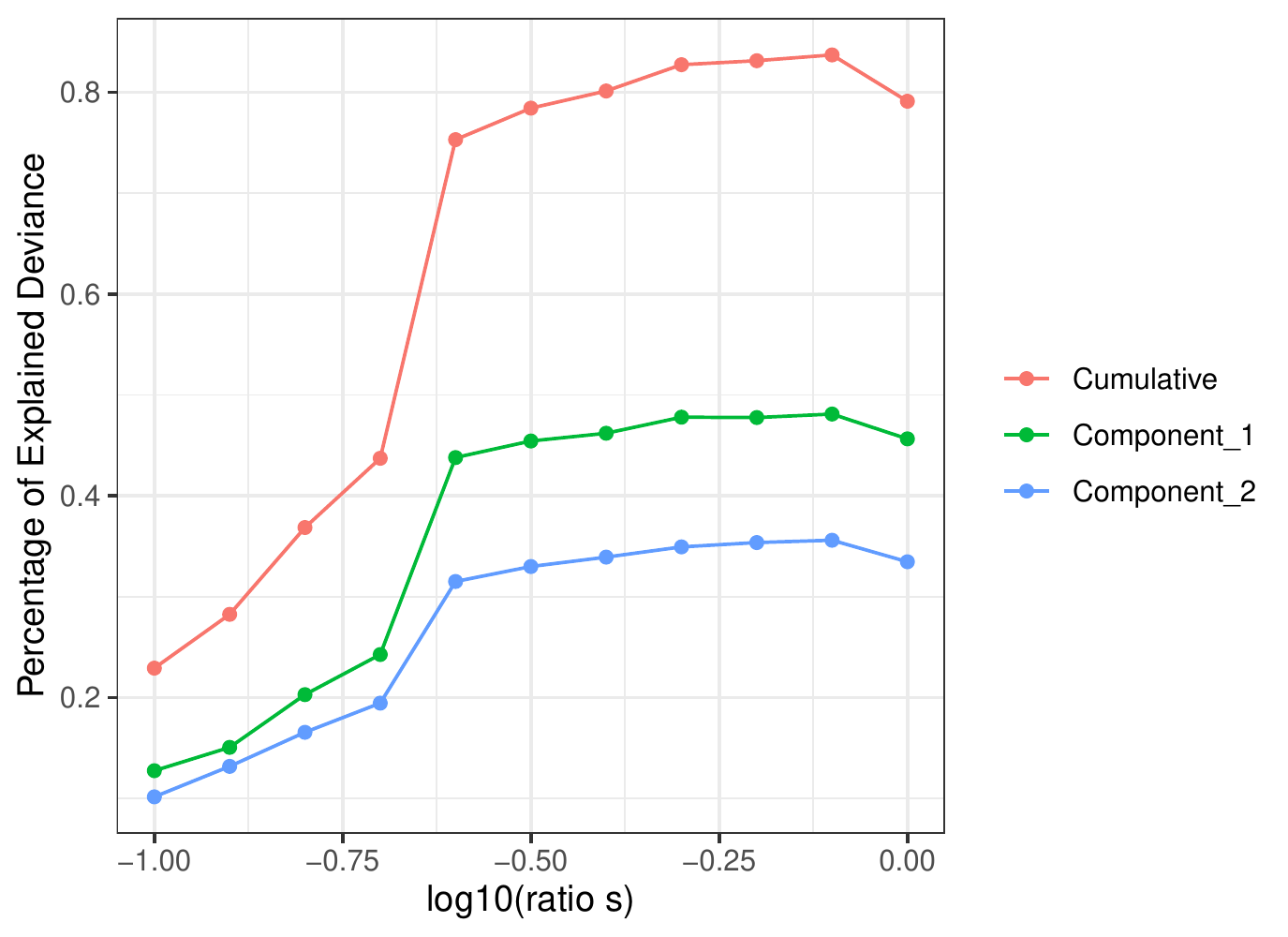}
  \end{minipage}\hfill
    \caption{The percentage of the deviance explained by the first two components in sparse logistic tensor decomposition constrained with  $\ell_{1}$-norm (left) and $\ell_{0}$-norm (right)
    for simulated data with $p_{1}=1000, p_{2}=100, p_{3}=10$ and $R=2$.     
    }\label{dev}  
\end{figure}

\begin{figure}[H]
  \begin{minipage}[b]{0.45\textwidth}
  \centering
  \includegraphics[height=60mm,width=70mm]{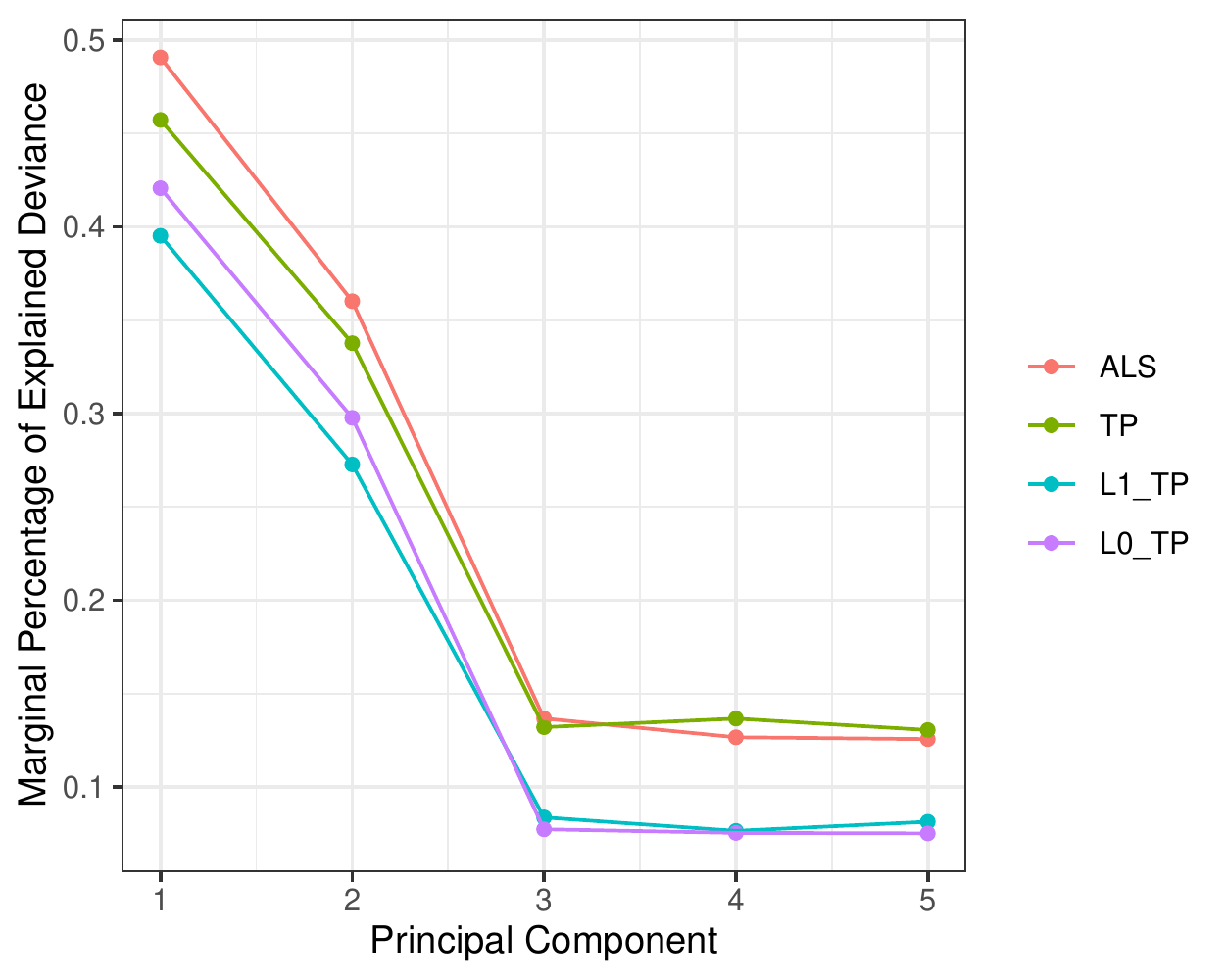}
  \end{minipage}
  \hfill
  \begin{minipage}[b]{0.45\textwidth}
  \centering
  \includegraphics[height=60mm,width=70mm]{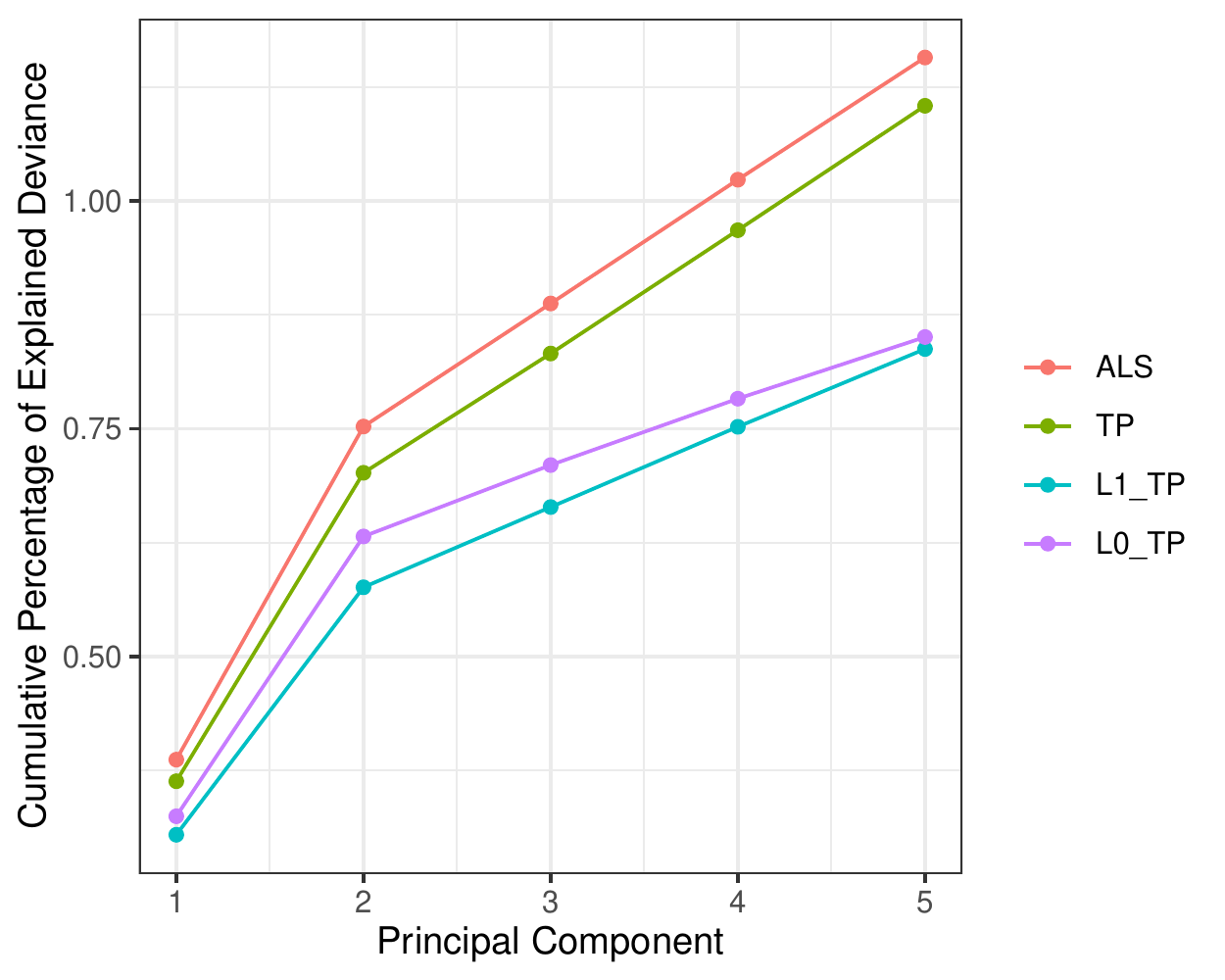}
  \end{minipage}  
    \caption{The percentage of explained deviance in logistic tensor decomposition by four methods for simulated data with $p_{1}=1000, p_{2}=100, p_{3}=10$ and $R=2$.  The marginal percentage of deviance explained by each of the five components (left) and the cumulative percentage of deviance explained by the five components (right).     
    }\label{dev1}  
\end{figure}

In general, there is a tradeoff between the sparsity of factor matrices and the explained deviance. We illustrate the tradeoff using the simulated data as before.
%we generate data $\mathcal{X}$ with $p_{1}=1000, p_{2}=100, p_{3}=10$ and rank $R^{*}=2$. 
Here we fit  $\ell_{1}$-norm and $\ell_{0}$-norm regularized rank-two models with varying tuning parameter values $c$ and $s$, respectively. When ratio $c\in [\underset{i}{\max}\frac{1}{\sqrt{p_{i}}},1]$ or ratio $s\in [\underset{i}{\max}\frac{1}{ p_{i}},1]$ decreases, the factor matrices become sparser and easier to interpret. However, as shown in Figure \ref{dev}, the cumulative percentage of explained deviance for the first two components and marginal percentage of explained deviance of the first component and second component also tend to be smaller compared to the un-regularized model. Besides, we fit a rank-five model with all methods including ALS and TP for logistic CP decomposition, $\ell_{0}$-TP and $\ell_{1}$-TP for sparse logistic CP decomposition. The cumulative percentage of explained deviance by the first 5 components is shown in Figure \ref{dev1}. 
%\red{Similar to the results shown in \cite{allen2012sparse}, the percentages below are very small in tensor data.}
The percentage is small due to the sparsity in $\Theta$. For comparison, we consider a subset $\Omega=\{(i,j,k): \theta_{ijk}^{*}\neq 0\}$, and calculate the explained deviance based on the partial data $\mathcal{X}_{\Omega}$ with nonzero $\theta_{ijk}^{*}$ only. The cumulative percentages of deviance explained by the five  components based on $\mathcal{X}_{\Omega}$ are $26.62, 45.05, 45.08, 45.17$ and $45.17.$  The marginal percentages of deviance explained by each of the five components are $26.62, 18.43,  0.035,  0.085$ and $0.0019.$ We find that the explained deviance for the partial data is much higher than the whole data. 
It's clear that TP method and ALS method produce a very close explained deviance. The same is true for TSP method and TTP method, but their solutions explain less deviance due to the sparsity in factors. 
As in standard PCA, a scree plot of marginal explained deviance can be used to determine the number of rank and degree of sparsity for tensor decomposition.  The left panel of Figure \ref{dev1} indicates an elbow point around $R=3$ and suggests the choice of rank 2, which is the same as the true rank.

\subsection{True Positive Rate and False Positive Rate}
\begin{figure}[H]
  \begin{minipage}[b]{0.4\textwidth}
  \centering
      \includegraphics[height=64mm,width=80mm]{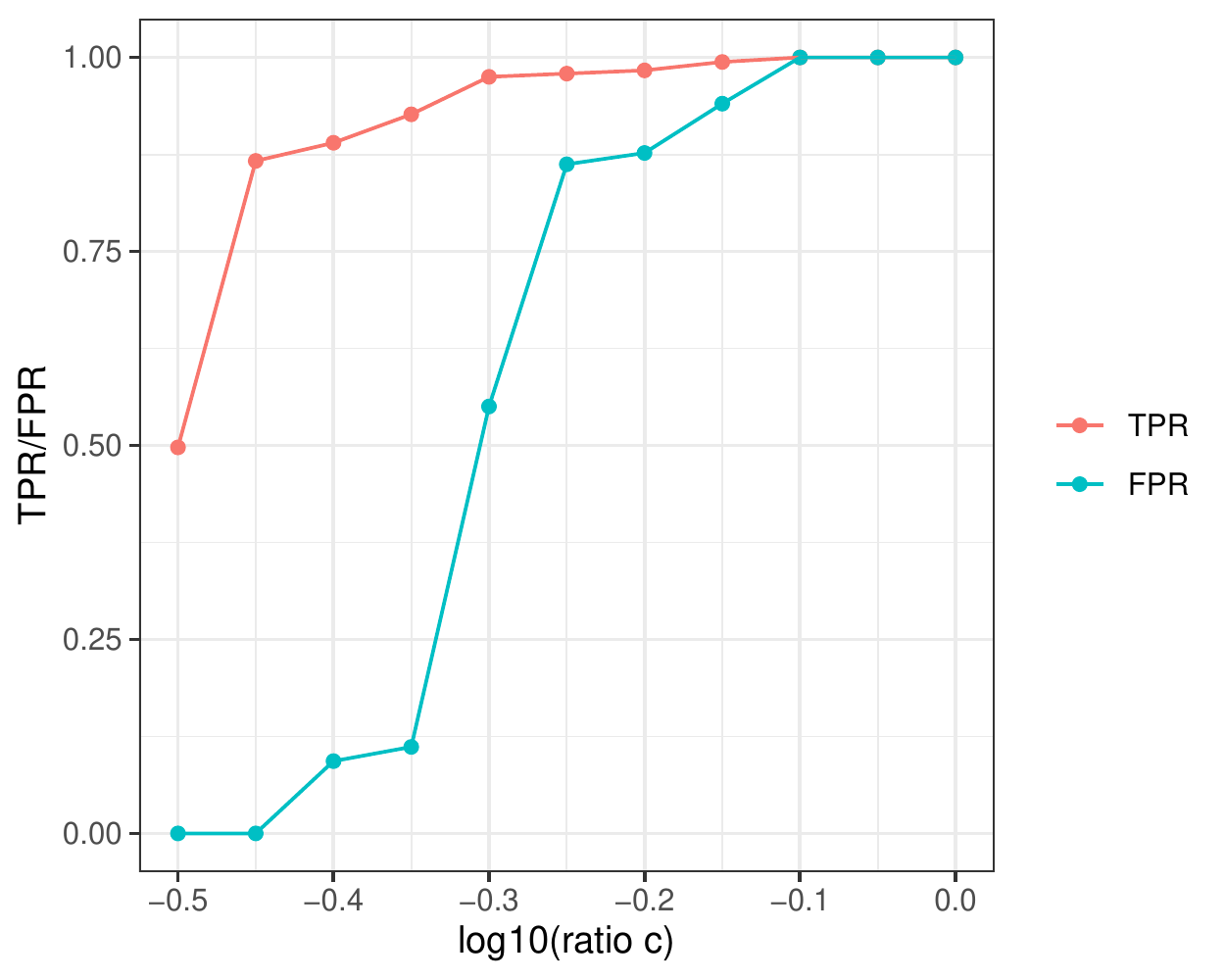}
  \end{minipage}
  \hfill
  \begin{minipage}[b]{0.4\textwidth}
  \centering  
    \includegraphics[height=64mm,width=80mm]{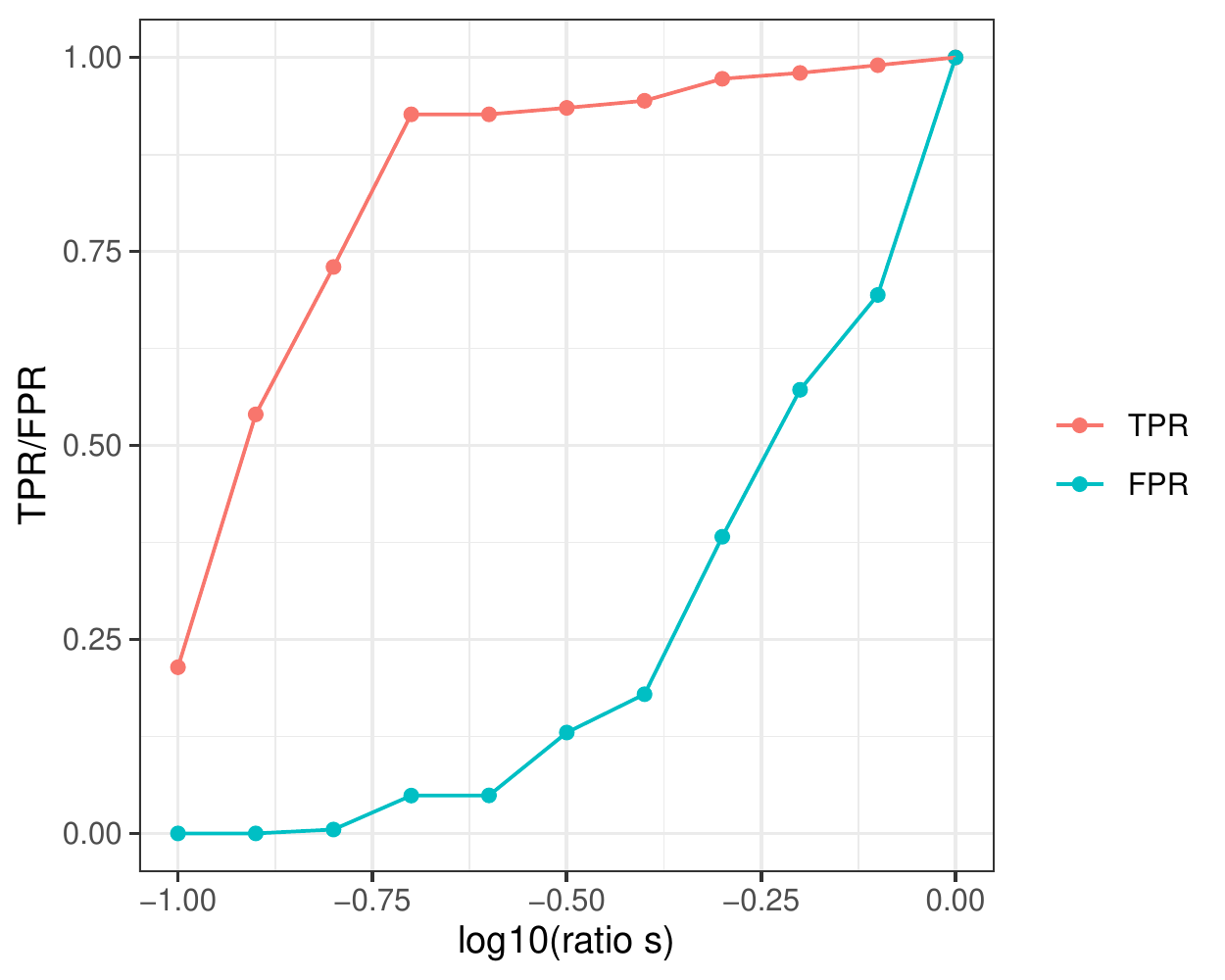}
  \end{minipage}
%  \hfill
%    \begin{minipage}[b]{0.3\textwidth}
%    \centering    
%    \includegraphics[height=64mm,width=80mm]{2l1l0.pdf}
%  \end{minipage}  
  \caption{TPR and FPR in $\ell_{1}$-norm constrained logistic tensor decomposition with ratio $c$ (left), and $\ell_{0}$-norm constrained decomposition with ratio $s$ (right) for simulated data with $p_{1}=1000, p_{2}=10, p_{3}=10$ and $R=2$. }
  \label{roc4}
  \end{figure}

\begin{figure}[H]
     \begin{minipage}[b]{1\textwidth}
    \centering    
    \includegraphics[height=64mm,width=80mm]{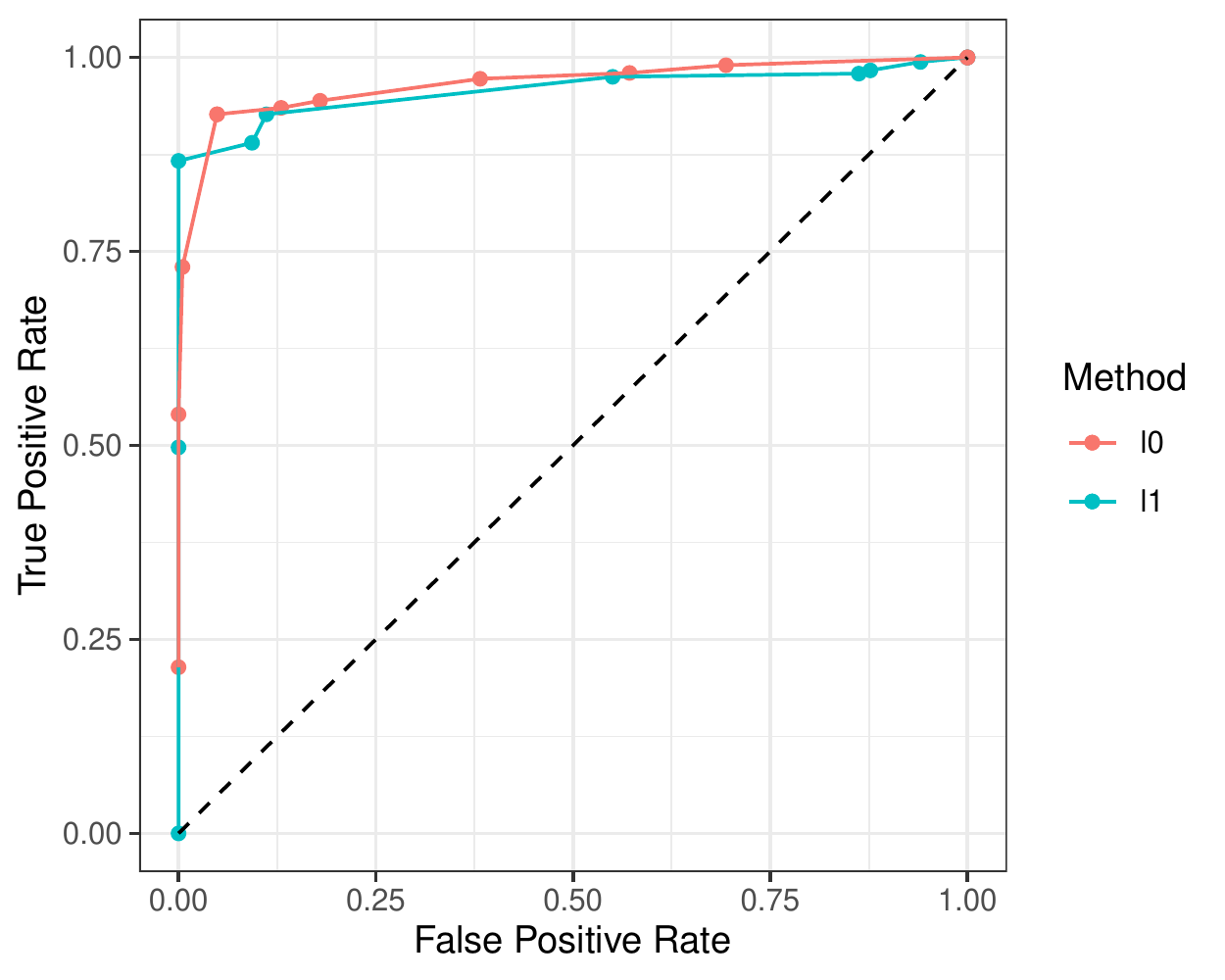}
  \end{minipage}  
  \caption{ROC curves of $\ell_{1}$-norm constrained logistic tensor decomposition and $\ell_{0}$-norm constrained decomposition for simulated data with $p_{1}=1000, p_{2}=10, p_{3}=10$ and $R=2$. }
  \label{rocc}
  \end{figure}

 Figure \ref{roc4} shows how the mean  TPR and FPR change with ratio $c\in [\underset{i}{\max}\frac{1}{\sqrt{p_{i}}},1]$ and ratio $s\in [\underset{i}{\max}\frac{1}{p_{i}},1]$ for simulated data from scenario 4. %which are shown in Figures \ref{roc1}, \ref{roc2}, \ref{roc3} and \ref{roc4}. 
Both $\ell_{1}$-norm and $\ell_{0}$-norm penalties can do selection of nonzero parameters in sparse logistic CP decomposition. The receiver operating characteristic (ROC) curves for both methods in Figure \ref{rocc} show that they  have comparable area under the curve (AUC) values. Results are similar for other settings and thus omitted.

  \subsection{Estimation Errors}

\begin{figure}[H]
  \begin{minipage}[b]{0.45\textwidth}
  \centering  
    \includegraphics[height=65mm,width=70mm]{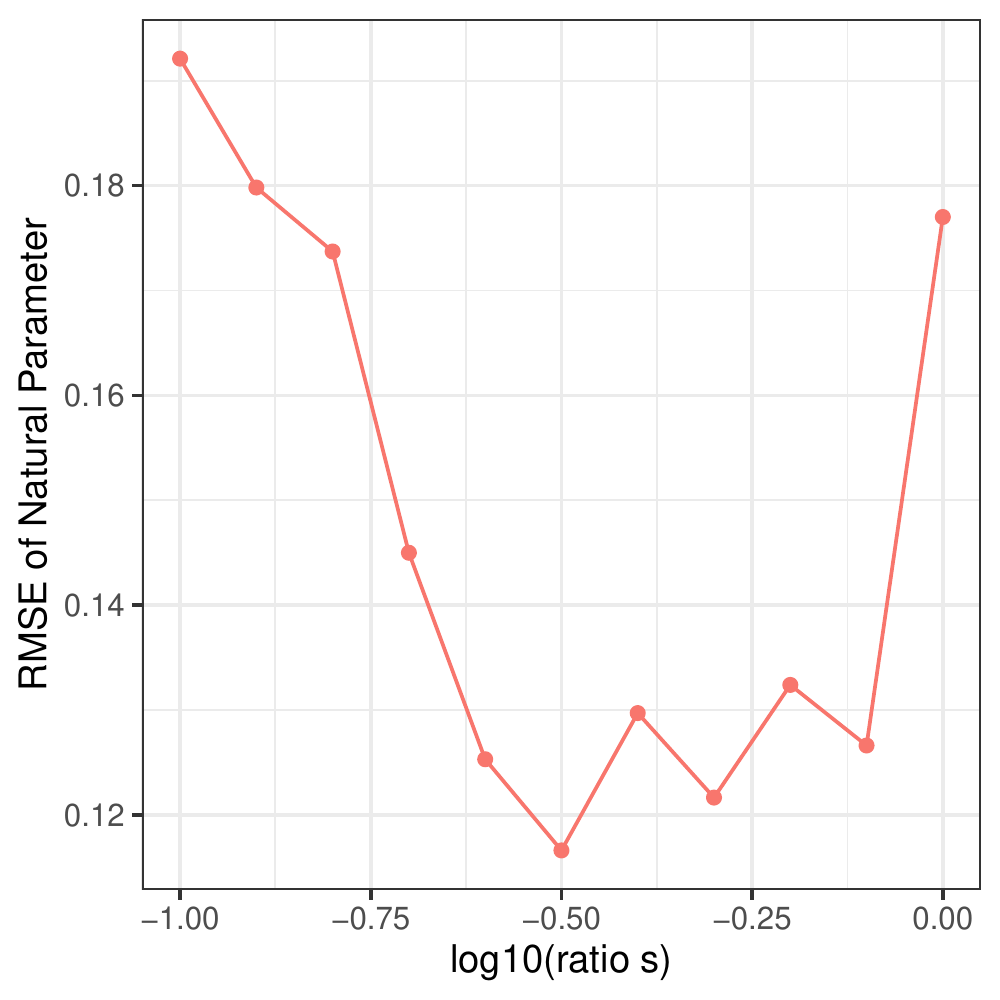}
    \end{minipage}
  \hfill
  \begin{minipage}[b]{0.45\textwidth}
  \centering  
    \includegraphics[height=65mm,width=70mm]{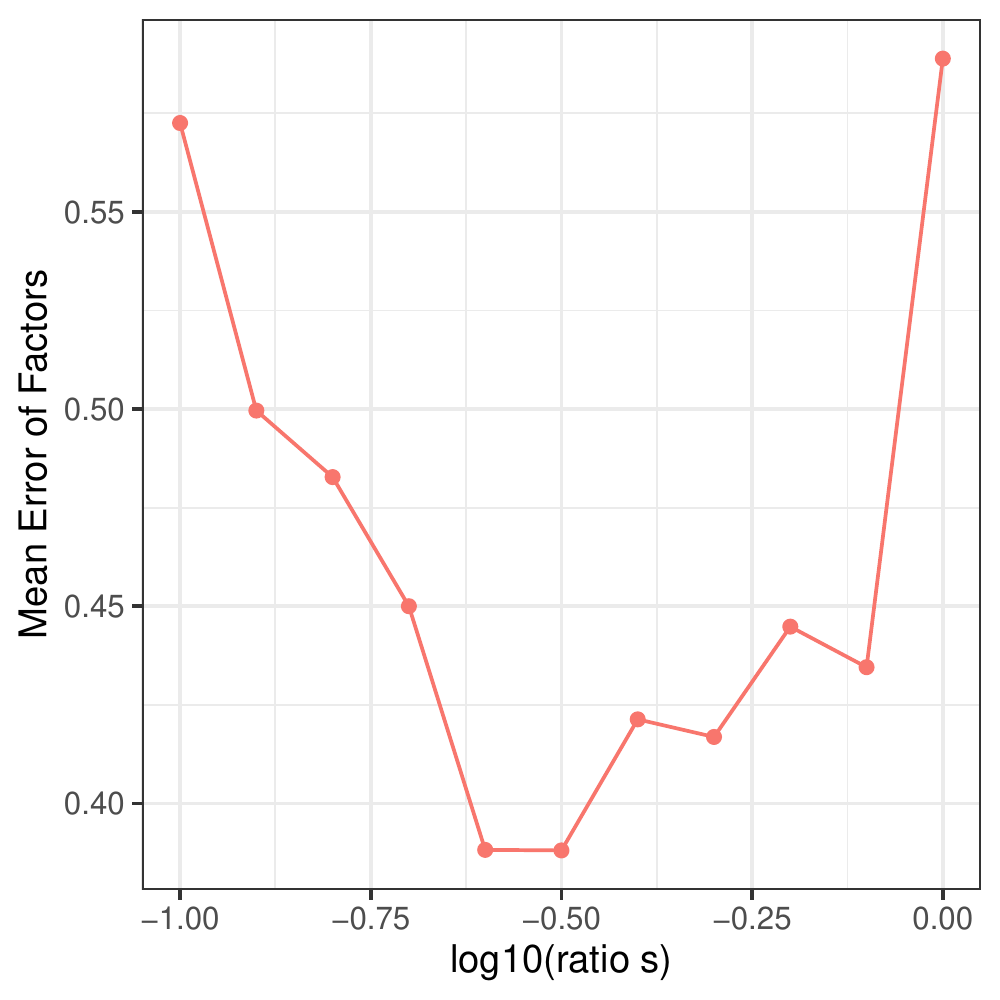}
  \end{minipage}
  \caption{RMSE and mean error in $\ell_{0}$-norm constrained logistic tensor decomposition with ratio $s$ for simulated data from scenario 3.}
  \label{me}      
    \end{figure}

  To illustrate the benefit of regularization, we generate data from scenario 3 and show the $\text{RMSE}$ and $\text{Mean Error}$ of $\ell_{1}$-norm and $\ell_{0}$-norm constrained logistic tensor decompositions, respectively,  as a function of the ratio parameter in Figure \ref{me}.
%where ratio $c\in [\frac{1}{\min\sqrt{p_{i}}},1]$ and ratio $s\in [\frac{1}{\min p_{i}},1]$.
 Clearly, $\ell_{1}$-norm and $\ell_{0}$-norm regularizations help to reduce both the overall estimation error and component estimation error if the true factors are sparse.
 Note that the estimation errors for un-regularized solutions at $c=1$ and $s=1$ differ slightly due to difference in initializations.

\begin{figure}[H]
  \begin{minipage}[b]{0.45\textwidth}
  \centering
    \includegraphics[height=65mm,width=70mm]{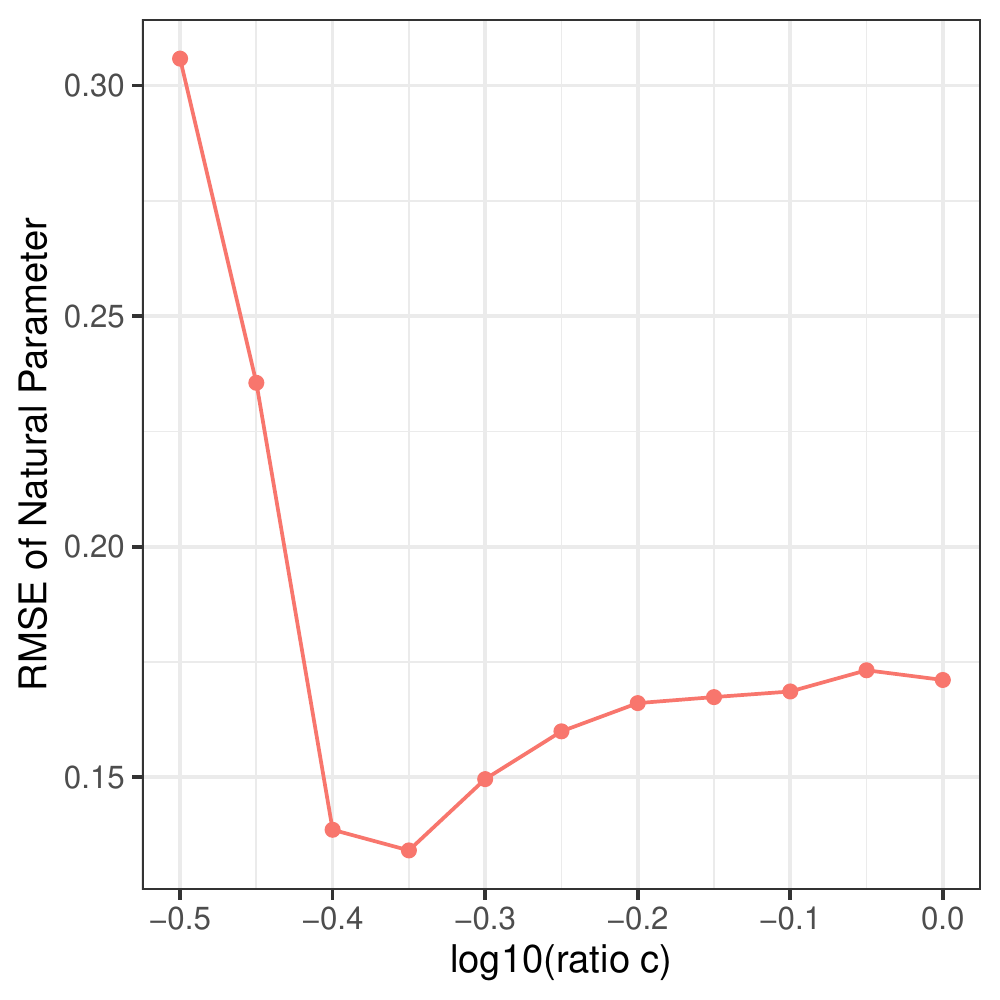}
    \end{minipage}
  \hfill
  \begin{minipage}[b]{0.45\textwidth}
  \centering  
    \includegraphics[height=65mm,width=70mm]{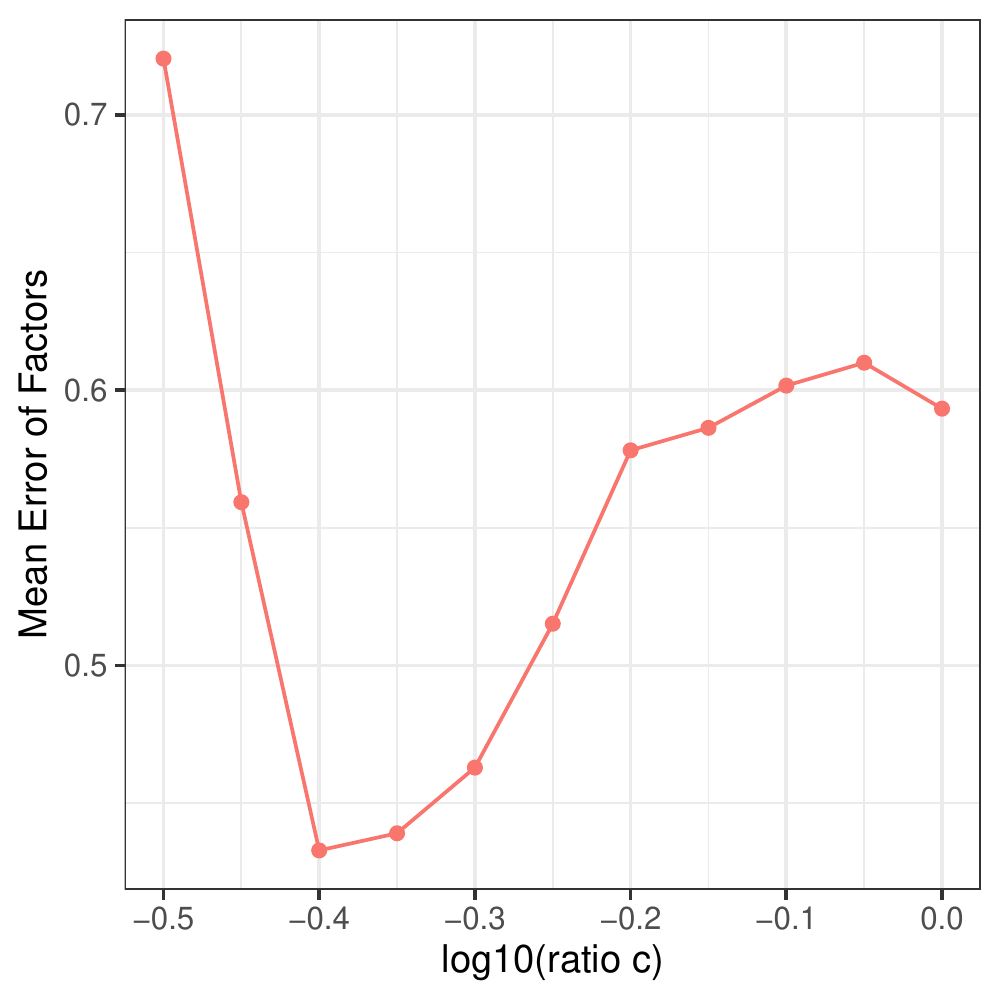}
  \end{minipage}\hfill
\caption{RMSE and mean error in $\ell_{1}$-norm constrained logistic tensor decomposition with ratio $c$ for simulated data from scenario 3.}
      \end{figure}

%\end{appendix}

\section{Choice of Parameters in Algorithms}\label{apppara}

In order to obtain accurate estimates of logit parameters, we need to choose control parameters in algorithms carefully in practice. In this section, we discuss the initialization and termination of the proposed algorithms as well as the clustering procedure. 

\subsection{Initialization}

There are two kinds of initialization for the proposed algorithms: one is spectral initialization and the other is random initialization. For Alternating Least Squares (ALS) method, 
\textit{spectral initialization} starts with the tensor $2\mathcal{X}-\one_{p_{1}p_{2}p_{3}}$, and obtains $\mu=\frac{1}{p_{1}p_{2}p_{3}}(2\mathcal{X}-\one_{p_{1}p_{2}p_{3}})\times_{1}\one_{p_{1}}^{T}\times_{2}\one_{p_{2}}^{T}\times_{3}\one_{p_{3}}^{T}$ and $\mathcal{Q}=(2\mathcal{X}-\one_{p_{1}p_{2}p_{3}})-\mu\one_{p_{1}p_{2}p_{3}}$. Then it extracts the left $R$ singular vectors of matricizations of $\mathcal{Q}$, say $Q_{(1)}$ and $Q_{(2)}$ as the initialization of factor matrices $U$ and $V$, and lets $W=Q_{(3)}[(U\odot V)^{T}]^{-1}$. \textit{Random initialization} starts with matrices $U$ and $V$ where their entries are independently generated from the standard Gaussian  distribution. Then it normalizes the matrices $U$ and $V$ to those with unit columns and lets $W=Q_{(3)}[(U\odot V)^{T}]^{-1}$. With $\mu, U, V$ and $W$, we could define $\Theta^{[0]}$ using $\eqref{theta}$ and then $\mathcal{Z}^{[0]}$ using \eqref{newx}.

%\widehat{\mu}^{[0]}

 For Tensor Power (TP) method, \textit{spectral initialization} of a rank-one model first generates $\theta\sim\text{N}(0,I_{p_{3}})$, and truncates $\theta$ to $\theta^{\prime}=T(\theta,\max\{s_{1},s_{2},s_{3}\})$. Then it calculates $\bu\in\mathbb{R}^{p_{1}}$ and $\bv\in\mathbb{R}^{p_{2}}$ as the leading left and right singular vectors of $\mathcal{Q}\times_{3}(\theta^{\prime})^{T}\in\mathbb{R}^{p_{1}\times p_{2}}$. Finally it computes sparse vectors $\bu^{\prime}=T(\bu, s_{1})$ and $\bv^{\prime}=T(\bv, s_{2})$,  normalizes them by $\bar{\bu}=\bu^{\prime}/\|\bu^{\prime}\|_{2}$ and $\bar{\bv}=\bv^{\prime}/\|\bv^{\prime}\|_{2}$ and calculates $\bar{\bw}=\mathcal{Q}\times_{1}\bar{\bu}^{T}\times_{2}\bar{\bv}^{T}\in\mathbb{R}^{p_{3}}$. $(\bar{\bu}, \bar{\bv}, \bar{\bw})$ is the initial value from this spectral initialization. On the other hand, \textit{random initialization} of a rank-one model first generates $\bu\sim\text{N}(0,I_{p_{1}})$ and $\bv\sim\text{N}(0,I_{p_{2}})$, and then computes the sparse vectors $\bu^{\prime}=T(\bu, s_{1})$ and $\bv^{\prime}=T(\bv, s_{2})$. Then it normalizes them by $\bar{\bu}=\bu^{\prime}/\|\bu^{\prime}\|_{2}$ and $\bar{\bv}=\bv^{\prime}/\|\bv^{\prime}\|_{2}$, and finally calculates $\bar{\bw}=\mathcal{Q}\times_{1}\bar{\bu}^{T}\times_{2}\bar{\bv}^{T}\in\mathbb{R}^{p_{3}}$. This gives  $(\bar{\bu}, \bar{\bv}, \bar{\bw})$ as the initial value.  
 With $\mu, \bar{\bu}, \bar{\bv}$ and $\bar{\bw}$, we could define  $\Theta^{[0]}$ and $\mathcal{Z}^{[0]}$ using \eqref{newtheta} and \eqref{newx}.

Overall, the above initializations are based on a fully observed data tensor. In practice, if the input data tensor $\mathcal{X}$ contains missing values, we could set all missing entries of $\mathcal{X}$ to $1/2$, or equivalently set all missing entries of $2\mathcal{X}- \one_{p_{1}p_{2}p_{3}}$ to zero.

\subsection{Termination}

For the tensor power methods  in Algorithms \ref{mm}, \ref{mmtp} and \ref{mmttp}, we terminate the inner loop when
$$\max\{\|\widehat{\bu}_{\tau}^{[n+1]}-\widehat{\bu}_{\tau}^{[n]}\|_{2},\|\widehat{\bv}_{\tau}^{[n+1]}-\widehat{\bv}_{\tau}^{[n]}\|_{2},\|\widehat{\bw}_{\tau}^{[n+1]}-\widehat{\bw}_{\tau}^{[n]}\|_{2}\}\leq 10^{-4}$$ is satisfied for some  iteration $n$.
For the alternating least squares method in Algorithm \ref{mmals} with rank $R$, we terminate the inner loop when
$$\max\{\|\widehat{U}^{[n+1]}-\widehat{U}^{[n]}\|_{F},\|\widehat{V}^{[n+1]}-\widehat{V}^{[n]}\|_{F},\|\widehat{W}^{[n+1]}-\widehat{W}^{[n]}\|_{F}\}\leq \sqrt{R}\cdot 10^{-4}$$ is satisfied for some  iteration $n$.

For the outer loop (MM algorithm) of Algorithms \ref{mmals}, \ref{mm}, \ref{mmtp} and \ref{mmttp}, the maximal number of iterations is usually less than $50$,  but it may take more iterations to converge in Algorithm \ref{mmtp}. We terminate the outer loop  (MM algorithm) when any of the following stopping criteria is satisfied.

 1. The change in the value of the objective function, i.e., the negative log likelihood function $-\ell(\mathcal{X};\Theta)$
%$\max\{\|\bu^{[m+1]}-\bu^{[m]}\|_{F}^{2}, \|\bv^{[m+1]}-\bv^{[m]}\|_{F}^{2}, \|\bw^{[m+1]}-\bw^{[m]}\|_{F}^{2}\}\leq 1$
is small: $$|\ell(\mathcal{X};\Theta^{[m+1]})-\ell(\mathcal{X};\Theta^{[m]})|<10^{-2}$$ or the relative change is small: $$|\{\ell(\mathcal{X};\Theta^{[m+1]})-\ell(\mathcal{X};\Theta^{[m]})\}/\ell(\mathcal{X};\Theta^{[m]})|<10^{-5}.$$

2. The change in factors $\bu,\bv$ and $\bw$ is small:
$$\max\{\|\widehat{\bu}_{\tau}^{[m+1]}-\widehat{\bu}_{\tau}^{[m]}\|_{2},\|\widehat{\bv}_{\tau}^{[m+1]}-\widehat{\bv}_{\tau}^{[m]}\|_{2},\|\widehat{\bw}_{\tau}^{[m+1]}-\widehat{\bw}_{\tau}^{[m]}\|_{2}\}\leq 10^{-4}.$$

Based on the general property of MM algorithm \citep{hunter2004tutorial},  the objective function value 
$-\ell(\mathcal{X};\Theta^{[m]})$ 
decreases as $m$ gets large and converges to a local minimum of $-\ell(\mathcal{X};\Theta)$ 
%over $\mathcal{N}$ ($\mathcal{N}_{1}$ or $\mathcal{N}_{0}$) 
as $m\rightarrow \infty$.
The global and local convergence of our proposed algorithms can be derived similarly as in the work of \cite{zhou2013tensor} and \cite{wang2020learning}.

\subsection{Clustering Procedure}
To avoid local optima, we have extended the clustering procedure from 
\cite{anandkumar2014guaranteed} and \cite{sun2017provable} to our problem \eqref{lcpd}, and suggested to extract $R$ components from $L$ tuples sequentially. We optimize $\ell(\mathcal{X};\Theta)$ over a rank-one region for $L$ times, and each time we start with either spectral initialization or random initialization. After specifying the number of components $R$, Algorithm \ref{clust} for clustering can identify $R$ components from $L$ estimated tuples of Algorithm \ref{mm} by ordering the weights $d_{\tau}$. To recover the true components $(\bu_{r}^{*}, \bv_{r}^{*},\bw_{r}^{*}), r\in[R]$, we need to focus on large estimates of $d_{\tau}$ and remove all tuples which are too similar to one another
%close to $(\bu_{\tau},\bv_{\tau},\bw_{\tau})$ 
because they will eventually lead to the same component. We terminate the clustering procedure after finding $R$ largest weights, and order the rank-one components according to the magnitude of $\widehat{d}_{j}$ at the end of Algorithm \ref{clust}.
The line 4 of Algorithm \ref{clust} is optional, but reestimation with initialization $(\widehat{\bu},\widehat{\bv},\widehat{\bw})$ will lead to a more accurate model.

In practice, we could choose the number of initializations, $L=\max\{10, R^{3}\}$, which works well in most cases. In general, we need large $L$ to avoid local minima for small data sets. The number of initializations $L$ depends on the size of problem $(p_{1},p_{2},p_{3},R)$ and signal-to-noise ratios (SNRs). 
The clustering procedure will certainly increase the computing time, but it could avoid many local optima for non-convex problems. When the difference between SNRs is smaller, sometimes the order of estimated factor matrices may be flipped when compared with the true factor matrices. 

Note that the default threshold $\nu$ to remove redundant tuples is set to $0.5$ in Algorithm \ref{clust}, which could be any number between $10^{-4}$ and $1$ as suggested by \cite{sun2017provable}. This threshold $\nu$ is proportional to the size of $\bu,\bv$ and $\bw$, and small $\nu$ will result in more remaining tuples in $S=\{(\widehat{d}_{\tau},\widehat{\bu}_{\tau},\widehat{\bv}_{\tau},\widehat{\bw}_{\tau}),\tau\in[L]\}$.
% and large $\nu$ will lead to less remaining tuples in $S$.  
 Therefore we should choose $\nu$ adaptively depending on the size of data.

\begin{algorithm}[] 
\small 
\caption{Clustering Procedure}
\label{clust}
\begin{algorithmic}[1]
\STATE \textbf{input:} set $S=\{(\widehat{d}_{\tau},\widehat{\bu}_{\tau},\widehat{\bv}_{\tau},\widehat{\bw}_{\tau}),\tau\in[L]\}$.

\FOR {$\tau=1$ {\bfseries to} $R$}
\STATE Find $(\widehat{\bu},\widehat{\bv},\widehat{\bw})=\underset{(\widehat{\bu}_{\tau},\widehat{\bv}_{\tau},\widehat{\bw}_{\tau}),\tau\in[L]}{\arg\max} \widehat{d}_{\tau}$.

\STATE Run Algorithm \ref{mm} (\ref{mmtp} or \ref{mmttp}) with initialization $(\widehat{\bu},\widehat{\bv},\widehat{\bw})$ and denote the final update as $(\widehat{d}_{j},\widehat{\bu}_{j},\widehat{\bv}_{j},\widehat{\bw}_{j}), j\in[R]$.

\STATE Remove all tuples in $S$ with $\min\{\|\widehat{\bu}_{\tau}\pm\widehat{\bu}\|_{2}, \|\widehat{\bv}_{\tau}\pm\widehat{\bv}\|_{2}, \|\widehat{\bw}_{\tau}\pm\widehat{\bw}\|_{2}\}\leq \nu=0.5$, $\tau\in[L]$.

\ENDFOR
 \STATE \textbf{output:} $\{(\widehat{d}_{j},\widehat{\bu}_{j},\widehat{\bv}_{j},\widehat{\bw}_{j}), j\in[R]\}$.
   
  \end{algorithmic} 
 \end{algorithm}

\section{Analysis of Nations Data}\label{appda}

%\label{appB}

\subsection{Choice of Rank and Tuning Parameters}

\begin{figure}[H]
    \begin{minipage}[b]{0.45\textwidth}
  \centering
   \includegraphics[height=65mm,width=70mm]{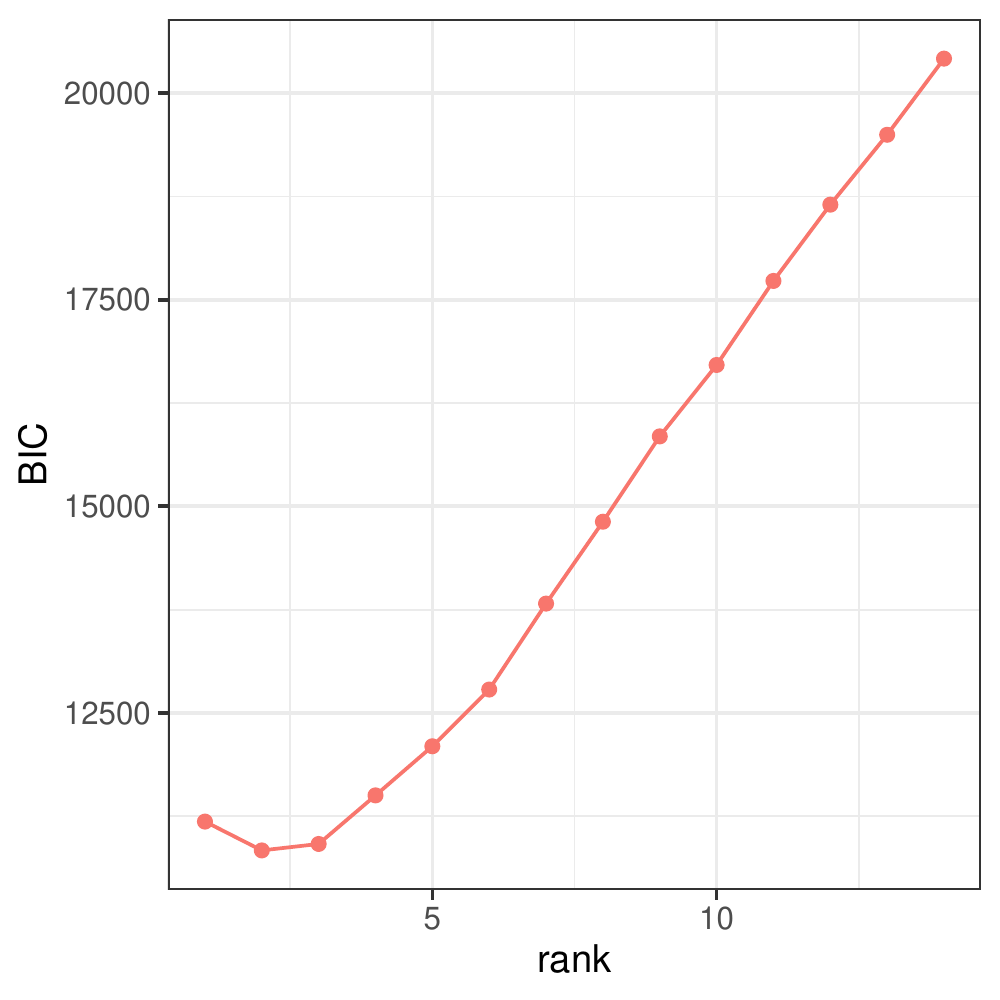}
  \end{minipage}\hfill
  \begin{minipage}[b]{0.45\textwidth}
  \centering
    \includegraphics[height=65mm,width=70mm]{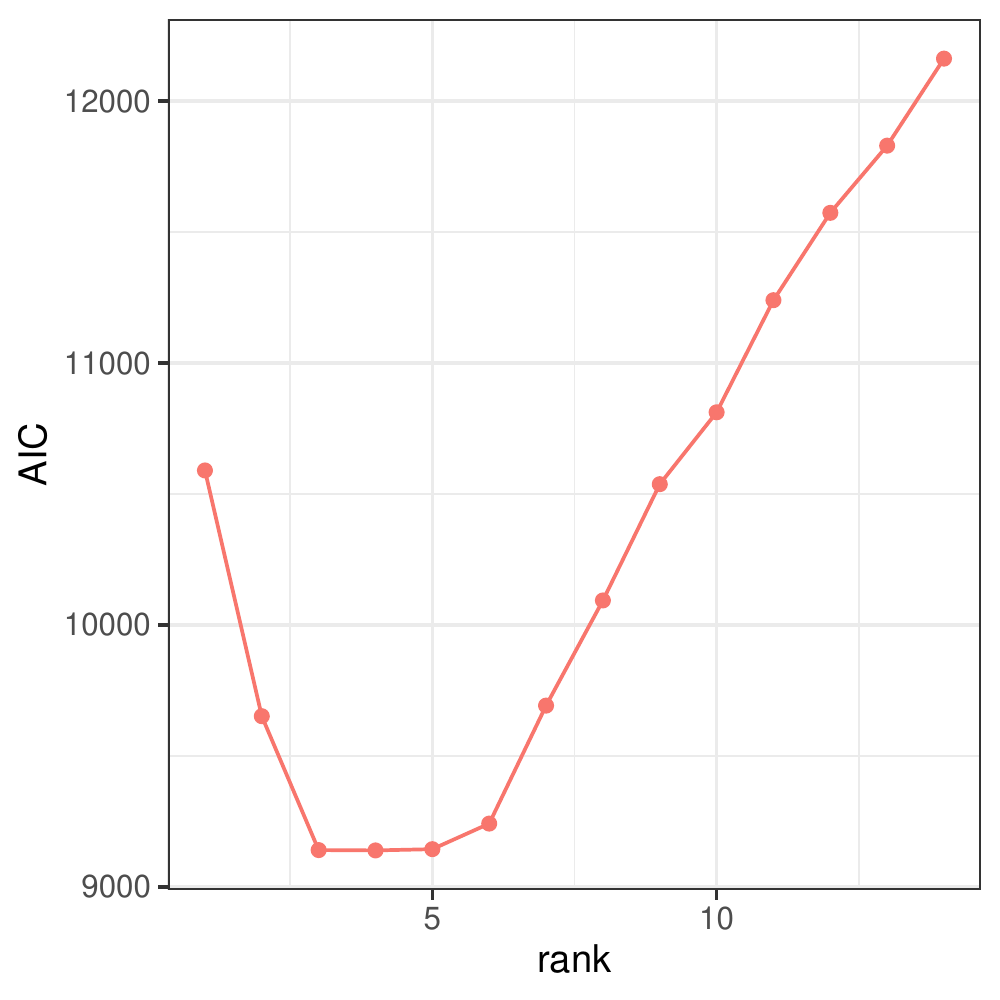}
  \end{minipage}
    \caption{BIC and AIC versus rank $R$ for the nations data.} 
    \label{da1}
\end{figure}

In addition to the choice of rank suggested by the marginal explained
deviance and weights for the initial unregularized logistic tensor decomposition, we can use the BIC and AIC. 
 As shown in Figure \ref{da1},  rank $R=2$ is chosen by BIC criterion, and $R=4$ is chosen by AIC criterion. 
 
\begin{figure}[H]
  %\centering
\begin{minipage}[b]{1\textwidth}
  \centering
  \includegraphics[height=64mm,width=80mm]{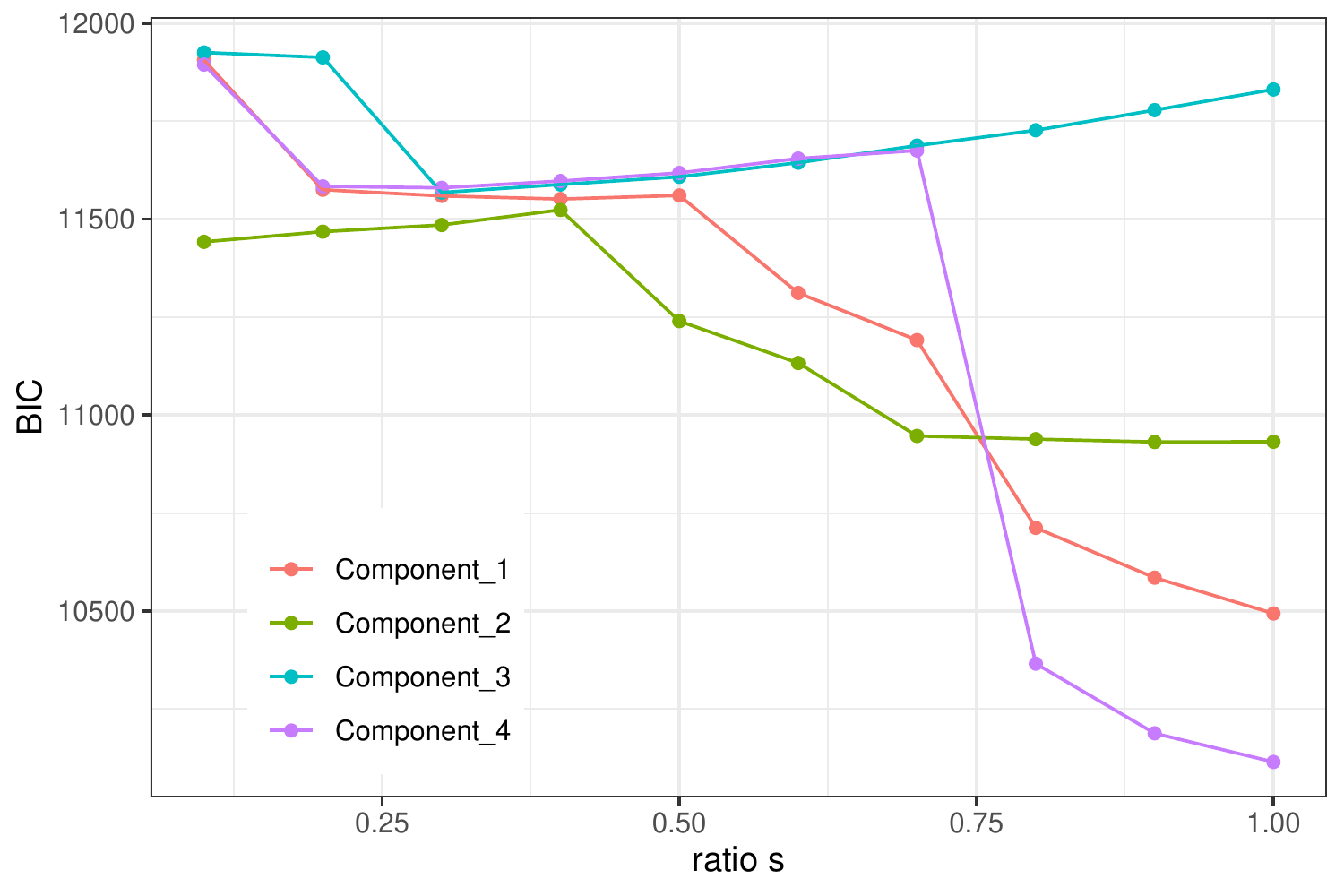}
 \end{minipage}
\caption{BIC versus ratio $s$ for four relation factors for the nations data.}  
\label{daCARD}
     \end{figure}

For the sparse logistic CP decomposition, the estimated offset term $\mu$ is $-2.86$, and the tuning parameters are chosen by BIC criterion. As shown in Figure \ref{daCARD}, four BIC curves versus ratio $s$ suggest 56, 39, 16 and 56 nonzero entries for the first four components respectively.

\subsection{Visualization of Components}     
\begin{figure}[H]
   \centering
    %\vspace{-5mm}
    \includegraphics[height=50mm,width=150mm]{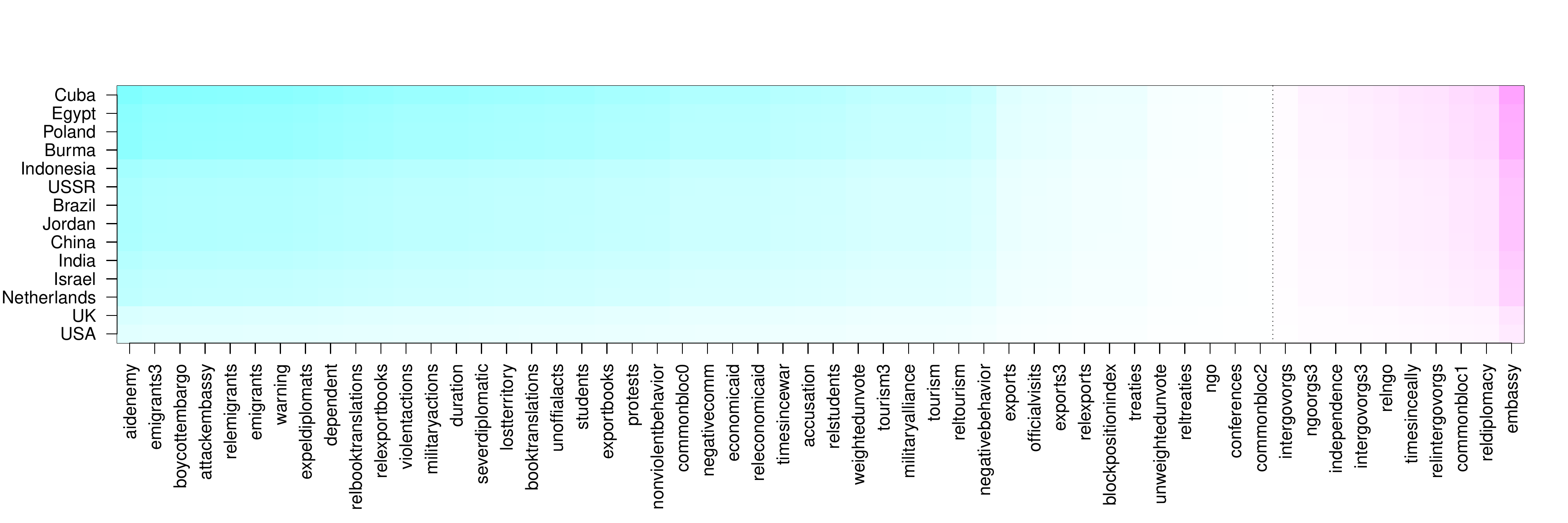}
   \caption{Heatmap of relation versus nation  for component 1 with regularization.} 
 \label{hmap_1}
     \end{figure}

\begin{figure}[H]
   \centering
   \vspace{-5mm}
    \includegraphics[height=50mm,width=150mm]{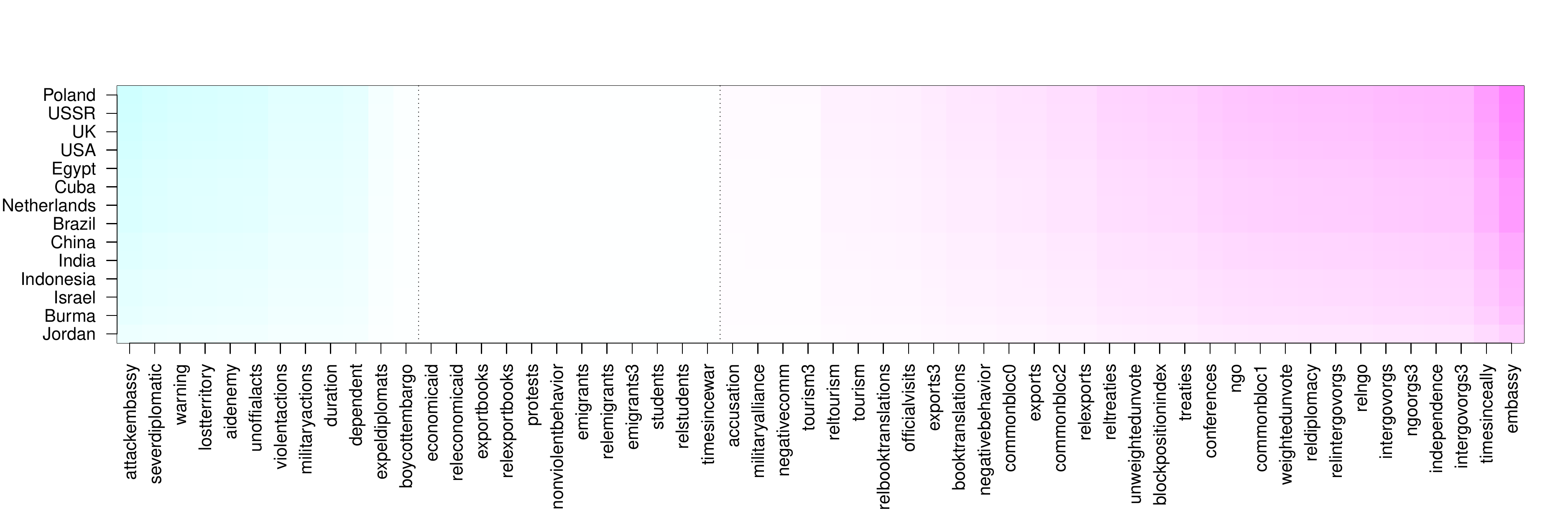}
   \caption{Heatmap of relation versus nation  for component 2 with regularization.} 
 \label{hmap_2}
     \end{figure}

\begin{figure}[H]
   \centering
   \vspace{-5mm}
    \includegraphics[height=50mm,width=150mm] {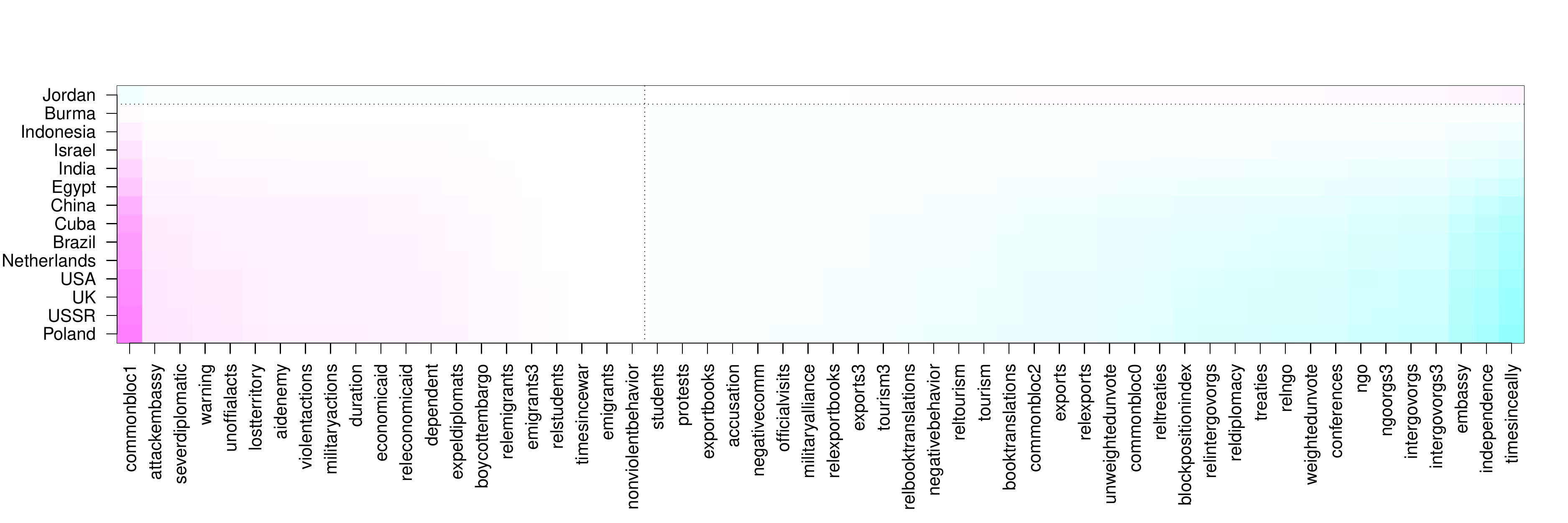}
  \caption{Heatmap of relation versus nation  for component 4 with regularization.} 
 \label{hmap_4}
     \end{figure}

For component 1 shown in Figure \ref{hmap_1}, 
positive values are associated with international partnerships and
diplomatic relations. All nations have the same signs, so this component can be viewed as a main effect of relations.
For component 2 shown in Figure \ref{hmap_2}, negative values are 
associated with hostile actions, zero values are related to exports
or population exchanges, and positive values are related to
international partnerships. Again, all nations have same the signs, so this component can be viewed as a main effect of relations.
For component 4 shown in Figure \ref{hmap_4}, negative or hostile actions are on the negative end of the relation factor
while international partnerships are on the positive end. The nations can
be grouped into two clusters: one with neutral countries and the other
with the rest countries. 
In order to plot the heatmaps using a common scale for visualization, all entries of the components are divided by the maximum absolute value of all  entries and then scaled to between $-1$ and $1$.

\end{appendix}

\end{document}